\documentclass[12pt]{article}
\usepackage{array}
\usepackage{tabularx}
\usepackage{makecell}
\usepackage{amsmath,amssymb,amsfonts,epsfig,cite,setspace,bigstrut,framed}
\usepackage[all]{xy}
\usepackage{color}
\usepackage{pifont}
\usepackage{xcolor}
\usepackage{tikz}
\usepackage{tikz-cd}
\usepackage[export]{adjustbox}
\usetikzlibrary{shapes.geometric}
\usepackage[colorlinks=true,urlcolor=blue,anchorcolor=blue,citecolor=blue,filecolor=blue,linkcolor=blue,menucolor=blue,linktocpage=true,pdfproducer=medialab,pdfa=true]{hyperref}	 
\usetikzlibrary{positioning}
\usepackage[perpage]{footmisc}
\usepackage[compat=1.1.0]{tikz-feynman}
\usepackage{colortbl}

\usepackage{todonotes}
\usepackage{draft,hyperref,tikz,cancel,subfig,float,commutative-diagrams,multirow}
\usetikzlibrary{snakes}
\usetikzlibrary{shapes.misc}
\usetikzlibrary{cd}
\usepackage{dsfont}

\definecolor{dgreen}{rgb}{0, 0.55, 0}
\definecolor{llightyellow}{rgb}{1.0, 0.95, 0.7}
\definecolor{llightblue}{rgb}{0.7, 0.9, 1.0}
\definecolor{llightpink}{rgb}{1.0, 0.85, 0.95}
\definecolor{llightgreen}{rgb}{0.7, 1.0, 0.4}
\colorlet{lightyellow}{llightyellow!50!white}
\colorlet{lightblue}{llightblue!50!white}
\colorlet{lightgreen}{llightgreen!50!white}
\colorlet{lightpink}{llightpink!50!white}

\usepackage{mathtools}

\newcommand{\dsl}{\pa \kern-0.5em /}

\newcommand{\pa}{\partial}

\newcommand{\bit}{\begin{itemize}}
\newcommand{\eit}{\end{itemize}}

\newcommand{\ba}{\begin{array}}
\newcommand{\ea}{\end{array}}

\makeatletter \@addtoreset{equation}{section} \makeatother

\newcommand{\comment}[1]{}

\newcommand{\bi}{\begin{itemize}}
\newcommand{\ei}{\end{itemize}}
\newcommand{\beq}{\begin{equation}}
\newcommand{\eeq}{\end{equation}}

\begin{document}

\begin{titlepage}
\hspace{\fill} USTC-ICTS/PCFT-24-46\\\\

\begin{center}

\title{SymTFT Approach to 2D Orbifold Groupoids:\\
`t Hooft Anomalies, Gauging, and Partition Functions}

\author{Jin Chen,$^{\dagger,\,\flat}$ and Qiang Jia$^{\natural}$}

\address{\small${}^\dagger$Department of Physics, Xiamen University, Xiamen, 361005, China}
\vspace{-.1in}
\address{\small${}^\flat$Peng Huanwu Center for Fundamental Theory, Hefei, Anhui 230026, China}
\vspace{-.1in}
\address{\small${}^\natural$Department of Physics,
Korea Advanced Institute of Science and Technology,\\
$~~ $Daejeon 34141, Korea}
\vspace{-.1in}

\email{zenofox@gmail.com, qjia1993@kaist.ac.kr}

\end{center}

\vfill

\begin{abstract}
We use the 3D SymTFT approach to study the generalized symmetries and partition functions of 2D CFTs in various orbifolded and fermionic phases. These phases can be realized by the sandwich construction in the associated 3D SymTFTs with different gapped boundaries that encode the data of symmetries in the 2D CFTs. We demonstrate that the gapped boundaries can all be identified with the (fermionic) Lagrangian algebra in the 3D SymTFT, and thus use them to establish webs of dualities of the boundary CFTs in different phases on the level of partition functions. In addition, we introduce the concept of ``para-fermionic Lagrangian algebra" which enables us to construct the partition functions of para-fermionized CFTs on the 2D boundary. Finally, we provide many important examples, including a 3D SymTFT viewpoint on gauging non-invertible symmetries in 2D CFTs.

\end{abstract}

\vfill
\end{titlepage}

\tableofcontents

\section{Introduction and Conclusion}
Global symmetries and the anomalies associated with them have been playing a distinguished role in the study of quantum field theories (QFTs) in strongly coupled regimes. From a modern perspective, global symmetries of a QFT can be regarded as extended topological operators supported on lower dimensional submanifolds of the spacetime\cite{Gaiotto:2014kfa}. They amount to rich topological data of the given QFT, see in ref\cite{Cordova:2018cvg, Benini:2018reh, Antinucci:2022eat,
Bartsch:2022mpm, Choi:2021kmx, Kaidi:2021xfk, Bhardwaj:2022yxj, Choi:2022zal, Kaidi:2022uux, Kaidi:2023maf}. More importantly, due to their topological nature, they are invariant under renormalization group flows, and thus provide valuable and effective tools to help people understand physics in the deep infrared\cite{Chang:2018iay, Choi:2022jqy, Gaiotto:2017yup, Gaiotto:2017tne, Choi:2022rfe, Komargodski:2020mxz}. An elegant approach to utilize the global symmetries is to turn on background gauge fields coupled to them. It enables us to compute 't Hooft anomalies in the weakly coupled regime to constrain the dynamics of the strongly interacting system, such as providing additional selection rules, predicting the possible phases in the routes of the RG-flows, etc. Furthermore, when the global symmetries of a given QFT are free of 't Hooft anomalies, one can perform the gauging operations by making the background gauge field dynamic. It is not only nature's preferred way to introduce fundamental interactions but also connects the old to new theories in the landscape of QFTs. Especially, when the to-be-gauged symmetries are discrete, the corresponding gauge fields are flat and thus have no dynamics. Such operations, known as orbifolding, are genuinely topological manipulations of the given QFTs. The gauged new theories share the same local dynamics as the old ones, but only different in global properties\cite{Gaiotto:2020iye, Bhardwaj:2017xup, Burbano:2021loy, Roumpedakis:2022aik,Bashmakov:2022jtl,Bashmakov:2022uek,Chen:2023qnv, Bashmakov:2023kwo, Chen:2024fno, Duan:2024xbb}.

In the context of two-dimensional QFTs, the ordinary global symmetries can be interpreted as one-dimensional topological defect lines (TDLs) in modern language. In this framework, the notion of symmetries can be easily generalized to the ``non-invertible" ones that are not necessary elements of symmetry groups. Rather, these non-invertible symmetries, together with ordinary ones and the compositions among them all, carry new mathematical structures known as fusion algebras\cite{Frohlich:2004ef,Fuchs:2007tx,Frohlich:2009gb,Petkova:2000ip, Bhardwaj:2017xup,Chang:2018iay,Thorngren:2019iar,Komargodski:2020mxz,Thorngren:2021yso,Choi:2021kmx,Kaidi:2021xfk,Choi:2022zal,Cordova:2022ieu,Choi:2022jqy,Chang:2022hud,Bashmakov:2022uek,Grover:2023loq,Seiberg:2023cdc,Seiberg:2024gek}. In fact, non-invertible symmetries are ubiquitous in two-dimensional conformal field theories as a large class of 2D QFTs. The investigation of these non-invertible symmetries can be traced back to the 80's, known as Verlinde lines in rational CFTs, and the collections of them are referred to as fusion category symmetries of the CFTs. Like ordinary invertible symmetries, one can study a generalized notion of 't Hooft anomalies of these non-invertible topological defects, and introduce corresponding gauge field to gauge them when they ease to anomalies. In terms of TDLs, an equivalent way to view the orbifolded phase $\mathfrak{T}/\mathcal A$ of a CFT $\mathfrak{T}$ respect to the gaugable set $\mathcal A$ of TDLs is to insert a fine-enough $\mathcal A$-meshed network on the 2D spacetime manifold, where the to-be-gauged set $\mathcal A$ is known as Frobenius algebra of the fusion category\cite{Diatlyk:2023fwf,Perez-Lona:2023djo,Perez-Lona:2024sds}. An interesting feature of $\mathfrak{T}/\mathcal A$ is that there are always new TDLs, due to the Wilson lines of the gauge fields, in the resulting orbifolded theory, and they furnish new fusion category symmetries therein. The generalization of symmetries and their orbifolding has extended the notion of gauging a group symmetry to a fusion category. In addition to the orbifolding procedure mentioned above, one can also fermionize a 2D bosonic system or vice versa. In the modern language, it is equivalent to stack a fermionic $\mathbb Z_2$-SPT phase and perform the $\mathbb Z_2$ orbifolding\cite{Hsieh:2020uwb, Kulp:2020iet}.  A paradigm of fermionization is the Jordan-Wigner transformation of a 2D Ising model to a free Majorana fermion. Such operations are recently generalized to the transformation from bosonic ones to para-fermionic systems by stacking para-fermionic $\mathbb Z_N$-SPT phases and orbifolding\cite{Thorngren:2019iar, Yao:2020dqx}. We denote generalized orbifoldings for these topological operations of a given CFT $\mathfrak{T}$ with fusion category symmetry $\mathcal C$. Therefore one would ask the following questions: 1. How to find all possible generalized orbifolding theories $\mathfrak{T}/\mathcal A$ for every gaugable set $\mathcal A\subset\mathcal C$; 2. To determine the fusion category symmetries of these $\mathfrak{T}/\mathcal A$'s; 3. Explicitly establish the relations between the orbifolded $\mathfrak{T}/\mathcal A$ and the original theory $\mathfrak{T}$ on the level of partition functions.

The above problems have been investigated throughout in recent years by a bootstrap style analysis on two-dimensional spacetime, see for example \cite{Thorngren:2021yso, Diatlyk:2023fwf}. On the other hand, however, a key observation to tackle these problems is to notice that the generalized orbifolding operations are topological and thus independent of the detailed dynamics of the underlying theories. Instead, the orbifolding properties essentially depend only on the fusion category symmetries and their associated 't Hooft anomalies that many different concrete theories may share. The data for generalized orbifoldings of 2D theories can be remarkably characterized by a three-dimensional topological field theory known as Symmetry Topological Field Theory (SymTFT). The SymTFT has played a key role in recent developments \cite{Gaiotto:2020iye,Apruzzi:2021nmk,Lin:2022dhv,Kaidi:2022cpf,Kaidi:2023maf,Zhang:2023wlu,Bhardwaj:2023ayw,Bartsch:2023wvv,Antinucci:2023ezl,Cordova:2023bja,Duan:2023ykn,Cao:2023rrb,Baume:2023kkf,Bhardwaj:2023idu,Brennan:2024fgj,Antinucci:2024zjp,Bonetti:2024cjk,GarciaEtxebarria:2024jfv,Apruzzi:2024htg,Antinucci:2024bcm,Heckman:2024zdo,Bhardwaj:2024xcx,Bhardwaj:2024kvy,Choi:2024tri,Choi:2024wfm,Antinucci:2024ltv,Bhardwaj:2024igy,Freed:2022qnc}, and arises naturally from geometric engineering or holography \cite{Apruzzi:2022rei,Heckman:2022xgu,Antinucci:2022vyk,vanBeest:2022fss,Bashmakov:2023kwo,Chen:2023qnv,Putrov:2024uor,Argurio:2024oym,Braeger:2024jcj,DelZotto:2024tae,GarciaEtxebarria:2024fuk,Bergman:2024aly,Yu:2023nyn,Franco:2024mxa,Tian:2024dgl}. 
A SymTFT strips off the symmetries from the local degrees of freedom of its associated 2D theories, and thus help us understand the general orbifolding procedures in a unified fashion. 

A typical example known a long time ago is to consider a 2D theory $\mathfrak{T}$ with invertible finite group symmetry $G$ with possible 't Hooft anomaly captured by an element $\mu_3\in H^3(G,{\rm U(1)})$. To gauge a subgroup $H\subset G$, one needs to find a trivialization of $\mu_3$ restricted to $H$, i.e. $\delta\nu_2=\mu_3|_{H}$, otherwise $H$ will suffer from anomalies and is obstructed to be gauged. The group $G$ together with $\mu_3$ define the twisted 3D Dijkgraaf-Witten (DW) gauge theory ${\rm DW}(G,\mu_3)$ as the SymTFT of the 2D theories with group symmetry $G$. Suppose that $H\subset G$ is gaugable, one can have orbifolded phases, denoted as $\mathfrak{T}/H$, of the original 2D theory $\mathfrak{T}$. 

\begin{figure}
    \centering
    \includegraphics[scale=0.7]{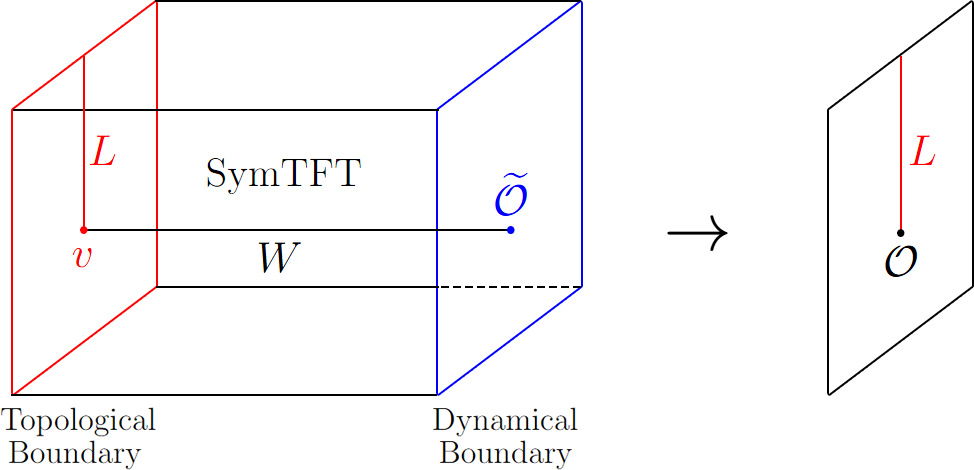}
    \caption{From the SymTFT point of view, a (disorder) operator attached to a symmetry generator $L\in \mathcal{S}$ can be expanded into a line operator $W$ stretching between two boundaries. It ends on $\widetilde{O}$ at the dynamical boundary, and transits to a symmetry defect $L$ via the topological junction $v$ at the topological boundary. This picture also applies to extended operators.}
    \label{fig:SymTFT}
\end{figure}

In the picture of SymTFT, both $\mathfrak{T}$ and $\mathfrak{T}/H$ can be described uniformly. The 2D theory $\mathfrak{T}$ on $M_2$ is equivalent to the 3D DW theory on $M_2 \times [0,1]$ with two boundaries. The information of symmetry $G$ and the dynamics of the 2d CFT are separately stored in the two boundaries. The left boundary is the topological (symmetry) boundary $\mathfrak{B}^{\textrm{sym}}_{G}$ supporting the symmetry $G$ and all symmetry manipulations take place on this boundary. The right boundary is the dynamical (physical) boundary $\mathfrak{B}^{\textrm{phys}}_{\mathfrak{T}}$ that depends on the details of $\mathfrak{T}_{\mathcal{S}}$. A (twisted) operator $\mathcal{O}$ in the 2d theory $\mathfrak{T}$ can be expanded as a line operator $W$ in the bulk stretching between the two boundaries as explained in figure \ref{fig:SymTFT}. The SymTFT provides a unified picture to describe the gauging by fixing the dynamical boundary $\mathfrak{B}^{\textrm{phys}}_{\mathfrak{T}}$ while changing the topological boundary $\mathfrak{B}^{\textrm{sym}}_{G}$ to $\mathfrak{B}^{\textrm{sym}}_{G/H}$. 

Therefore the question about gauging is converted to the question about classifying topological boundaries in the SymTFT. Each topological boundary corresponds to a line operator $\mathcal L$ written as
    \begin{equation}
        \mathcal{L} = \bigoplus_{\alpha} \mathcal{N}_{\alpha} W_{\alpha}\notag\, ,
    \end{equation}
where $W_{\alpha}$ are the simple line operators, namely the line operators that cannot be decomposed, in DW theory and $\mathcal{N}_{\alpha}$ are non-negative positive integers. It defines a gapped boundary that only the line operators with zero coefficient $\mathcal{N}_{\alpha}$ can end on without introducing the symmetry defect $L$ in figure \ref{fig:SymTFT}. Mathematically, such a line operator is the algebraic object of the Lagrangian algebra whose defining properties will be reviewed in the next section. Suppose $M_2$ is a compact Riemann surface, torus for example. We can prepare the vacuum of the topological boundary state $|\Omega\rangle_{\mathcal L}$ and furnish a complete basis $|i\rangle_{\mathcal{L}}$ (with $|0\rangle_{\mathcal{L}} \equiv |\Omega\rangle_{\mathcal L}$) by acting line operators and performing modular transformation. A consequence in SymTFT is that one can neatly organize the 2D partition functions of $\mathfrak{T}$ in twisted sectors as coefficients of the basis $|i\rangle_{\mathcal L}$. It thus leads to a physical boundary state
\begin{align}
|\chi\rangle_{\mathcal L}=\sum_{i,j} g^{i j} Z_{\mathfrak{T}}[i]\,|j\rangle_{\mathcal L}\notag\, ,
\end{align}
where the metric is defined as $g_{i j} = \null_{\mathcal{L}}\langle i | j \rangle_{\mathcal{L}}$ and $g^{i j}$ is the inverse.
When one sandwiches the physical boundary $|\chi\rangle_{\mathcal L}$ with a topological boundary state $|i\rangle_{\mathcal L}$, it by construction gives the desired 2D partition functions
\begin{align}
_{\mathcal L}\langle i |\chi\rangle_{\mathcal L}=Z_{\mathfrak{T}}[i]\notag \,.
\end{align}
One can prepare other topological boundary states $|\Omega\rangle_{\widetilde{\mathcal L}}$ with $\widetilde{\mathcal{L}}$, the descended line operators $\mathcal W_\alpha$ on the boundary are thus different accordingly, but still furnish a complete set of the basis of the same 3D SymTFT. Applying the sandwiching procedure again, one can then establish the relations between different orbifolded 2D theories.

So far we have discussed the 3D SymTFT implementing a group-like symmetry. One can also apply the SymTFT method to arbitrary dimensional theory $\mathfrak{T}_{\mathcal{S}}$ with a global symmetry $\mathcal{S}$. Here $\mathcal{S}$ is not necessarily a finite group-like symmetry and one can consider finite higher-group, non-invertible, or subsystem symmetry, etc. There are also proposals for the extension to continuous symmetries\cite{Brennan:2024fgj,Antinucci:2024zjp,Bonetti:2024cjk,Jia-Wang-Tian-Zhang,Yu:2024jtk}. In the following, we shall use the phrase \emph{symmetry category} $\mathcal{S}$ as suggested in \cite{Bhardwaj:2017xup} which covers all kinds of generalized symmetries. 

In this paper, we will apply the powerful 3D SymTFT machinery to understand various generalized orbifolding operations of 2D CFTs\footnote{The 3D SymTFT approach can be applied to generic QFTs with different symmetry categories. We here restrict ourselves to the case of CFTs only for convenience to compute their partition functions}. Although the sandwich construction has been known in the literature for quite a while, to the authors' knowledge, the explicit connections of a CFT $\mathfrak{T}$ in different generalized orbifolded phases, especially for the gauging of categorical symmetries, have yet to be discussed on the level of partition functions. One of the aims of the paper is to close this gap. Furthermore, we also demonstrate that many topological data of the boundary CFTs, such as F-moves and half-braidings of the TDLs can be easily computed in the framework of 3D SymTFT.

For a 2D CFT with the symmetry given $\mathcal{S}$, one can construct its corresponding SymTFT known as the Turaev-Viro TQFT. Furthermore, the concept of Lagrangian algebras has also been generalized to fermionic Lagrangian algebras very recently. Utilizing it one can prepare fermionic topological boundary states, and thus capture the super-fusion category symmetries of a fermionic CFT which can be either fermionized from a bosonic CFT, or orbifolded from another fermionic one. For sure all these 2D CFTs share the same 3D SymTFT. However, on the boundary CFT, the symmetry category in fermionic theory allows for fermionic zero modes existing both on the 3-way junctions of TDLs, and some special TDLs known as $q$-type ones, e.g. $(-1)^{F_L}$ in Majarona fermions \cite{Aasen:2017ubm}. The symmetry category equipped with such structures is known as the super-fusion category. It is very tempting to understand such a symmetry category from the perspective of its 3D SymTFTs and the associated fermionic Lagrangian algebras, which is one of the future directions to pursue. Another straightforward application of the machinery we developed in this note is to study various (fermionic) modular invariant partition functions in the 2D rational CFTs with $A_k$-type affine Kac-Moody algebras, where the $A_1$ and $A_2$ cases was just classified very recently from a pure 2D perspective\cite{Kawabata:2024hzx}. We will investigate its classification problem from a 3D SymTFTs viewpoint in the near future.

Besides, one can also generalize the machinery to higher dimensions like the case of 3D/4D setups\cite{Argurio:2024oym,Cui:2024cav,Bergman:2024its}, that is to construct 4D SymTFTs to capture the symmetry categories of 3D QFTs which contain both 0-form and 1-form generalized symmetries. In this situation, orbifolding the 0-form symmetries of one theory will result in another theory with emergent 1-form symmetries, or vice versa. It's very interesting to develop the 4D SymTFTs systematically and establish the duality webs for their corresponding 3D QFTs. More examples of 6D dualities and 7D SymTFT can be found in \cite{Gukov:2020btk,Hsin:2021qiy,Apruzzi:2024cty}.

The plan of the paper is as follows. In section 2, we will review the quantization of SymTFT on a torus and illustrate how to implement dualities. We will consider the duality web of 2D $\mathbb{Z}_2$ symmetry as a toy example. In section 3, we include many important and non-trivial examples to demonstrate the generalized orbifoldings in 2D CFTs from a 3D SymTFT perspective. Especially, we discussed the generalized orbifolding structures of a 2D CFT with potentially anomalous $\mathbb Z_N$ symmetry. When the $\mathbb Z_N$-symmetry is non-anomalous, we classify various Lagrangian algebras in it. They allow us to establish many relations among different orbifolded partition functions on the boundary CFT. On the other hand, for the case of even $N$, when $\mathbb Z_2\subset \mathbb Z_N$ is non-anomalous but $\mathbb Z_N$ could be anomalous, we classify the structures of symmetries on the 2D boundaries in both bosonic and fermionic phases corresponding to gauging $\mathbb{Z}_2$ symmetry. In addition, we also generalized the concept of fermionic Lagrangian algebra to define the \emph{para-fermionic Lagrangian algebra}, which on the boundary demonstrates the procedure of para-fermionization of the 
original bosonic CFT\cite{Fateev:1985mm, Yao:2020dqx, Thorngren:2019iar, Duan:2023ykn, Chen:2023jht}. Besides those invertible symmetry cases, we also include examples of SymTFTs for fusion categories containing non-invertible symmetries, e.g. $\mathbb S_3$, Ising and Lee-Yang categories. Especially, in our framework, we explain the simplest non-trivial example of gauging a fusion category, say the Lee-Yang category, from the viewpoint of SymTFT.

At the final stage of this work, we become aware of another work \cite{Huang:2024ror} in the condensed matter field which shares similar ideas to ours.

\section{SymTFT and topological boundary states}

In this section, we will begin with a review of the quantization of SymTFT on tours, Lagrangian algebras, and topological boundaries. After that, we will develop a general method to construct the topological boundary states corresponding to a given Lagrangian algebras on the torus using the modular data of the SymTFT. It can be applied to extract useful information such as dualities, F-moves, and half-braidings of the underlying 2D theory. At the end of this section, we will illustrate the construction for the simplest example, the SymTFT for $2d$ $\mathbb{Z}_2$ symmetry, and discuss the dualities. We will provide more examples in the next sections.

\begin{figure}
        \centering
    \includegraphics[scale=1]{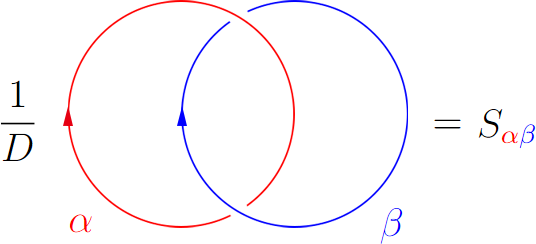}
    \caption{The $S$-matrix between two line operators is represented as the correlation function of two line operators linked with each other. Notice that the way of linking in this paper is chosen differently from that in \cite{Barkeshli:2014cna}.}
        \label{fig:S-matrix}
\end{figure}

We will focus on a 2D theory $\mathfrak{T}_{\mathcal{S}}$ whose global symmetry is given by the category $\mathcal{S}$, and the SymTFT is the 3-dimensional Turaev-Viro TQFT $\textrm{TV}(\mathcal{S})$. The line operators (anyons) in the Turaev-Viro TQFT form a modular tensor category called the Drinfeld center $\mathcal{Z}(\mathcal{S})$ (or quantum double). One may consult \cite{Barkeshli:2014cna} for a systematic review of anyon system. We label the simple line operators as $W_{\alpha}$ with $\alpha=0,\cdots,N-1$. Here $N$ is the number of simple line operators and $\alpha=0$ is the identity line. We will also use $W_{\bar{\alpha}}$ to denote the reversal of $W_{\alpha}$. The modular data of the 3D TQFT can be described by $S$ and $T$ matrices and we will denote the components by
    \begin{equation}
        S_{\alpha \beta} , T_{\alpha \beta},\quad \alpha,\beta = 0,\cdots, N-1,
    \end{equation}
$S$-matrix is related to the correlation function of line operators shown in figure \ref{fig:S-matrix} and $T$-matrix is a diagonal matrix
    \begin{equation}
        T = \textrm{Diag} \left(\theta_0,\theta_1,\cdots,\theta_{N-1} \right)
    \end{equation}
where $\theta_{\alpha}$ is the topological spin of each line operator.
The line operators satisfy the fusion rule
    \begin{equation}
        W_{\alpha} \times W_{\beta} = \sum_{\gamma} N_{\alpha \beta}^{\gamma} W_{\gamma},
    \end{equation}
where the fusion coefficients $N_{\alpha \beta}^{\gamma}$ are non-negative integers. According to the Verlinde formula, the fusion coefficients $N_{\alpha \beta}^{\gamma}$ can be given as
    \begin{equation}\label{Fusion-coefficient}
        N_{\alpha \beta}^{\gamma} = \sum_{\delta=0}^{N-1} \frac{S^*_{\delta \alpha} S^*_{\delta \beta} S_{\delta \gamma}}{S_{0\delta}},
    \end{equation}
or reversely as
    \begin{equation}\label{S-matrix}
        S^*_{\alpha \beta} = \frac{1}{D} \sum_{\gamma} N_{\bar{\alpha} \beta}^{\gamma} \frac{\theta_{\gamma}}{\theta_{\alpha} \theta_{\beta}} d_{\gamma}.
    \end{equation}
Here $d_{\gamma}$ is the quantum dimension of $\gamma$-th line operator (which can be negative for non-unitary theory) and $D$ is the total quantum dimension of $\mathcal{Z}(\mathcal{S})$ defined as
    \begin{equation}\label{total-quantum-dimension}
        D = \sqrt{\sum_{\alpha} d^2_{\alpha}}.
    \end{equation}
Setting $\alpha=0$ one can read out the quantum dimensions from the first line (or row) from $S$-matrix
    \begin{equation}
        d_{\beta} = D S_{0\beta}.
    \end{equation}

\begin{figure}
    \centering
    \includegraphics[width=0.5\linewidth]{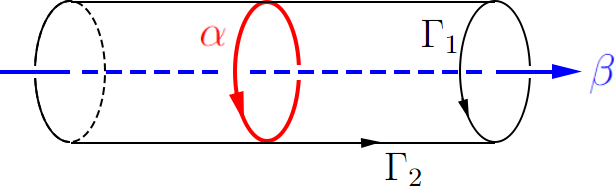}
    \caption{A segment of the solid torus $D^2 \times S^1$ where $\Gamma_1$ is shrinkable. The state $|\beta\rangle$ is prepared by inserting $W_{\beta}[\Gamma_2]$ along the $\Gamma_2$ cycle inside the solid torus $D^2 \times S^1$ and the action of $W_{\alpha}[\Gamma_1]$ can be read from the $S$-matrix.}
    \label{fig-state}
\end{figure}

Let us put the theory on a $\mathcal{M}_g \times \mathbb{R}$ where $\mathcal{M}_g$ is a 2D Riemann surface. For simplicity, we will consider $g=1$ such that the space is a torus. We will also denote the two 1-cycles on the torus $S^1 \times S^1$ as $\Gamma_1$ and $\Gamma_2$, and denote the $\alpha$-th line operator wrapping along $\Gamma_1,\Gamma_2$ separately as $W_{\alpha}[\Gamma_1]$ and $W_{\alpha}[\Gamma_2]$. Consider the solid torus $D^2 \times S^1$ where the first 1-cycle $\Gamma_1$ is shrinkable. The Hilbert space $\mathcal{H}(T^2)$ on the torus is constructed by inserting a line operator along $\Gamma_2$ at the center of $D^2$ and do the path-integral, see figure \ref{fig-state} for the illustration and the appendix of \cite{Gang:2021hrd} for a review. 
Therefore a natural basis is labelled as $|\alpha \rangle$ with $\alpha = 0,\cdots,N-1$ and they are orthogonal to each other
\begin{equation}
    \langle \alpha | \beta \rangle = \delta_{\alpha \beta}.
\end{equation}
In particular, $|0\rangle$ is the vacuum with nothing inserted in the bulk. The line operators act on Hilbert space $\mathcal{H}(T^2)$ as
    \begin{equation}\label{action-of-line-operators}
        W_{\alpha}[\Gamma_1] |\beta \rangle = \frac{S_{\beta \alpha}}{S_{\beta 0}} |\beta \rangle,\quad W_{\alpha}[\Gamma_2] |\beta \rangle = \sum_{\gamma} N_{\beta \alpha}^{\gamma} |\gamma \rangle.
    \end{equation}
The $S$ and $T$ matrices are given by
    \begin{equation}
        S_{\alpha \beta} = \langle \alpha| \hat{\mathbb{S}} |\beta\rangle,\quad T_{\alpha \beta} = \langle \alpha| \hat{\mathbb{T}} |\beta\rangle
    \end{equation}
where $\hat{\mathbb{S}}$ and $\hat{\mathbb{T}}$ are two canonical generators of $SL(2,\mathbb{Z})$ acting on the two 1-cycles according to
    \begin{equation}
        \hat{\mathbb{S}} = \left( \begin{array}{cc}
            0 & -1 \\
            1 & 0
        \end{array} \right),\quad \hat{\mathbb{T}} = \left(\begin{array}{cc}
            1 & 1 \\
            0 & 1
        \end{array} \right)\, ,
    \end{equation}
where $S$-transformation switch $\Gamma_1$ with $\Gamma_2$ and $T$-transformation shift $\Gamma_2 \rightarrow \Gamma_2 + \Gamma_1$.

\begin{figure}
    \centering
    \includegraphics[scale=0.8]{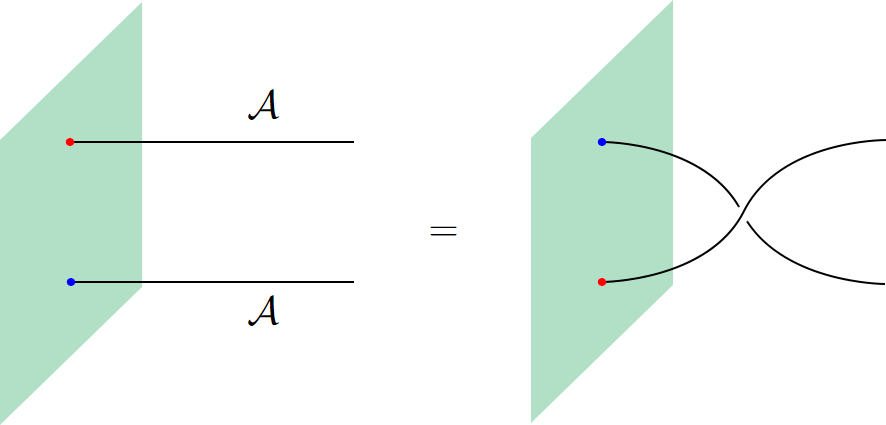}
    \caption{The Lagrangian algebra should be commutative as line operators.}
    \label{fig-Lagrangian-algebra}
\end{figure}

In SymTFT the topological boundary $\mathfrak{B}^{\textrm{sym}}_{\mathcal{S}}$ is defined by specifying line operators that can end on the boundary $\mathfrak{B}^{\textrm{sym}}_{\mathcal{S}}$, see section 4.5.2 in \cite{Bhardwaj:2023ayw} for a short review. In mathematics, the collection of those line operators is denoted as the object of a Lagrangian algebra. To be explicit, denote $\mathcal{N}_{\alpha}$ as the dimension of vector space carried by the endpoint of the simple line operator $W_{\alpha}$ ending on the topological boundary. In particular, if $\mathcal{N}_{\alpha}=0$ that means the line operator $W_{\alpha}$ cannot end at the boundary. We can consider the non-simple line operator
    \begin{equation}
        \mathcal{L} = \bigoplus_{\alpha} \mathcal{N}_{\alpha} W_{\alpha}\, ,
    \end{equation}
which is required to be associative. The key feature for Lagrangian algebra is that $\mathcal{L}$ must be commutative so that it can end at the boundary unambiguously as shown in figure \ref{fig-Lagrangian-algebra}.
Another important condition that is extremely helpful is that the quantum dimension of $\mathcal{L}$ should equal the total quantum dimension\eqref{total-quantum-dimension}
    \begin{equation}\label{L-dimension}
        \dim (\mathcal{L}) = \sum_{\alpha} \mathcal{N}_{\alpha} d_{\alpha} = D.
    \end{equation}
In the following, we will also call the line operator $\mathcal{L}$ as Lagrangian algebra without introducing any misunderstanding. To construct the states on the torus for a given topological boundary corresponding to the Lagrangian algebra $\mathcal{L}$, we insert the whole line operator $\mathcal{L} = \bigoplus_{\alpha} \mathcal{N}_{\alpha} W_{\alpha}$ along $\Gamma_2$-cycle and sitting at the origin of $D^2$. After path-integral it will produce a state $|\Omega\rangle_{\mathcal{L}} \in \mathcal{H}(T^2)$ which reads
    \begin{equation}
        |\Omega\rangle_{\mathcal{L}} = \sum_{\alpha} \mathcal{N}_{\alpha} |\alpha\rangle.
    \end{equation}
We propose that $|\Omega\rangle_{\mathcal{L}}$ is the vacuum of the topological boundary state corresponding to the Lagrangian algebra $\mathcal{L}$.

It is usually not easy to find the Lagrangian algebra for a given SymTFT based on the definition. However, there exist some necessary conditions that $|\Omega\rangle_{\mathcal{L}}$ should satisfy if we wish to identify it as the vacuum of the topological boundary state. Firstly, the vacuum should remain invariant under $S$-transformation so that
    \begin{equation}
        \hat{\mathbb{S}} |\Omega\rangle_{\mathcal{L}} = |\Omega\rangle_{\mathcal{L}}.
    \end{equation}
Secondly, under $T^2$-transformation the vacuum should also be invariant
    \begin{equation}
        \hat{\mathbb{T}}^2 |\Omega\rangle_{\mathcal{L}} = |\Omega\rangle_{\mathcal{L}}.
    \end{equation}
For bosonic vacuum $|\Omega\rangle_{\mathcal{L}}$ should be invariant under $T$-transformation. On the other hand, for fermionic vacuum since $T$-transformation will switch between (NS,NS) and (NS,R) sectors, where the first and second NS/R in the bracket stands for the boundary condition along $\Gamma_1$ and $\Gamma_2$. So one expects the the vacuum $|\Omega\rangle_{\mathcal{L}}$ is only invariant under $T^2$-transformation. Combined with \eqref{L-dimension} we can write down the candidates for the Lagrangian algebra.

Given the topological boundary, if we restrict a simple line operator $W_{\alpha}$ in the bulk onto the topological boundary there might be two possibilities
    \begin{itemize}
        \item $W_{\alpha}$ can be identified as a simple object in the symmetry category $\mathcal{S}_{\mathcal{L}}$ corresponding to $\mathcal{L}$.
        \item $W_{\alpha}$ is not simple in the symmetry category $\mathcal{S}_{\mathcal{L}}$ and is composed of several simple objects.
    \end{itemize}
To examine which case it belongs to, one needs the following identity
    \begin{equation}
        \textrm{Hom}_{\mathcal{L}}(W_{\alpha},W_{\beta}) = \textrm{Hom}_{\mathcal{Z}(\mathcal{S})} (W_{\alpha},W_{\beta} \times \mathcal{L})
    \end{equation}
where $W_{\alpha},W_{\beta}$ are any two simple line operators in the bulk SymTFT and we use $\textrm{Hom}_{\mathcal{L}}$ to denote the homomorphism on the topological boundary defined by $\mathcal{L}$, see appendix B in \cite{Bhardwaj:2023idu}. If the dimension of $\textrm{Hom}_{\mathcal{L}}(W_{\alpha},W_{\alpha})$ is one, then it implies $W_{\alpha}$ is simple restricting to the boundary, otherwise it composed several simple objects. 

In terms of the vacuum state $|\Omega\rangle_{\mathcal{L}}$, we can examine the action of line operators upon the Hilbert space $\mathcal{H}_{T^2}$ following \eqref{action-of-line-operators} and it helps us retrieve lots of information encoded in the SymTFT easily. If two of them give the same result, for example
    \begin{equation}
        W_{\alpha}[\Gamma_i]|\Omega\rangle_{\mathcal{L}} = W_{\beta}[\Gamma_i]|\Omega\rangle_{\mathcal{L}}
    \end{equation}
for both $i=1,2$, then it indicates $W_{\alpha}$ and $W_{\beta}$ project to the same TDLs in $\mathcal{S}_{\mathcal{L}}$ on the topological boundary. It is easy to check
    \begin{equation}\label{Hom}
        \textrm{Hom}_{\mathcal{L}}(W_{\alpha},W_{\beta}) = \langle \alpha | W_{\beta}[\Gamma_2] |\Omega\rangle_{\mathcal{L}},
    \end{equation}
and it determines whether the line operator is simple or not on the topological boundary. We can obtain the fusion rules by acting two line operators along the same 1-cycle upon the topological boundary, which helps us deduce the symmetry $\mathcal{S}_{\mathcal{L}}$ on the topological boundary. Moreover, by acting two line operators along different 1-cycles we can read out the F-move and half-braiding structures.

Before moving on we need to clarify a possible misunderstanding. Two bulk line operators project to the same line operators at the topological boundary does not mean they are equivalent at the boundary. They give rise to operators attaching to the same symmetry defect but transforming in a different way under the symmetry category $\mathcal{S}_{\mathcal{L}}$ in the sandwich picture depicted in figure \ref{fig:SymTFT}. In particular, if $S_{\mathcal{L}}$ is a group-like symmetry the two operators are in different irreducible representations.

Based on the vacuum $|\Omega\rangle_{\mathcal{L}}$ we can construct another basis of the Hilbert space $\mathcal{H}_{T^2}$ labeled as $|i\rangle_{\mathcal{L}}$ with $i=0,\cdots,N-1$, where we use Latin characters and subscript $\mathcal{L}$ in order to distinguish with the basis $|\alpha\rangle$ in \eqref{action-of-line-operators}. We will denote $|0\rangle_{\mathcal{L}} \equiv |\Omega\rangle_{\mathcal{L}}$ and other states are constructed by acting bulk line operators and performing modular transformation. Introduce the metric $g_{ij}$ as
    \begin{equation}
        g_{ij} = \,_{\mathcal{L}}\langle i | j \rangle_{\mathcal{L}},\quad i,j=0,\cdots,N-1.
    \end{equation}
The states $|i\rangle_{\mathcal{L}}$ are one-to-one correspondent to the background of the symmetry $S_{\mathcal{L}}$ on the torus. 

Given any 2d theory $\mathfrak{T}_{\mathcal{S}_{\mathcal{L}}}$ whose symmetry category $\mathcal{S}_{\mathcal{L}}$, we can construct the dynamical boundary state (partition vector) on the torus as
    \begin{equation}
        |\chi\rangle_{\mathfrak{T}_{\mathcal{S}_{\mathcal{L}}}} = \sum_{i,j} g^{ij} Z_{\mathfrak{T}_{\mathcal{S}_{\mathcal{L}}}} [i] |j\rangle_{\mathcal{L}},
    \end{equation}
where the coefficients are the partition functions of $\mathfrak{T}_{\mathcal{S}_{\mathcal{L}}}$ for different sectors on the torus and $g^{ij}$ is the inverse of the metric. We can recover the partition functions trivially as
    \begin{equation}
        Z_{\mathfrak{T}_{\mathcal{S}_{\mathcal{L}}}}[i] = \,_{\mathcal{L}}\langle i |\chi\rangle.
    \end{equation}
If we have another Lagrangian algebra $\widetilde{\mathcal{L}}$ and the corresponding topological boundary states $|\tilde{i}\rangle_{\widetilde{\mathcal{L}}}$. We can obtain a new set of partition functions $Z_{\mathfrak{T}_{\mathcal{S}_{\widetilde{\mathcal{L}}}}}[\tilde{i}]$ by fixing the dynamical boundary state $|\chi\rangle$ while changing the topological boundary state
    \begin{equation}
        Z_{\mathfrak{T}_{\mathcal{S}_{\widetilde{\mathcal{L}}}}}[\tilde{i}] = \,_{\widetilde{\mathcal{L}}}\langle \tilde{i} |\chi\rangle.
    \end{equation}
Physically, they are the partition functions of the dual theory $\mathfrak{T}_{\mathcal{S}_{\widetilde{\mathcal{L}}}}$ obtained by gauging the symmetry $\mathcal{S}_{\mathcal{L}}$ or part of $\mathcal{S}_{\mathcal{L}}$. 

It is best to illustrate the basic idea of SymTFT and the construction of topological boundary states via a simple example. Consider a $(1+1)$d theory $\mathfrak{T}_{\mathbb{Z}_2}$ with $\mathbb{Z}_2$ symmetry. The corresponding SymTFT is the $(2+1)$-dimensional BF theory with level $2$
    \begin{equation}
        S_{BF} = \frac{2}{2\pi} \int \widetilde{A} \wedge d A ,
        \end{equation}
where $\widetilde{A},A$ are 1-form gauge fields. The gauge invariant operators are Wilson loops defined as
    \begin{equation}
        W[\Gamma] = \exp \left(i \oint_{\Gamma} A \right),\quad \widetilde{W}[\Gamma] = \exp \left(i \oint_{\Gamma} \widetilde{A} \right),
    \end{equation}
for $A$ and $\widetilde{A}$. The $S-$matrix element between two kinds of line operators can be obtained by computing the correlation function using path integral
    \begin{equation}
        \langle W[\Gamma] \widetilde{W}[\Gamma'] \rangle = \int \mathcal{D}A \mathcal{D}\widetilde{A} e^{i \frac{2}{2\pi} \int \widetilde{A}\wedge d A + i \int \eta(\Gamma) \wedge A + i \int \eta(\Gamma') \wedge \widetilde{A}},
    \end{equation}
where $\eta(\Gamma)$ is a 2-form and is the Poincare dual of $\Gamma$. Using the relation
    \begin{equation}
        \int \eta(\Gamma) \wedge A = \int d \eta(D_{\Gamma}) \wedge A = \int \eta(D_{\Gamma}) \wedge d A,
    \end{equation}
where $D_{\Gamma}$ is the 2-surface bounded by $\Gamma$ and $d \eta(D_{\Gamma}) = \eta(\partial D_{\Gamma}) = \eta(\Gamma)$. The phase on the exponent can be written as
    \begin{align}
        &\frac{2}{2\pi} \int \widetilde{A}\wedge d A + \int \eta(\Gamma) \wedge A + \int \eta(\Gamma') \wedge \widetilde{A} \nonumber \\
        =& \frac{2}{2\pi} \int \left(\widetilde{A} + \frac{2\pi}{2} \eta(D_{\Gamma}) \right)\wedge d \left(A + \frac{2\pi}{2} \eta(D_{\Gamma'})\right) - \frac{2\pi}{2} \int \eta(D_{\Gamma}) \wedge d \eta(D_{\Gamma'}).
    \end{align}
Absorb the first two factors into $A$ and $\widetilde{A}$, one gets
\begin{equation}
    \langle W[\Gamma] \widetilde{W}[\Gamma'] \rangle = \langle W[\Gamma]\rangle \langle \widetilde{W}[\Gamma'] \rangle e^{-\frac{2\pi i}{2} \int \eta(D_{\Gamma}) \wedge d \eta(D_{\Gamma'})},
\end{equation}
where the second term gives the linking number between $\Gamma$ and $\Gamma'$. Assuming both $\Gamma,\Gamma'$ are shrinkable so that $\langle W[\Gamma]\rangle= \langle \widetilde{W}[\Gamma'] \rangle=1$, we can immediately read the $S$-matrix element (up to $D^{-1}$) between $W$ and $\widetilde{W}$ is $-1$. Moreover, since $W^2$ and $\widetilde{W}^2$ link other line operators trivially, we can identify
    \begin{equation}
        W[\Gamma]^2 = \widetilde{W}[\Gamma]^2 = 1.
    \end{equation}
Moreover, both $W$ and $\widetilde{W}$ are bosonic and the topological spins are zero.

We will denote generic line operators using a pair of $\mathbb{Z}_2$-valued indices $\alpha,\tilde{\alpha} = 0,1$
    \begin{equation}
        W_{(\alpha,\tilde{\alpha})} [\Gamma] = \exp \left(i \oint_{\Gamma} \alpha A + \tilde{\alpha} \widetilde{A} \right) = W[\Gamma]^{\alpha} \widetilde{W}[\Gamma]^{\tilde{\alpha}}.
    \end{equation}
The quantum dimension of each line operator is one so the total quantum dimension is $D=2$. They satisfy the fusion rule
    \begin{equation}
        W_{(\alpha,\tilde{\alpha})} \times W_{(\beta,\tilde{\beta})} = W_{(\alpha+\beta,\tilde{\alpha}+\tilde{\beta})},
    \end{equation}
where $\alpha+\beta$ is also evaluated within $\mathbb{Z}_2$ . The elements of $S$ and $T$ matrices are
    \begin{equation}\label{Z2-S-and-T}
        S_{(\alpha,\tilde{\alpha}),(\beta,\tilde{\beta})} = \frac{1}{2}(-1)^{\alpha \tilde{\beta} + \tilde{\alpha} \beta},\quad T_{(\alpha,\tilde{\alpha})(\beta,\tilde{\beta})} = (-1)^{\alpha \tilde{\alpha}} \delta_{\alpha,\beta} \delta_{\tilde{\alpha},\tilde{\beta}}.
    \end{equation}
Here both the rows and columns of the $S$ and $T$ matrices are labeled by a pair of $\mathbb Z_2$-valued indices, say $(\alpha, \tilde\alpha
)$ for example. So they are $4\times 4$ matrices.

Let us put the theory on the torus and we will denote the state in the Hilbert space $\mathcal{H}(T^2)$ as $|(\alpha,\tilde{\alpha})\rangle$. They satisfy
    \begin{equation}
        W_{(\beta,\tilde{\beta})}[\Gamma_1] |(\alpha,\tilde{\alpha})\rangle = (-1)^{\alpha \tilde{\beta} + \tilde{\alpha} \beta}|(\alpha,\tilde{\alpha})\rangle,\quad W_{(\beta,\tilde{\beta})}[\Gamma_2] |(\alpha,\tilde{\alpha})\rangle = |(\alpha+\beta,\tilde{\alpha}+\tilde{\beta})\rangle.
    \end{equation}
There exist three kinds of Lagrangian algebras, two of them are bosonic
    \begin{equation}
        \mathcal{L}_{\textrm{Dir}} = W_{(0,0)} \oplus W_{(1,0)},\quad \mathcal{L}_{\textrm{Neu}} = W_{(0,0)} \oplus W_{(0,1)}
    \end{equation}
where Dir(Neu) is short for Dirichlet(Neumann) boundary condition. The other one is fermionic
    \begin{equation}
        \mathcal{L}_f = W_{(0,0)} \oplus W_{(1,1)}
    \end{equation}
where $W_{(1,1)}$ has topological spin half.

\subsubsection*{Topological boundary state for $\mathcal{L}_{\textrm{Dir}}$}
For $\mathcal{L}_{\textrm{Dir}}$ the topological boundary state is
    \begin{equation}
        |\Omega\rangle_{\mathcal{L}_{\textrm{Dir}}} = |(0,0)\rangle + |(1,0)\rangle
    \end{equation}
and it is invariant under both $S$ and $T$ transformations
    \begin{equation}
        \hat{\mathbb{S}} |\Omega\rangle_{\mathcal{L}_{\textrm{Dir}}} = |\Omega\rangle_{\mathcal{L}_{\textrm{Dir}}},\quad \hat{\mathbb{T}} |\Omega\rangle_{\mathcal{L}_{\textrm{Dir}}} = |\Omega\rangle_{\mathcal{L}_{\textrm{Dir}}}.
    \end{equation}
One can check the action of $W_{(\alpha,\tilde{\alpha})}$ along $\Gamma_1$ and $\Gamma_2$ gives
    \begin{equation}
        W_{(\alpha,\tilde{\alpha})}[\Gamma_1]|\Omega\rangle_{\mathcal{L}_{\textrm{Dir}}} = |(0,0)\rangle + (-1)^{\tilde{\alpha}}|(1,0)\rangle,\quad W_{(\alpha,\tilde{\alpha})}[\Gamma_2]|\Omega\rangle_{\mathcal{L}_{\textrm{Dir}}} = |(0,\tilde{\alpha})\rangle + |(1,\tilde{\alpha})\rangle.
    \end{equation}
Therefore $W_{(0,\tilde{\alpha})}$ and $W_{(1,\tilde{\alpha})}$ are equivalent when acting on the vacuum of the topological boundary represented by $|\Omega\rangle_{\mathcal{L}_{\textrm{Dir}}}$. Choose $W_{(0,\tilde{\alpha})}$ as a representative, acting the operators twice we can read the fusion rule of $W_{(0,\beta)}$
    \begin{equation}
        W_{(0,\tilde{\alpha})} \times_{\mathcal{L}_{\textrm{Dir}}} W_{(0,\tilde{\beta})} = W_{(0,\tilde{\alpha}+\tilde{\beta})}
    \end{equation}
where we use $\times_{\mathcal{L}_{\textrm{Dir}}}$ to emphasize it is the fusion rule on the topological boundary. Therefore the line operators restricted on the topological boundary generate a $\mathbb{Z}_2$ symmetry. 

Let us label the twisted sectors of the $\mathbb{Z}_2$ symmetry as $|a_1,a_2\rangle$ where $a_1,a_2$ are the holonomies along $\Gamma_1$ and $\Gamma_2$ cycles. The trivial sector is the vacuum $|0,0\rangle \equiv |\Omega\rangle_{\mathcal{L}_{\textrm{Dir}}}$ and the non-trivial sector can be obtained by acting the symmetry generator $W_{(0,1)}$
    \begin{equation}
        \begin{split}
        |0,1\rangle \equiv W_{(0,1)}[\Gamma_1] |\Omega\rangle_{\mathcal{L}_{\textrm{Dir}}} = |(0,0)\rangle - |(1,0)\rangle, \\
        |1,0\rangle \equiv W_{(0,1)}[\Gamma_2]|\Omega\rangle_{\mathcal{L}_{\textrm{Dir}}} = |(0,1)\rangle + |(1,1)\rangle,
        \end{split}
    \end{equation}
and $|1,1\rangle$ sector is obtained by performing $T$-transformation on the state $|1,0\rangle$
    \begin{equation}
        |1,1\rangle = \hat{\mathbb{T}} |1,0\rangle = |(0,1)\rangle - |(1,1)\rangle
    \end{equation}
or acting $W_{(0,1)}[\Gamma_1] W_{(0,1)}[\Gamma_2]$ on $|0,0\rangle$. 
It is easy to check the topological boundary state $|a_1,a_2\rangle$ is the eigenstate of $W_{(1,0)}$ operators
    \begin{equation}
        W_{(1,0)}[w_1 \Gamma_1 + w_2 \Gamma_2] |a_1,a_2\rangle = (-1)^{w_1 a_1 + w_2 a_2} |a_1,a_2\rangle
    \end{equation}
where $w_1,w_2$ are winding numbers of the line operators along $\Gamma_1$ and $\Gamma_2$.

\subsubsection*{Topological boundary states for $\mathcal{L}_{\textrm{Neu}}$}
The topological boundary state for $\mathcal{L}_{\textrm{Neu}}$ can be obtained similarly. The vacuum is
\begin{equation}
        |\Omega\rangle_{\mathcal{L}_{\textrm{Neu}}} = |(0,0)\rangle + |(0,1)\rangle
    \end{equation}
and it is also invariant under both $S$ and $T$ transformations
    \begin{equation}
        \hat{\mathbb{S}} |\Omega\rangle_{\mathcal{L}_{\textrm{Neu}}} = |\Omega\rangle_{\mathcal{L}_{\textrm{Neu}}},\quad \hat{\mathbb{T}} |\Omega\rangle_{\mathcal{L}_{\textrm{Neu}}} = |\Omega\rangle_{\mathcal{L}_{\textrm{Neu}}}.
    \end{equation}
The action of $W_{(\alpha,\tilde{\alpha})}$ are
    \begin{equation}
        W_{(\alpha,\tilde{\alpha})}[\Gamma_1]|\Omega\rangle_{\mathcal{L}_{\textrm{Neu}}} = |(0,0)\rangle + (-1)^{\alpha}|(0,1)\rangle,\quad W_{(\alpha,\tilde{\alpha})}[\Gamma_2]|\Omega\rangle_{\mathcal{L}_{\textrm{Neu}}} = |(\alpha,0)\rangle + |(\alpha,1)\rangle.
    \end{equation}
Therefore $W_{(\alpha,0)}$ and $W_{(\alpha,1)}$ are equivalent on the topological boundary. Acting the operators twice we can read the fusion rule
    \begin{equation}
        W_{(\alpha,0)} \times_{\mathcal{L}_{\textrm{Neu}}} W_{(\beta,0)} = W_{(\alpha+\beta,0)}.
    \end{equation}
and they generate a $\mathbb{Z}_2$ symmetry. 

Similarly, let us label the twisted sectors of the $\mathbb{Z}_2$ symmetry as $|\tilde{a}_1,\tilde{a}_2\rangle$ where $\tilde{a}_1,\tilde{a}_2$ are holonomies along $\Gamma_1$ and $\Gamma_2$ cycles. The trivial sector is $|\tilde{0},\tilde{0}\rangle \equiv |\Omega\rangle_{\mathcal{L}_{\textrm{Neu}}}$ and the non-trivial sectors can be obtained by acting the symmetry generator $W_{(1,0)}$ as
    \begin{equation}
        \begin{split}
        |\tilde{0},\tilde{1}\rangle \equiv W_{(1,0)}[\Gamma_1] |\Omega\rangle_{\mathcal{L}_{\textrm{Neu}}} = |(0,0)\rangle - |(0,1)\rangle, \\
        |\tilde{1},\tilde{0}\rangle \equiv W_{(1,0)}[\Gamma_2]|\Omega\rangle_{\mathcal{L}_{\textrm{Neu}}} = |(1,0)\rangle + |(1,1)\rangle,
        \end{split}
    \end{equation}
and $|\tilde{1},\tilde{1}\rangle$ sector is obtained by performing $T$-transformation on the state $|\tilde{1},\tilde{0}\rangle$
    \begin{equation}
        |\tilde{1},\tilde{1}\rangle = \hat{\mathbb{T}} |\tilde{1},\tilde{0}\rangle = |(1,0)\rangle - |(1,1)\rangle
    \end{equation}
or acting $W_{(1,0)}[\Gamma_1] W_{(1,0)}[\Gamma_2]$ on $|\tilde{0},\tilde{0}\rangle$. One can also check the topological boundary state $|\tilde{a}_1,\tilde{a}_2\rangle$ is the eigenstate of $W_{(0,1)}$ operators
    \begin{equation}
        W_{(0,1)}[w_1 \Gamma_1 + w_2 \Gamma_2] |\tilde{a}_1,\tilde{a}_2\rangle = (-1)^{w_1 \tilde{a}_1 + w_2 \tilde{a}_2} |\tilde{a}_1,\tilde{a}_2\rangle.
    \end{equation}

\subsubsection*{Topological boundary states for $\mathcal{L}_f$}
Finally, let us move to the topological boundary corresponding to $\mathcal{L}_f$ and the vacuum is 
    \begin{equation}
        |\Omega\rangle_{\mathcal{L}_f} = |(0,0)\rangle + |(1,1)\rangle,
    \end{equation}
which is invariant under $S$-transformation. However, it is not invariant under $T$-transformation and one has
    \begin{equation}
        \hat{\mathbb{T}}|\Omega\rangle_{\mathcal{L}_f} = |(0,0)\rangle - |(1,1)\rangle.
    \end{equation}
As mentioned earlier, this is because the ground state is fermionic such that the $T$-transformation will change the boundary condition along $\Gamma_2$ cycle from Neveu–Schwarz to Ramond. We will return to this point soon. The action of $W_{(\alpha,\tilde{\alpha})}$ are
\begin{equation}
    \begin{split}
    W_{(\alpha,\tilde{\alpha})}[\Gamma_1] |\Omega\rangle_{\mathcal{L}_f} &= |(0,0)\rangle + (-1)^{\alpha + \tilde{\alpha}}|(1,1)\rangle\\
    W_{(\alpha,\tilde{\alpha})}[\Gamma_2] |\Omega\rangle_{\mathcal{L}_f}& = |(\alpha,\tilde{\alpha})\rangle + |(1+\alpha,1+\tilde{\alpha})\rangle
    \end{split}
\end{equation}
and one can see $W_{(\alpha,\tilde{\alpha})}$ is identified with $W_{(\alpha+1,\tilde{\alpha}+1)}$ on the topological boundary. Therefore one can choose $W_{(\alpha,0)}$ (or $W_{(0,\widetilde{\alpha})}$) as the generator at the boundary, and the fusion rule is
    \begin{equation}
        W_{(\alpha,0)} \times_{\mathcal{L}_f} W_{(\beta,0)} = W_{(\alpha+\beta,0)}
    \end{equation}
which generates a $\mathbb{Z}_2$ symmetry.

We will label the twisted sectors as $|s_1,s_2\rangle$ where $s_1,s_2$ are holonomies along $\Gamma_1$ and $\Gamma_2$ cycles. The trivial sector is $|0,0\rangle \equiv |\Omega\rangle_{\mathcal{L}_f}$ and the non-trivial sectors can be obtained by acting the symmetry generator $W_{(1,0)}$ as
    \begin{equation}
        \begin{split}
        |0,1\rangle \equiv W_{(1,0)}[\Gamma_1] |\Omega\rangle_{\mathcal{L}_f} = |(0,0)\rangle - |(1,1)\rangle, \\
        |1,0\rangle \equiv W_{(1,0)}[\Gamma_2]|\Omega\rangle_{\mathcal{L}_f} = |(1,0)\rangle + |(0,1)\rangle,
        \end{split}
    \end{equation}
and $|1,1\rangle$ sector is obtained by applying both $W_{(1,0)}[\Gamma_1]$ and $W_{(1,0)}[\Gamma_2]$ on the state $|0,0\rangle$ 
    \begin{equation}
        |1,1\rangle \equiv  W_{(1,0)}[\Gamma_2] W_{(1,0)}[\Gamma_1]|0,0\rangle = |(1,0)\rangle - |(0,1)\rangle.
    \end{equation}
Notice that $|0,1\rangle$ is the same as $\hat{\mathbb{T}}|0,0\rangle$ which implies the $T$-transformation maps (NS,NS) sector to (NS,R) sector. One can also check the topological boundary state $|s_1,s_2\rangle$ are eigenstates of $W_{(1,1)}$ operators
    \begin{equation}
        W_{(1,1)}[w_1 \Gamma_1 + w_2 \Gamma_2] |s_1,s_2\rangle = (-1)^{w_1 s_1 + w_2 s_2} |s_1,s_2\rangle .
    \end{equation}
\subsubsection*{Dualities from SymTFT}
In summary, we can construct three topological boundary states, the bosonic topological boundary states $|a_1,a_2\rangle, |\tilde{a}_1,\tilde{a}_2\rangle$ and the fermionic topological boundary state $|s_1,s_2\rangle$. They are not independent and one can check the transformation
    \begin{equation}
|\tilde{a}_1,\tilde{a}_2\rangle = \frac{1}{2} \sum_{a_1,a_2} (-1)^{\tilde{a}_1 a_2 + \tilde{a}_2 a_1} |a_1,a_2\rangle, \quad |s_1,s_2\rangle = \frac{1}{2} \sum_{a_1,a_2} (-1)^{(a_1+s_1)(a_2+s_2)-s_1 s_2}|a_1,a_2\rangle.
    \end{equation}
Given any 2d theory $\mathfrak{T}_{\mathbb{Z}_2}$ with non anomalous $\mathbb{Z}_2$ symmetry, denote the partition function as $Z_{\mathfrak{T}_{\mathbb{Z}_2}}[a_1,a_2]$ for each twisted sector labelled by $(a_1,a_2)$. One can build a dynamical boundary state according to
    \begin{equation}
        |\chi\rangle_{\mathfrak{T}_{\mathbb{Z}_2}} = \sum_{a_1,a_2} Z_{\mathfrak{T}_{\mathbb{Z}_2}}[a_1,a_2] |a_1,a_2\rangle.
    \end{equation}
Then the dualities in $\mathfrak{T}_{\mathbb{Z}_2}$ can be interpreted as switching the topological boundary states between $|a_1,a_2\rangle$, $|\tilde{a}_1,\tilde{a}_2\rangle$ and $|s_1,s_2\rangle$. The partition function of the original theory is $Z_{\mathfrak{T}_{\mathbb{Z}_2}} = \frac{1}{2}\langle a_1,a_2|\chi\rangle_{\mathfrak{T}_{\mathbb{Z}_2}}$, and the partition function of its $\mathbb{Z}_2$-gauging (Kramers–Wannier duality) is obtained by
    \begin{equation}
        Z_{\mathfrak{T}_{\mathbb{Z}_2} / \mathbb{Z}_2}[\tilde{a}_1,\tilde{a}_2] = \frac{1}{2} \langle \tilde{a}_1,\tilde{a}_2| \chi\rangle_{\mathfrak{T}_{\mathbb{Z}_2}},
    \end{equation}
and the partition function of its fermionization (Jordan-Wigner transformation) is obtained by
    \begin{equation}
        Z_{\mathfrak{T}_{\mathbb{Z}_2} / \mathbb{Z}_{2}^F} [s_1,s_2] = \frac{1}{2} \langle s_1,s_2|\chi\rangle_{\mathfrak{T}_{\mathbb{Z}_2}}.
    \end{equation}

\section{Examples}
In this section, we will present some examples of SymTFT and the constructions of topological boundary states on the torus for both group-like and non-invertible symmetries in 2D. We will mainly focus on the dualities at the level of partition functions.
\subsection{Non-anomalous $\mathbb{Z}_N$ symmetry}
The first example is the $2D$ $\mathbb{Z}_N$ theory with anomaly characterized by $H^3(\mathbb{Z}_N,U(1)) = \mathbb{Z}_N$ and labeled by $k=0,\cdots,N-1$. The SymTFT is the 3D Dijkgraaf-Witten theory
    \begin{equation}
        S = \frac{N}{2\pi} \int \widetilde{A}\wedge d A  + \frac{2 k}{4\pi} \int A \wedge d A.
    \end{equation}
The gauge transformations are
    \begin{equation}
        A \rightarrow A + d \Lambda,\quad \widetilde{A} \rightarrow \widetilde{A} + d \widetilde{\Lambda}
    \end{equation}
and we can consider the gauge invariant Wilson loops $W_{(\alpha,\tilde{\alpha})}[\Gamma]$ 
    \begin{equation}
        W_{(\alpha,\tilde{\alpha})} [\Gamma] = \exp \left(i \oint_{\Gamma} \alpha A + \tilde{\alpha} \widetilde{A} \right)
    \end{equation}
with $\alpha,\tilde{\alpha} = 0,1,\cdots,N-1$. Each line operator has quantum dimension one so that the total quantum dimension is $N$. The element of $S$ and $T$ matrices can be derived as
    \begin{equation}\label{ZN-S-and-T}
        S_{(\alpha,\tilde{\alpha}),(\beta,\tilde{\beta})} = \frac{1}{N}\omega^{-\alpha \tilde{\beta} - \tilde{\alpha} \beta + \frac{2k}{N} \tilde{\alpha} \tilde{\beta} },\quad T_{(\alpha,\tilde{\alpha}),(\beta,\tilde{\beta})} = \omega^{-\alpha \tilde{\alpha} + \frac{k}{N} \tilde{\alpha}^2} \delta_{\alpha,\beta} \delta_{\tilde{\alpha},\tilde{\beta}}
    \end{equation}
with $\omega = e^{\frac{2\pi i}{N}}$. The fusion coefficient is obtained via \eqref{Fusion-coefficient} as
    \begin{equation}\label{ZN-fusion}
        N_{(\alpha,\tilde{\alpha}),(\beta,\tilde{\beta})}^{(\gamma,\tilde{\gamma})} = \delta_{\gamma,\alpha+\beta - 2k \left[\frac{\tilde{\alpha} + \tilde{\beta}}{N}\right]} \delta_{\tilde{\gamma},\tilde{\alpha}+\tilde{\beta}},
    \end{equation}
where the delta function is defined modulo $N$ and $[...]$ is the floor function. We present a detailed derivations in the appendix.

Let us focus on the non-anomalous case where $k=0$. The $S$ and $T$ matrices are
    \begin{equation}
        S_{(\alpha,\tilde{\alpha}),(\beta,\tilde{\beta})} = \frac{1}{N}\omega^{-\alpha \tilde{\beta} - \tilde{\alpha} \beta},\quad T_{(\alpha,\tilde{\alpha}),(\beta,\tilde{\beta})} = \omega^{-\alpha \tilde{\alpha}} \delta_{\alpha,\beta} \delta_{\tilde{\alpha},\tilde{\beta}}
    \end{equation}
and the fusion rule is simply
    \begin{equation}
        N_{(\alpha,\tilde{\alpha}),(\beta,\tilde{\beta})}^{(\gamma,\tilde{\gamma})} = \delta_{\gamma,\alpha+\beta} \delta_{\tilde{\gamma},\tilde{\alpha}+\tilde{\beta}}.
    \end{equation}
The state vectors satisfy
    \begin{equation}
        W_{(\alpha,\tilde{\alpha})}[\Gamma_1] |(\beta,\tilde{\beta})\rangle = \omega^{-\alpha \tilde{\beta} - \tilde{\alpha} \beta} |(\beta,\tilde{\beta})\rangle,\quad 
        W_{(\alpha,\tilde{\alpha})}[\Gamma_2] |(\beta,\tilde{\beta})\rangle = |(\beta+\alpha,\tilde{\beta}+\tilde{\alpha})\rangle,
    \end{equation}
with the periodicity
    \begin{equation}
        |(\alpha+N,\tilde{\alpha})\rangle = |(\alpha,\tilde{\alpha})\rangle,\quad |(\alpha,\tilde{\alpha}+N)\rangle = |(\alpha,\tilde{\alpha})\rangle.
    \end{equation}

In the present case, the Lagrangian algebras can be found by finding the maximal commuting set of operators. First of all, we have two basic Lagrangian algebras, the Dirichlet one and the Neumann one
    \begin{equation}
        \mathcal{L}_{\textrm{Dir}} = \bigoplus_{\alpha} W_{(\alpha,0)},\quad \mathcal{L}_{\textrm{Neu}} =  \bigoplus_{\tilde{\alpha}} W_{(0,\tilde{\alpha})}.
    \end{equation}
In general if $N$ is not prime, for any positive integers $P,Q$ satisfying $N=P\times Q$ one has the following Lagrangian algebra
    \begin{equation}
        \mathcal{L}_{P,Q} = \bigoplus_{\alpha\in \mathbb{Z}_Q,\tilde{\alpha}\in \mathbb{Z}_P} W_{(P\alpha,Q\tilde{\alpha})},
    \end{equation}
and the Dirichlet (Neumann) algebra is the special case when $P=1$ ($Q=1$). If $N$ is even, we can also consider two fermionic Lagrangian algebras
    \begin{equation}
        \mathcal{L}_{f} = \bigoplus_{\alpha} W_{\left(\alpha,\frac{N}{2}\alpha\right)},\quad \mathcal{L}_{\widetilde{f}} =  \bigoplus_{\tilde{\alpha}} W_{\left(\frac{N}{2}\tilde{\alpha},\tilde{\alpha}\right)}.
    \end{equation}
Finally, for each $k\in \mathbb{Z}_N$ and $k\neq N/2$ when $N$ is even, we will also propose the algebras
    \begin{equation}
        \mathcal{L}_{pf;k} = \bigoplus_{\alpha} W_{\left(\alpha,k\alpha\right)},\quad \mathcal{L}_{\widetilde{pf};k} = \bigoplus_{\alpha} W_{\left(k\tilde{\alpha},\tilde{\alpha}\right)}
    \end{equation}
Strictly speaking, they are not Lagrangian algebras. Later we shall see the algebras for $k$ and $-k$ are pairwise and their corresponding topological boundary supports parafermionic structure. Moreover, when $k$ is prime the two kinds of Lagrangian algebras are equivalent
    \begin{equation}
        \mathcal{L}_{pf;k} = \mathcal{L}_{\widetilde{pf};\frac{1}{k}},
    \end{equation}
where $\frac{1}{k}$ is the inverse of $k$ in $\mathbb{Z}_N$.
\subsubsection*{Dirichlet/Neumann topological boundary}
The topological boundary states for Dirichlet and Neumann Lagrangian algebras $\mathcal{L}_{\textrm{Dir}}$ and $\mathcal{L}_{\textrm{Neu}}$ are constructed as follows,
    \begin{equation}
        |\Omega\rangle_{\mathcal{L}_{\textrm{Dir}}} = \sum_{\alpha} |(\alpha,0)\rangle,\quad |\Omega\rangle_{\mathcal{L}_{\textrm{Neu}}} = \sum_{\tilde{\alpha}} |(0,\tilde{\alpha})\rangle.
    \end{equation}
For the Dirichlet boundary, the operators $W_{(\alpha,\tilde{\alpha})}$ act as
    \begin{equation}
        W_{(\alpha,\tilde{\alpha})}[\Gamma_1] |\Omega\rangle_{\mathcal{L}_{\textrm{Dir}}} = \sum_{\beta}\omega^{-\tilde{\alpha} \beta} |(\beta,0)\rangle,\quad W_{(\alpha,\tilde{\alpha})}[\Gamma_2] |\Omega\rangle_{\mathcal{L}_{\textrm{Dir}}} = \sum_{\beta} |(\beta,\tilde{\alpha})\rangle
    \end{equation}
therefore the line operators $W_{(\alpha,\tilde{\alpha})}$ with different $\alpha$ are equivalent when restricting onto the boundary. Choose $W_{(0,\tilde{\alpha})}$ as representatives, acting the operators twice will give the fusion rule
    \begin{equation}
        W_{(0,\tilde{\alpha})} \times_{\mathcal{L}_{\textrm{Dir}}} W_{(0,\tilde{\beta})} = W_{(0,\tilde{\alpha}+\tilde{\beta})}
    \end{equation}
which generates the $\mathbb{Z}_N$ symmetry. We can build the topological boundary states for twisted sectors by acting the $\mathbb{Z}_N$ generator. Since the S-matrix elements among the operators $W_{(0,\tilde{\alpha})}$ are trivial, they commute with each other when restricting to the boundary and the twisted states are independent of the order of the action. This reflects that the $\mathbb{Z}_N$ symmetry is anomaly-free. To be concrete, we introduce the topological boundary states for twisted sectors as
    \begin{equation}
        |a_1,a_2\rangle = \left(W_{(0,1)}[\Gamma_2]\right)^{a_1} \left(W_{(0,1)}[\Gamma_1]\right)^{a_2} |\Omega\rangle_{\mathcal{L}_{\textrm{Dir}}} = \sum_{\alpha} \omega^{-a_2 \alpha} |(\alpha,a_1)\rangle
    \end{equation}
such that they are the eigenstates of $W_{(1,0)}$ operators
    \begin{equation}
        W_{(1,0)}[w_1 \Gamma_1 + w_2 \Gamma_2]|a_1,a_2\rangle = \omega^{-w_1 a_1 + w_2 a_2} |a_1,a_2\rangle.
    \end{equation}

For the Neumann boundary, the operators $W_{(\alpha,\tilde{\alpha})}$ act like
    \begin{equation}
        W_{(\alpha,\tilde{\alpha})}[\Gamma_1] |\Omega\rangle_{\mathcal{L}_{\textrm{Neu}}} = \sum_{\tilde{\beta}}\omega^{-\alpha\tilde{\beta}}|(0,\tilde{\beta})\rangle,\quad W_{(\alpha,\tilde{\alpha})}[\Gamma_2] |\Omega\rangle_{\mathcal{L}_{\textrm{Neu}}} = \sum_{\tilde{\beta}}|(\alpha,\tilde{\beta})\rangle
    \end{equation}
therefore the line operators $W_{(\alpha,\tilde{\alpha})}$ with different $\tilde{\alpha}$ are equivalent restricting on the boundary. Choose $W_{(\alpha,0)}$ as representatives, acting the operators twice will give the fusion rule
\begin{equation}
    W_{(\alpha,0)} \times_{\mathcal{L}_{\textrm{Neu}}} W_{(\beta,0)} = W_{(\alpha+\beta,0)}
\end{equation}
which generates another copy of $\mathbb{Z}_N$ symmetry. The topological boundary states for the twisted sectors are again obtained by acting $\mathbb{Z}_N$ generators and we can define
    \begin{equation}
        |\tilde{a}_1,\tilde{a}_2\rangle = \left(W_{(1,0)}[\Gamma_2]\right)^{\tilde{a}_1} \left(W_{(1,0)}[\Gamma_1]\right)^{\tilde{a}_2}|\Omega\rangle_{\mathcal{L}_{\textrm{Neu}}} = \sum_{\tilde{\alpha}} \omega^{-\tilde{a}_2 \tilde{\alpha}}|(\tilde{a}_1,\tilde{\alpha})\rangle,
    \end{equation}
and they are the eigenstates of $W_{(0,1)}$ operators
    \begin{equation}
        W_{(0,1)}[w_1 \Gamma_1 + w_2 \Gamma_2]|\tilde{a}_1,\tilde{a}_2\rangle = \omega^{-w_1 \tilde{a}_1 + w_2 \tilde{a}_2} |\tilde{a}_1,\tilde{a}_2\rangle.
    \end{equation}

Moreover, it is easy to check that $\langle \tilde{a}_1 ,\tilde{a}_2 |a_1,a_2\rangle = \omega^{-\tilde{a}_1 a_2 + \tilde{a}_2 a_1}$ and the two kinds of topological boundary states are related by the discrete Fourier transformation
    \begin{equation}
        |\tilde{a}_1,\tilde{a}_2\rangle = \frac{1}{N}\sum_{a_1,a_2} \omega^{-a_1 \tilde{a}_2 + a_2 \tilde{a}_1} |a_1,a_2\rangle.
    \end{equation}
For any theory $\mathfrak{T}_{\mathbb{Z}_N}$ with a non-anomalous $\mathbb{Z}_N$ symmetry, we can construct the dynamical boundary states as
    \begin{equation}
        |\chi\rangle_{\mathfrak{T}_{\mathbb{Z}_N}} = \sum_{a_1,a_2} Z_{\mathfrak{T}_{\mathbb{Z}_N}}[a_1,a_2] |a_1,a_2\rangle
    \end{equation}
such that the partition function of $\mathfrak{T}_{\mathbb{Z}_N}$ and its KW dual $\mathfrak{T}_{\mathbb{Z}_N}/\mathbb{Z}_N$ after gauging $\mathbb{Z}_N$ symmetry is
    \begin{equation}
        Z_{\mathfrak{T}_{\mathbb{Z}_N}}[a_1,a_2]=\frac{1}{N}\langle a_1,a_2 | \chi\rangle_{\mathfrak{T}_{\mathbb{Z}_N}},\quad Z_{\mathfrak{T}_{\mathbb{Z}_N}/\mathbb{Z}_N}[\tilde{a}_1,\tilde{a}_2] = \frac{1}{N}\langle \tilde{a}_1,\tilde{a}_2 |\chi\rangle_{\mathfrak{T}_{\mathbb{Z}_N}}.
    \end{equation}

\subsubsection*{Topological boundary states for $\mathcal{L}_{(P,Q)}$}
The topological boundary states on torus corresponding to $\mathcal{L}_{(P,Q)}$ is
    \begin{equation}
        |\Omega\rangle_{\mathcal{L}_{(P,Q)}} = \sum_{\alpha\in \mathbb{Z}_Q,\tilde{\alpha}\in \mathbb{Z}_P} |(P\alpha,Q\tilde{\alpha})\rangle,
    \end{equation}
which is invariant under S/T-transformation. One can check the operators $W_{(\alpha,\tilde{\alpha})}$ act as
    \begin{equation}
        \begin{split}
        W_{(\alpha,\tilde{\alpha})}[\Gamma_1]|\Omega\rangle_{\mathcal{L}_{(P,Q)}} =& \sum_{\beta\in \mathbb{Z}_Q,\tilde{\beta}\in \mathbb{Z}_P} \omega^{-Q\alpha \tilde{\beta} - P\tilde{\alpha}\beta} |P\beta,Q\tilde{\beta}\rangle,\\
        W_{(\alpha,\tilde{\alpha})}[\Gamma_2]|\Omega\rangle_{\mathcal{L}_{(P,Q)}} =& \sum_{\beta\in \mathbb{Z}_Q,\tilde{\beta}\in \mathbb{Z}_P} |P\beta + \alpha,Q\tilde{\beta}+\tilde{\alpha}\rangle,
        \end{split}
    \end{equation}
where the line operators are identified on the topological boundary according to
    \begin{equation}
        W_{(\alpha,\tilde{\alpha})} \sim W_{(\alpha+P,\tilde{\alpha})} \sim W_{(\alpha,\tilde{\alpha}+Q)}.
    \end{equation}
Therefore we can pick two generators $W_{(1,0)}$ and $W_{(0,1)}$ on the topological boundary and they generate the symmetry group $\mathbb{Z}_P \times \mathbb{Z}_Q$. In particular, one can easily check that
    \begin{equation}
        \frac{1}{N} \langle \Omega |_{\mathcal{L}_{(P,Q)}}| \chi\rangle_{\mathfrak{T}_{\mathbb{Z}_N}} = \frac{1}{P} \sum_{c_1,c_2\in\mathbb{Z}_P} Z_{\mathfrak{T}_{\mathbb{Z}_N}}[Q c_1, Qc_2],
    \end{equation}
which corresponds to gauging the $\mathbb{Z}_P$ subgroup of $\mathbb{Z}_N$. Since the generators $W_{(1,0)}$ and $W_{(0,1)}$ do not commute with each other, the twist sectors are defined up to phase ambiguities which is a signal of the mixed 't Hooft anomaly between $\mathbb{Z}_P$ and $\mathbb{Z}_Q$.

\subsubsection*{Topological boundary states for $\mathcal{L}_{f}$ and $\mathcal{L}_{\widetilde{f}}$}

The topological boundary state on torus corresponding to $\mathcal{L}_{f}$ and $\mathcal{L}_{\widetilde{f}}$ are separately,
    \begin{equation}
        |\Omega\rangle_{\mathcal{L}_{f}} = \sum_{\alpha} |(\alpha,\frac{N}{2}\alpha)\rangle,\quad |\Omega\rangle_{\mathcal{L}_{\widetilde{f}}} = \sum_{\alpha}|(\frac{N}{2}\tilde{\alpha},\tilde{\alpha})\rangle,
    \end{equation}
which are invariant under S-transformation but not under T-transformation due to the fermionic nature. For $|\Omega\rangle_{\mathcal{L}_{f}}$, the operators $W_{(\alpha,\tilde{\alpha})}$ act on the vacuum as
    \begin{equation}
        \begin{split}
            W_{(\alpha,\tilde{\alpha})}[\Gamma_1] |\Omega\rangle_{\mathcal{L}_{f}} =& \sum_{\beta} \omega^{-\frac{N}{2}\alpha \beta - \tilde{\alpha}\beta} |(\beta,\frac{N}{2}\beta)\rangle,\\
            W_{(\alpha,\tilde{\alpha})}[\Gamma_2] |\Omega\rangle_{\mathcal{L}_{f}} =& \sum_{\beta}|(\beta+\alpha,\frac{N}{2}\beta + \tilde{\alpha})\rangle,
        \end{split}
    \end{equation}
and one can deduce the following identification on the topological boundary
    \begin{equation}
        W_{(\alpha,\tilde{\alpha})} \sim W_{(\alpha+1,\tilde{\alpha}+\frac{N}{2})}
    \end{equation}
and one can choose $W_{(0,1)}$ as a generator. It is clear that it generates a $\mathbb{Z}_N$ symmetry. One has the dual partition function
    \begin{equation}
        \frac{1}{N} \langle \Omega|_{\mathcal{L}_{f}}|\chi\rangle_{\mathfrak{T}_{\mathbb{Z}_N}} = \frac{1}{2} \sum_{c_1,c_2 \in \mathbb{Z}_2} (-1)^{c_1 c_2} Z_{\mathfrak{T}_{\mathbb{Z}_N}}[\frac{N}{2} c_1,\frac{N}{2} c_2],
    \end{equation}
which corresponds to doing a Jordan-Wigner transformation to the $\mathbb{Z}_2$ subgroup.

On the other hand, for $|\Omega\rangle_{\mathcal{L}_{\widetilde{f}}}$ the operators $W_{(\alpha,\tilde{\alpha})}$ act on the vacuum as
    \begin{equation}
        \begin{split}
        W_{(\alpha,\tilde{\alpha})}[\Gamma_1]|\Omega\rangle_{\mathcal{L}_{\widetilde{f}}} =& \sum_{\tilde{\beta}} \omega^{-\alpha \tilde{\beta} - \frac{N}{2} \tilde{\alpha}\tilde{\beta}} |(\frac{N}{2}\tilde{\beta},\tilde{\beta})\rangle\\
        W_{(\alpha,\tilde{\alpha})}[\Gamma_1]|\Omega\rangle_{\mathcal{L}_{\widetilde{f}}}=& \sum_{\tilde{\beta}}|\frac{N}{2}\tilde{\beta}+\alpha,\tilde{\beta}+\tilde{\alpha}\rangle
        \end{split}
    \end{equation}
and one can identify
    \begin{equation}
        W_{(\alpha,\tilde{\alpha})} = W_{(\alpha+\frac{N}{2},\tilde{\alpha}+1)}
    \end{equation}
on the topological boundary. The $\mathbb{Z}_N$ generator can be chosen as $W_{(1,0)}$ and the dual partition function is
    \begin{equation}
        \frac{1}{N}\langle \Omega|_{\mathcal{L}_{\widetilde{f}}}|\chi\rangle_{\mathfrak{T}_{\mathbb{Z}_N}} = \frac{1}{N} \sum_{a_1,a_2\in \mathbb{Z}_N} (-1)^{a_1 a_2} Z_{\mathfrak{T}_{\mathbb{Z}_N}}[a_1,a_2],
    \end{equation}
where one gauges the $\mathbb{Z}_N$ symmetry with an phase $(-1)^{a_1 a_2}$ inserted. The RHS can also be written as
    \begin{equation}
        \frac{1}{2} \sum_{c_1,c_2\in \mathbb{Z}_2} \frac{1}{N}\sum_{b_1,b_2 \in \mathbb{Z}_N} (-1)^{c_1 c_2} \omega^{\frac{N}{2}c_1 b_2 - \frac{N}{2}c_2 b_1} Z_{\mathfrak{T}_{\mathbb{Z}_N}}[b_1,b_2].
    \end{equation}
which can be interpreted as doing a Jordan-Wigner transformation in the dual theory.

\subsubsection*{Topological boundary states for $\mathcal{L}_{pf;k}$ and $\mathcal{L}_{\widetilde{pf};k}$}

Let us move to the topological boundary state on torus corresponding to $\mathcal{L}_{pf;k}$ and $\mathcal{L}_{\widetilde{pf};k}$. For $\mathcal{L}_{pf;k}$ let us consider the pair of states
    \begin{equation}
        |\Omega\rangle_{\mathcal{L}_{pf;k}} = \sum_{\alpha} |(\alpha,k\alpha)\rangle,\quad |\Omega\rangle_{\mathcal{L}_{pf;-k}} = \sum_{\alpha} |(\alpha,-k\alpha)\rangle,
    \end{equation}
with $0<k\leq \left[\frac{N}{2}\right]$ ($k\neq \frac{N}{2}$ when $N$ is even), one can check the S-transformation will exchange the two states
    \begin{equation}
        \hat{\mathbb{S}}|\Omega\rangle_{\mathcal{L}_{pf;k}} = |\Omega\rangle_{\mathcal{L}_{pf;-k}},\quad \hat{\mathbb{S}}|\Omega\rangle_{\mathcal{L}_{pf;-k}} = |\Omega\rangle_{\mathcal{L}_{pf;k}}.
    \end{equation}
Similarly, we have another pair of states for $\mathcal{L}_{\widetilde{pf};k}$
    \begin{equation}
        |\Omega\rangle_{\mathcal{L}_{\widetilde{pf};k}} = \sum_{\tilde{\alpha}} |(k\tilde{\alpha},\tilde{\alpha})\rangle,\quad |\Omega\rangle_{\mathcal{L}_{\widetilde{pf};-k}} = \sum_{\tilde{\alpha}} |(-k\tilde{\alpha},\tilde{\alpha})\rangle,
    \end{equation}
and the S-transformation also exchanges the two states
    \begin{equation}
        \hat{\mathbb{S}}|\Omega\rangle_{\mathcal{L}_{\widetilde{pf};k}} = |\Omega\rangle_{\mathcal{L}_{\widetilde{pf};-k}},\quad \hat{\mathbb{S}}|\Omega\rangle_{\mathcal{L}_{\widetilde{pf};-k}} = |\Omega\rangle_{\mathcal{L}_{\widetilde{pf};k}}.
    \end{equation}

Let us first consider the states $|\Omega\rangle_{\mathcal{L}_{pf;k}}$, one can check the action of line operators
    \begin{equation}
        \begin{split}
            W_{(\alpha,\tilde{\alpha})}[\Gamma_1] |\Omega\rangle_{\mathcal{L}_{pf;k}} =& \sum_{\beta} \omega^{-k \alpha \beta - \tilde{\alpha}\beta} |(\beta,k\beta)\rangle,\\
            W_{(\alpha,\tilde{\alpha})}[\Gamma_2] |\Omega\rangle_{\mathcal{L}_{pf;k}} =& \sum_{\beta}|(\beta+\alpha,k\beta + \tilde{\alpha})\rangle,
        \end{split}
    \end{equation}
which indicates
    \begin{equation}
        W_{(\alpha,\tilde{\alpha})}[\Gamma_1] \sim W_{(\alpha+1,\tilde{\alpha}-k)}[\Gamma_1],\quad W_{(\alpha,\tilde{\alpha})}[\Gamma_2] \sim W_{(\alpha+1,\tilde{\alpha}+k)}[\Gamma_2]
    \end{equation}
on the topological boundary for line operators along $\Gamma_1$ and $\Gamma_2$ and we can use that to set $\alpha=0$. Therefore the symmetry is $\mathbb{Z}_N$ and one can choose the generator to be $W_{(0,1)}$. Projecting $|\chi\rangle_{\mathfrak{T}_{\mathbb{Z}_N}}$ onto the topological boundary states one gets
    \begin{equation}
        \begin{split}
        &\frac{1}{N}\langle \Omega |_{\mathcal{L}_{pf;k}}|\chi\rangle_{\mathfrak{T}_{\mathbb{Z}_N}}\\
        =& \frac{\gcd(k,N)}{N} \sum_{c_1,c_2 \in \mathbb{Z}_{N/\gcd(k,N)}} \left(\omega^{\gcd(k,N)}\right)^{-\left(\frac{k}{\gcd(k,N)}\right)^{-1}c_1 c_2} Z_{\mathfrak{T}_{\mathbb{Z}_N}}[\gcd(k,N)c_1,\gcd(k,N)c_2]            
        \end{split}
    \end{equation}
where $c_1,c_2 = 0,\cdots,\frac{N}{\gcd(k,N)}-1$ are $\mathbb{Z}_{N/\gcd(k,N)}$ valued and the inverse of $\frac{k}{\gcd(k,N)}$ is understood within $\mathbb{Z}_{N/\gcd(k,N)}$. In particular, when $k$ is coprime with $N$, the dual partition function is
    \begin{equation}
        \frac{1}{N}\langle \Omega |_{\mathcal{L}_{pf;k}}|\chi\rangle_{\mathfrak{T}_{\mathbb{Z}_N}} = \frac{1}{N} \sum_{c_1,c_2\in \mathbb{Z}_N} \omega^{-\frac{c_1 c_2}{k}} Z_{\mathfrak{T}_{\mathbb{Z}_N}}[c_1,c_2]
    \end{equation}
which gives the parafermionization with respect to $\mathbb{Z}_N$. For general $k$ one get the parafermionization with respect to $\mathbb{Z}_{N/\gcd(k,N)}$.
Finally, since S-transformation flips $k$ to $-k$ in the topological boundary state $|\Omega\rangle_{pf;k}$, which indicates the parafermionic theory is also not invariant under S-transformation.

For the topological boundary state $|\Omega\rangle_{\mathcal{L}_{\widetilde{pf};k}}$ the action of line operators are
    \begin{equation}
        \begin{split}
            W_{(\alpha,\tilde{\alpha})}[\Gamma_1] |\Omega\rangle_{\mathcal{L}_{\widetilde{pf};k}} =& \sum_{\tilde{\beta}} \omega^{-\alpha \tilde{\beta} - k \tilde{\alpha} \tilde{\beta}} |(k\tilde{\beta},\tilde{\beta})\rangle, \\
            W_{(\alpha,\tilde{\alpha})}[\Gamma_2] |\Omega\rangle_{\mathcal{L}_{\widetilde{pf};k}} =& \sum_{\tilde{\beta}} |(k\tilde{\beta}+\alpha,\tilde{\beta}+\tilde{\alpha})\rangle ,
        \end{split}
    \end{equation}
which indicates
    \begin{equation}
        W_{(\alpha,\tilde{\alpha})}[\Gamma_1] \sim W_{(\alpha-k,\tilde{\alpha}+1)}[\Gamma_1],\quad W_{(\alpha,\tilde{\alpha})}[\Gamma_2] \sim W_{(\alpha+k,\tilde{\alpha}+1)}[\Gamma_2]. 
    \end{equation}
for line operators along $\Gamma_1$ and $\Gamma_2$. Similarly, we can set $\tilde{\alpha}=0$ and the symmetry is $\mathbb{Z}_N$ generated by $W_{(1,0)}$. Projection $|\chi\rangle_{\mathfrak{T}_{\mathbb{Z}_N}}$ onto the topological boundary states one gets
    \begin{equation}
        \frac{1}{N}\langle\Omega|_{\mathcal{L}_{\widetilde{pf};k}}|\chi\rangle_{\mathfrak{T}_{\mathbb{Z}_N}} = \frac{1}{N} \sum_{a_1,a_2\in \mathbb{Z}_N} \omega^{-k a_1 a_2} Z_{\mathfrak{T}_{\mathbb{Z}_N}}[a_1,a_2],
    \end{equation}
which can also be written as
    \begin{equation}
        \frac{\gcd(k,N)}{N} \sum_{c_1,c_2 \in \mathbb{Z}_{N/\gcd(k,N)}} \frac{1}{N} \sum_{b_1,b_2\in \mathbb{Z}_N} \left(\omega^{\gcd(k,N)}\right)^{-\left(\frac{k}{\gcd(k,N)}\right)^{-1}c_1 c_2} \omega^{-\gcd(k,N)c_1 b_2 + \gcd(k,N)c_2 b_1} Z_{\mathfrak{T}_{\mathbb{Z}_N}}[b_1,b_2]
    \end{equation}
and it be interpreted as parafermionization in the dual theory. Here $\omega^{\gcd(k,N)}$ is the $\left(\frac{N}{\gcd(k,N)}\right)$-root of unity and the inverse of $\frac{k}{\gcd(k,N)}$ is understood within $\mathbb{Z}_{\frac{N}{\gcd(k,N)}}$ since it is coprime with $\frac{N}{\gcd(k,N)}$ by definition. 

\subsection{$\mathbb{Z}_N$ symmetry with non-anomalous $\mathbb{Z}_2$ subgroup}
Let us consider another case when both $N$ and $k$ are even and $k\neq 0$. The element of $S$ and $T$ matrices and the fusion rule are still given by \eqref{ZN-S-and-T} and \eqref{ZN-fusion}
    \begin{equation}
        S_{(\alpha,\tilde{\alpha}),(\beta,\tilde{\beta})} = \frac{1}{N}\omega^{-\alpha \tilde{\beta} - \tilde{\alpha} \beta + \frac{2k}{N} \tilde{\alpha} \tilde{\beta} },\quad T_{(\alpha,\tilde{\alpha}),(\beta,\tilde{\beta})} = \omega^{-\alpha \tilde{\alpha} + \frac{k}{N} \tilde{\alpha}^2} \delta_{\alpha,\beta} \delta_{\tilde{\alpha},\tilde{\beta}}
    \end{equation}
with $\omega = e^{\frac{2\pi i}{N}}$ and
    \begin{equation}
        N_{(\alpha,\tilde{\alpha}),(\beta,\tilde{\beta})}^{(\gamma,\tilde{\gamma})} = \delta_{\gamma,\alpha+\beta - 2k \left[\frac{\tilde{\alpha} + \tilde{\beta}}{N}\right]} \delta_{\tilde{\gamma},\tilde{\alpha}+\tilde{\beta}},
    \end{equation}
The Hilbert space is constructed as
    \begin{equation}
        \begin{split}
        W_{(\beta,\tilde{\beta})}[\Gamma_1] |(\alpha,\tilde{\alpha})\rangle =& \omega^{-\beta \tilde{\alpha} - \tilde{\beta} \alpha + \frac{2k}{N} \tilde{\alpha} \tilde{\beta}} |(\alpha,\tilde{\alpha})\rangle,\\
        W_{(\beta,\tilde{\beta})}[\Gamma_2] |(\alpha,\tilde{\alpha})\rangle =& |(\alpha+\beta-2k \left[\frac{\tilde{\alpha} + \tilde{\beta}}{N}\right],\tilde{\alpha}+\tilde{\beta})\rangle,
        \end{split}
    \end{equation}
with the periodicity
    \begin{equation}
        |(\alpha+N,\tilde{\alpha})\rangle = |(\alpha,\tilde{\alpha})\rangle,\quad |(\alpha,\tilde{\alpha}+N)\rangle = |(\alpha-2k,\tilde{\alpha})\rangle,
    \end{equation}
where $\alpha,\alpha'=0,\cdots,N-1$.

We still have the Dirichlet Lagrangian algebras
    \begin{equation}
        \mathcal{L}_{\textrm{Dir}} = \bigoplus_{\alpha \in \mathbb{Z}_N} W_{(\alpha,0)}
    \end{equation}
but the Neumann one is no longer Lagrangian. We will also consider another two kinds of Lagrangian algebras given by
    \begin{equation}
        \mathcal{L}_1 = \bigoplus_{\alpha\in \mathbb{Z}_{\frac{N}{2}} , \tilde{\alpha}\in\mathbb{Z}_2} W_{(2\alpha,\frac{N}{2}\tilde{\alpha})},\quad \mathcal{L}_{2} = \bigoplus_{\alpha\in \mathbb{Z}_{\frac{N}{2}} , \tilde{\alpha}\in\mathbb{Z}_2} W_{(2\alpha+\tilde{\alpha},\frac{N}{2}\tilde{\alpha})}.
    \end{equation}
One can check the $S$-matrix elements between each line operator in the algebras are all one up to $1/N$ so they form a maximally commuting set of operators. The $T$-matrix gives the spin of each line operator
    \begin{equation}
        (\theta_1)_{(2\alpha,\frac{N}{2}\widetilde{\alpha})} = (-1)^{\frac{k}{2}\widetilde{\alpha}^2},\quad (\theta_2)_{(2\alpha+\widetilde{\alpha},\frac{N}{2}\widetilde{\alpha})} = (-1)^{\left(\frac{k}{2}-1\right)\tilde{\alpha}^2}.
    \end{equation}
Therefore when $k$ is 0 mod $4$, all the line operators in $\mathcal{L}_1$ are spin-0 and $\mathcal{L}_2$ contains spin-$\frac{1}{2}$ line operators; when $k$ is $2$ mod $4$ the situation is opposite.

\subsubsection*{Dirichlet topological boundary state}
The vacuum for the Dirichlet topological boundary state is given by
    \begin{equation}
        |\Omega\rangle_{\mathcal{L}_{\textrm{Dir}}} = \sum_{\alpha} |(\alpha,0)\rangle,
    \end{equation}
and the operators $W_{(\alpha,\tilde{\alpha})}$ still acts like
    \begin{equation}
        W_{(\alpha,\tilde{\alpha})}[\Gamma_1] |\Omega\rangle_{\mathcal{L}_{\textrm{Dir}}} = \sum_{\beta}\omega^{-\tilde{\alpha} \beta} |(\beta,0)\rangle,\quad W_{(\alpha,\tilde{\alpha})}[\Gamma_2] |\Omega\rangle_{\mathcal{L}_{\textrm{Dir}}} = \sum_{\beta} |(\beta,\tilde{\alpha})\rangle,
    \end{equation}
and we can choose the $\mathbb{Z}_N$ generator as $W_{(0,1)}$. Acting $W_{(0,1)}[\Gamma_1]$ and $W_{(0,1)}[\Gamma_2]$ on the topological boundary state $|\Omega\rangle_{\mathcal{L}_{\textrm{Dir}}}$ will generate the states for twisted sectors. However, since $W_{(0,1)}$ has a non-trivial S-matrix element with itself, the action is ambiguous which implies the $\mathbb{Z}_N$ is anomalous. Nevertheless, since the $\mathbb{Z}_2$ generator $W_{(0,\frac{N}{2})}$ braid trivially with itself, the $\mathbb{Z}_2$ subgroup is still anomaly free. 

Let us define the topological boundary states for twisted sectors as
    \begin{equation}
        |a_1,a_2\rangle = W_{(0,1)}[a_1\Gamma_2 + a_2 \Gamma_1]|\Omega\rangle_{\mathcal{L}_{\textrm{Dir}}}
    \end{equation}
with $a_1,a_2 = 0,\cdots,N-1$. Use the commutation relation
    \begin{equation}
        W_{(0,1)}[\Gamma_2] W_{(0,1)}[\Gamma_1] = \omega^{-\frac{2k}{N}} W_{(0,1)}[\Gamma_1] W_{(0,1)}[\Gamma_2]
    \end{equation}
and also the BCH formula, one has
    \begin{equation}
        W_{(0,1)}[a_1\Gamma_2 + a_2 \Gamma_1] = \omega^{\frac{k}{N}a_1 a_2} \left(W_{(0,1)}[\Gamma_2]\right)^{a_1} \left(W_{(0,1)}[\Gamma_1]\right)^{a_2}
    \end{equation}
and the topological boundary states can be written as
    \begin{equation}
        |a_1,a_2\rangle = \omega^{\frac{k}{N}a_1 a_2} \left(W_{(0,1)}[\Gamma_2]\right)^{a_1} \left(W_{(0,1)}[\Gamma_1]\right)^{a_2} |\Omega\rangle_{\mathcal{L}_{\textrm{Dir}}}= \sum_{\alpha} \omega^{-a_2 \alpha + \frac{k}{N}a_1 a_2} |(\alpha,a_1)\rangle.
    \end{equation}
They are the eigenstates of $W_{(1,0)}$ operators
    \begin{equation}
        W_{(1,0)}[w_1 \Gamma_1 + w_2 \Gamma_2]|a_1,a_2\rangle = \omega^{-w_1 a_1 + w_2 a_2} |a_1,a_2\rangle.
    \end{equation}
The topological boundary states are not invariant under the shift $a_1 \rightarrow a_1 + N$ and $a_2 \rightarrow a_2 + N$. One can check
    \begin{equation}
        |a_1+N,a_2\rangle = \omega^{-k a_2} |a_1,a_2\rangle,\quad|a_1,a_2+N\rangle = \omega^{k a_1} |a_1,a_2\rangle,
    \end{equation}
where the phase ambiguity is due to the anomaly of $\mathbb{Z}_N$ symmetry.
    
For a given 2D theory $\mathfrak{T}_{\mathbb{Z}^{(k)}_{N}}$ with anomalous $\mathbb{Z}_N$ symmetry labelled by $k$, the dynamical boundary state $|\chi\rangle_{\mathbb{Z}^{(k)}_{N}}$ is defined as
    \begin{equation}
        |\chi\rangle_{\mathbb{Z}^{(k)}_{N}} = \sum_{a_1,a_2} Z_{\mathbb{Z}^{(k)}_{N}}[a_1,a_2] |a_1,a_2\rangle,
    \end{equation}
and we require the dynamical boundary state to be well-defined under the shift $a_1\rightarrow a_1+N$ and $a_2\rightarrow a_2+N$, therefore we will assume the partition function satisfying
    \begin{equation}
        Z_{\mathbb{Z}^{(k)}_{N}}[a_1+N,a_2] = \omega^{k a_2} Z_{\mathbb{Z}^{(k)}_{N}}[a_1,a_2],\quad Z_{\mathbb{Z}^{(k)}_{N}}[a_1,a_2+N] = \omega^{-k a_1} Z_{\mathbb{Z}^{(k)}_{N}}[a_1,a_2].
    \end{equation}
In other words, the anomaly of the boundary theory is canceled by the inflow of the bulk Dijkgraaf-Witten theory.

\subsubsection*{Topological boundary states for $\mathcal{L}_1$ and $\mathcal{L}_2$}
    
For the Lagrangian algebras $\mathcal{L}_1$ and $\mathcal{L}_2$ the vacua are separately given by
    \begin{equation}
        |\Omega\rangle_{\mathcal{L}_1} = \sum_{\alpha\in \mathbb{Z}_{\frac{N}{2}},\tilde{\alpha}\in\mathbb{Z}_2} |(2\alpha,\frac{N}{2}\tilde{\alpha})\rangle,\quad |\Omega\rangle_{\mathcal{L}_2} = \sum_{\alpha\in \mathbb{Z}_{\frac{N}{2}},\tilde{\alpha}\in\mathbb{Z}_2} |(2\alpha+\tilde{\alpha},\frac{N}{2}\tilde{\alpha})\rangle,
    \end{equation}
and we will consider the line operators acting on the topological vacuum. For $|\Omega\rangle_{\mathcal{L}_1}$ one has
    \begin{equation}
        \begin{split}
            W_{(\alpha,\tilde{\alpha})}[\Gamma_1] |\Omega\rangle_{\mathcal{L}_1} =& \sum_{\beta\in \mathbb{Z}_{\frac{N}{2}},\tilde{\beta}\in\mathbb{Z}_2} \omega^{-\frac{N}{2} \alpha \tilde{\beta} - 2 \tilde{\alpha}\beta + k \tilde{\alpha} \tilde{\beta}} |(2\beta,\frac{N}{2}\tilde{\beta})\rangle, \\
            W_{(\alpha,\tilde{\alpha})}[\Gamma_2] |\Omega\rangle_{\mathcal{L}_1} =& \sum_{\beta\in \mathbb{Z}_{\frac{N}{2}},\tilde{\beta}\in\mathbb{Z}_2} |(2\beta + \alpha,\frac{N}{2}\tilde{\beta} + \tilde{\alpha}) \rangle,
        \end{split}
    \end{equation}
which implies the following identification on the topological boundary
    \begin{equation}
        W_{(\alpha,\tilde{\alpha})} \sim W_{(\alpha+2,\tilde{\alpha})} \sim W_{(\alpha,\tilde{\alpha}+\frac{N}{2})}.
    \end{equation}
We can choose $W_{(1,0)}$ and $W_{(0,1)}$ as generators and they generate the group $\mathbb{Z}_2 \times \mathbb{Z}_{\frac{N}{2}}$. On the other hand, for $|\Omega\rangle_{\mathcal{L}_2}$ one has
    \begin{equation}
        \begin{split}
            W_{(\alpha,\tilde{\alpha})}[\Gamma_1] |\Omega\rangle_{\mathcal{L}_2} =& \sum_{\beta\in \mathbb{Z}_{\frac{N}{2}},\tilde{\beta}\in\mathbb{Z}_2} \omega^{-\frac{N}{2} \alpha \tilde{\beta} - 2 \tilde{\alpha}\beta + (k-1) \tilde{\alpha} \tilde{\beta}} |(2\beta+\tilde{\beta},\frac{N}{2}\tilde{\beta})\rangle \\
            W_{(\alpha,\tilde{\alpha})}[\Gamma_2] |\Omega\rangle_{\mathcal{L}_2} =& \sum_{\beta\in \mathbb{Z}_{\frac{N}{2}},\tilde{\beta}\in\mathbb{Z}_2} |(2\beta + \tilde{\beta}+\alpha,\frac{N}{2}\tilde{\beta} + \tilde{\alpha}) \rangle,
        \end{split}
    \end{equation}
and we have the following identification on the topological boundary
    \begin{equation}
        W_{(\alpha,\tilde{\alpha})}\sim W_{(\alpha+2,\tilde{\alpha})} \sim W_{(\alpha+1,\tilde{\alpha}+\frac{N}{2})}.
    \end{equation}
We can choose $W_{(0,1)}$ as the generator and it generates the group $\mathbb{Z}_N$. Actually, when $N$ is $2$ mod $4$ the $\mathbb{Z}_N$ group is the same to $\mathbb{Z}_2 \times \mathbb{Z}_{\frac{N}{2}}$, we only need to distinguish them when $N$ is 0 mod $4$.

Let us compute the projection of the dynamical boundary state $|\chi\rangle_{\mathbb{Z}^{(k)}_{N}}$ onto the topological boundary state $|\Omega\rangle_{\mathcal{L}_1}$ and $|\Omega\rangle_{\mathcal{L}_2}$. For $|\Omega\rangle_{\mathcal{L}_1}$ one has
    \begin{equation}
        \begin{split}
            &\frac{1}{N} \langle \Omega|_{\mathcal{L}_1} |\chi\rangle_{\mathbb{Z}^{(k)}_{N}}\\
            =& \frac{1}{N} \sum_{\beta,a_1,a_2\in \mathbb{Z}_N} \sum_{\substack{\alpha \in \mathbb{Z}_{N/2}\\ \tilde{\alpha}\in \mathbb{Z}_2}} \omega^{-a_2 \beta + \frac{k}{N}a_1 a_2} Z_{\mathbb{Z}^{(k)}_{N}}[a_1,a_2] \langle (2\alpha,\frac{N}{2} \tilde{\alpha})|(\beta,a_1)\rangle\\
            =&\frac{1}{N} \sum_{\beta,a_1,a_2\in \mathbb{Z}_N} \sum_{\substack{\alpha \in \mathbb{Z}_{N/2}\\ \tilde{\alpha}\in \mathbb{Z}_2}} \omega^{-a_2 \beta + \frac{k}{N}a_1 a_2} Z_{\mathbb{Z}^{(k)}_{N}}[a_1,a_2] \delta_{\beta,2\alpha} \delta_{a_1,\frac{N}{2}\tilde{\alpha}}\\
            =&\frac{1}{N}\sum_{a_2\in \mathbb{Z}_N}\sum_{\substack{\alpha \in \mathbb{Z}_{N/2}\\ \tilde{\alpha}\in \mathbb{Z}_2}} \omega^{-2 \alpha a_2 + \frac{k}{2}\tilde{\alpha}a_2} Z_{\mathbb{Z}^{(k)}_{N}}[\frac{N}{2}\tilde{\alpha},a_2]
        \end{split}
    \end{equation}
Sum over $\alpha$ will impose $a_2=\frac{N}{2} c_2$ for $c_2=0,1$. Relabel $\tilde{\alpha} = c_1$ we get
    \begin{equation}
        \frac{1}{N} \langle \Omega|_{\mathcal{L}_1} |\chi\rangle_{\mathbb{Z}^{(k)}_{N}} = \frac{1}{2} \sum_{c_1,c_2 \in \mathbb{Z}_2} (-1)^{\frac{k}{2} c_1 c_2} Z_{\mathbb{Z}^{(k)}_{N}}[\frac{N}{2}c_1,\frac{N}{2}c_2].
    \end{equation}
On the other hand, for $|\Omega\rangle_{\mathcal{L}_2}$ one has
    \begin{equation}
        \begin{split}
            &\frac{1}{N} \langle \Omega|_{\mathcal{L}_2} |\chi\rangle_{\mathbb{Z}^{(k)}_{N}}\\
            =& \frac{1}{N} \sum_{\beta,a_1,a_2\in \mathbb{Z}_N} \sum_{\substack{\alpha \in \mathbb{Z}_{N/2}\\ \tilde{\alpha}\in \mathbb{Z}_2}} \omega^{-a_2 \beta + \frac{k}{N}a_1 a_2} Z_{\mathbb{Z}^{(k)}_{N}}[a_1,a_2] \langle (2\alpha+\tilde{\alpha},\frac{N}{2} \tilde{\alpha})|(\beta,a_1)\rangle\\
            =&\frac{1}{N} \sum_{\beta,a_1,a_2\in \mathbb{Z}_N} \sum_{\substack{\alpha \in \mathbb{Z}_{N/2}\\ \tilde{\alpha}\in \mathbb{Z}_2}} \omega^{-a_2 \beta + \frac{k}{N}a_1 a_2} Z_{\mathbb{Z}^{(k)}_{N}}[a_1,a_2] \delta_{\beta,2\alpha+\tilde{\alpha}} \delta_{a_1,\frac{N}{2}\tilde{\alpha}}\\
            =&\frac{1}{N}\sum_{a_2\in \mathbb{Z}_N}\sum_{\substack{\alpha \in \mathbb{Z}_{N/2}\\ \tilde{\alpha}\in \mathbb{Z}_2}} \omega^{-2 \alpha a_2 + \left(\frac{k}{2}-1\right)\tilde{\alpha}a_2} Z_{\mathbb{Z}^{(k)}_{N}}[\frac{N}{2}\tilde{\alpha},a_2]
        \end{split}
    \end{equation}
Sum over $\alpha$ will impose $a_2=\frac{N}{2} c_2$ for $c_2=0,1$. Relabel $\tilde{\alpha} = c_1$ we get
    \begin{equation}
        \frac{1}{N} \langle \Omega|_{\mathcal{L}_2} |\chi\rangle_{\mathbb{Z}^{(k)}_{N}} = \frac{1}{2} \sum_{c_1,c_2 \in \mathbb{Z}_2} (-1)^{\left(\frac{k}{2}-1\right) c_1 c_2} Z_{\mathbb{Z}^{(k)}_{N}}[\frac{N}{2}c_1,\frac{N}{2}c_2].
    \end{equation}

In summary, the spin of each line operator in the two kinds of Lagrangian algebras are
    \begin{equation}
        (\theta_1)_{(2\alpha,\frac{N}{2}\widetilde{\alpha})} = (-1)^{\frac{k}{2}\widetilde{\alpha}^2},\quad (\theta_2)_{(2\alpha+\widetilde{\alpha},\frac{N}{2}\widetilde{\alpha})} = (-1)^{\left(\frac{k}{2}-1\right)\tilde{\alpha}^2},
    \end{equation}
and when $k$ is $0$ mod $4$, $\mathcal{L}_1$ is bosonic and $\mathcal{L}_2$ is fermionic; when $k$ is $2$ mod $4$ the situation is opposite. On the other hand, the dual partition functions for $|\Omega\rangle_{\mathcal{L}_1}$ and $|\Omega\rangle_{\mathcal{L}_2}$ are separately given by
    \begin{equation}
        \begin{split}
        \frac{1}{N} \langle \Omega|_{\mathcal{L}_1} |\chi\rangle_{\mathbb{Z}^{(k)}_{N}} =& \frac{1}{2} \sum_{c_1,c_2 \in \mathbb{Z}_2} (-1)^{\frac{k}{2} c_1 c_2} Z_{\mathbb{Z}^{(k)}_{N}}[\frac{N}{2}c_1,\frac{N}{2}c_2]\\
        \frac{1}{N}\langle \Omega |_{\mathcal{L}_2} |\chi\rangle_{\mathbb{Z}^{(k)}_{N}} =& \frac{1}{2} \sum_{c_1,c_2 \in \mathbb{Z}_2} (-1)^{(\frac{k}{2}-1)c_1 c_2} Z_{\mathbb{Z}^{(k)}_{N}}[\frac{N}{2}c_1,\frac{N}{2} c_2].
        \end{split}
    \end{equation}
Therefore when $k$ is 0 mod $4$, the first line corresponds to ordinary $\mathbb{Z}_2$ gauging and the second line corresponds to fermionization due to the fermionic SPT phase $(-1)^{c_1 c_2}$ inserted; when $k$ is $2$ mod $4$ the situation is opposite. Moreover, we have seen the symmetry group on $|\Omega\rangle_{\mathcal{L}_1}$ and $|\Omega\rangle_{\mathcal{L}_2}$ are separetely $\mathbb{Z}_2 \times \mathbb{Z}_{\frac{N}{2}}$ and $\mathbb{Z}_N$, which indicates the dual symmetry after $\mathbb{Z}_2$ gauging depends on both the $\mathbb{Z}_N$ anomaly and also on how we gauge the $\mathbb{Z}_2$.


\subsection{Discrete non-abelian symmetry and $S_3$ group}
For a general discrete group $G$, the elements of its Drinfeld center $\mathcal{Z}(G)$ are described by the pair $\alpha = (A,r)$ where $A$ labels the conjugacy class $K_A$ of $G$ and $r$ labels the irreducible representation under the centralizer group of $K_A$. To be explicit, let us pick a representative $a \in K_A$ and denote $[a]\equiv K_A$. Then $r$ labels the irreducible representation of the centralizer $C(a)$ of $a$, where the centralizer $C(a)$ is defined as
    \begin{equation}
        C(a) = \left\{ g \in G | g a = a g \right\}.
    \end{equation}
The character $\chi^{a}_r$ associated to $r$ is given by
    \begin{equation}
        \chi^a_r (g) = \textrm{Tr} r(g),\quad g \in C(a).
    \end{equation}
If we choose another representative $a'$ of $K_A$ such that $a' = x a x^{-1}$ for some $x \in G$, then the centralizer $C(a')$ is
    \begin{equation}
        C(a') = C(x a x^{-1}) = x C(a) x^{-1}
    \end{equation}
and the representations are induced by $r$ according to $r(x^{-1} g' x)$ for any $g' \in C(a')$. The character is
    \begin{equation}
        \chi^{a'}_{r}(g') = \chi^{x a x^{-1}}_r(g') = \chi^a_r(x^{-1} g' x).
    \end{equation}
With those notations, the $S$ and $T$ matrices are separately given by
    \begin{equation}
        S_{(A,r),(\tilde{A},\tilde{r})} = \frac{1}{|G|} \sum_{g \in K_A, h \in K_{\tilde{A}} \cap C(g)} \chi_r^{g}(h)^* \chi_{\tilde{r}}^{h}(g)^*,
    \end{equation}
where $|G|$ is the number of group elements in $G$ and 
    \begin{equation}
        T_{(A,r),(\tilde{A},\tilde{r})} = \delta_{A,\tilde{A}} \delta_{r,\tilde{r}} \frac{\chi_r^{a}(a)}{\chi_r^{a}(e)}.
    \end{equation}
We refer to \cite{Coste:2000tq} for the discussion when $G$ is anomalous.

We will consider the non-anomalous order-3 cyclic group $S_3 = \{e,a,a^2,b,ab,a^2 b \}$ as an illustration. $a$ and $b$ are the generators of $\mathbb{Z}_3$ and $\mathbb{Z}_2$ satisfying $a^3=b^2=1$ and the constraint $ba = a^2 b$. There are three conjugacy classes represented by $[e],[a],[b]$
    \begin{equation}
        [e] = \{ e \},\quad [a] = \{ a , a^2 \},\quad [b] = \{ b , ab, a^2 b\}
    \end{equation}
and the centralizer group for $e,a,b$ are separately
    \begin{equation}
        C(e) = S_3,\quad C(a) = \mathbb{Z}_3, \quad C(b) = \mathbb{Z}_2 
    \end{equation}
where $\mathbb{Z}_3$ and $\mathbb{Z}_2$ are also generated by $a$ and $b$. Recall that the irreducible representations for $\mathbb{Z}_N$ are characterized by an $\mathbb{Z}_N$-integer $k$ with
    \begin{equation}
        r_k (g) = e^{\frac{2\pi i}{N} k},\quad \textrm{($g$ is the generator of $\mathbb{Z}_N$)}
    \end{equation}
and the irreducible representations for $S_3$ contain the trivial representation $r_+$, sign representation $r_-$ and the 2-dimensional standard representation $r_E$. There are in total 8 line operators labeled by
    \begin{equation}
        ([e],r_+),([e],r_-),([e],r_E),([b],r_0),([b],r_1),([a],r_0),([a],r_1),([a],r_2)
    \end{equation}
and we shall label them as $W_0,\cdots,W_7$ in the following. The $S$ and $T$ matrices are separately given by
\begin{equation}\label{S3-S}
    S = \frac{1}{6}\left( \begin{array}{cccccccc}
    1&1&2&3&3&2&2&2\\
    1&1&2&-3&-3&2&2&2\\
    2&2&4&0&0&-2&-2&-2\\
    3&-3&0&3&-3&0&0&0\\
    3&-3&0&-3&3&0&0&0\\
    2&2&-2&0&0&4&-2&-2\\
    2&2&-2&0&0&-2&-2&4\\
    2&2&-2&0&0&-2&4&-2
    \end{array} \right)
\end{equation}
and 
    \begin{equation}\label{S3-T}
        T = \textrm{Diag}(1,1,1,1,-1,1,\omega,\omega^2).
    \end{equation}
The fusion rule is derived from $S$ and $T$ matrices and is shown in the following table.

    \begin{table}[!h]
    \footnotesize
    \hspace{-0.5cm}
    \begin{tabular}{|c|c|c|c|c|c|c|c|c|}
    \hline
    $\otimes$& $W_0$ & $W_1$ &$W_2$&$W_3$&$W_4$&$W_5$&$W_6$&$W_7$\\
    \hline
    $W_0$& $W_0$ & $W_1$ &$W_2$&$W_3$&$W_4$&$W_5$&$W_6$&$W_7$\\
    \hline
    $W_1$&$W_1$&$W_0$&$W_2$&$W_4$&$W_3$&$W_5$&$W_6$&$W_7$\\
    \hline
    $W_2$&$W_2$&$W_2$&$W_0\oplus W_1\oplus W_2$&$W_3\oplus W_4$&$W_3\oplus W_4$&$W_6\oplus W_7$&$W_5\oplus W_7$&$W_5\oplus W_6$\\
    \hline
    $W_3$&$W_3$&$W_4$&$W_3\oplus W_4$&$\makecell{W_0\oplus W_2 \oplus W_5\\ \oplus W_6 \oplus W_7}$ & $\makecell{W_1\oplus W_2 \oplus W_5\\ \oplus W_6 \oplus W_7}$&$W_3\oplus W_4$&$W_3\oplus W_4$&$W_3\oplus W_4$\\
    \hline
    $W_4$&$W_4$&$W_3$&$W_3\oplus W_4$&$\makecell{W_1\oplus W_2 \oplus W_5\\ \oplus W_6 \oplus W_7}$&$\makecell{W_0\oplus W_2 \oplus W_5\\ \oplus W_6 \oplus W_7}$&$W_3\oplus W_4$&$W_3\oplus W_4$&$W_3\oplus W_4$\\
    \hline
    $W_5$&$W_5$&$W_5$&$W_6\oplus W_7$&$W_3\oplus W_4$&$W_3\oplus W_4$&$W_0\oplus W_1 \oplus W_5$&$W_7 \oplus W_2$&$W_6\oplus W_2$\\
    \hline
    $W_6$&$W_6$&$W_6$&$W_5\oplus W_7$&$W_3\oplus W_4$&$W_3\oplus W_4$&$W_7\oplus W_2$&$W_0 \oplus W_1 \oplus W_6$&$W_5\oplus W_2$\\
    \hline
    $W_7$&$W_7$&$W_7$&$W_5\oplus W_6$&$W_3\oplus W_4$&$W_3\oplus W_4$&$W_6\oplus W_2$&$W_5 \oplus W_2$&$W_0\oplus W_1\oplus W_7$\\
    \hline
    \end{tabular}
    \caption{The fusion table of $\mathcal{Z}(S_3)$.}
    \label{tab:my_label}
\end{table}

The Hilbert space is constructed according to the method given in the previous sections. Inserting line operators $W_{\alpha}$ inside the solid torus will create $8$ states and generate a basis $|\alpha\rangle$ of the Hilbert space $\mathcal{H}_{T^2}$. The line operators act on the Hilbert space according to 
    \begin{equation}
        W_{\beta}[\Gamma_1] |\alpha \rangle = \frac{S_{\alpha \beta}}{S_{\alpha 0}} |\alpha \rangle,\quad W_{\beta}[\Gamma_2] |\alpha \rangle = \sum_{\gamma} N_{\alpha \beta}^{\gamma} |\gamma \rangle.
    \end{equation}
It is convenient to write down the states using 8-dimensional vectors such that
    \begin{equation}
        |\alpha\rangle\quad  \equiv \underbrace{|0,\cdots 0,1,0,\cdots\rangle}_\text{The $(\alpha+1)$-th component is $1$}
    \end{equation}
and then $W_{\alpha}[\Gamma_1]$ and $W_{\alpha}[\Gamma_2]$ are $8\times 8$ matrices and can be read from the $S$-matrix and the fusion rule.

There are six Lagrangian algebras, four of them are bosonic and two of them are fermionic\cite{Bhardwaj:2024ydc}. The bosonic algebras are
    \begin{gather}
        \begin{split}
        \mathcal{L}_1 = W_0 \oplus W_1 \oplus 2 W_2,\quad
        \mathcal{L}_2 = W_0 \oplus W_1 \oplus 2W_5\\
        \mathcal{L}_3 = W_0 \oplus W_2 \oplus W_3,\quad
        \mathcal{L}_4 = W_0 \oplus W_3 \oplus W_5
        \end{split}            
    \end{gather}
and the fermionic algebras are
    \begin{equation}
        \mathcal{L}_5 = W_0 \oplus W_2 \oplus W_4,\quad
        \mathcal{L}_6 = W_0 \oplus W_4 \oplus W_5            
    \end{equation}
where $W_4$ carries topological spin $\frac{1}{2}$. There are another two algebras corresponding to parafermion states and they are
    \begin{equation}
        \begin{split}
            \mathcal{L}_7 = W_0 \oplus W_1 \oplus 2W_6,\quad \mathcal{L}_8 = W_0 \oplus W_1 \oplus 2W_7.
        \end{split}
    \end{equation}
We will determine the topological boundary states for them in the following.

\subsubsection*{Topological boundary states for $\mathcal{L}_1$}

The vacuum is
    \begin{equation}
        |\Omega\rangle_{\mathcal{L}_1} = |0\rangle + |1\rangle + 2 |2\rangle = |1,1,2,0,0,0,0,0\rangle,
    \end{equation}
and one can check the following actions of line operators on the topological vacuum $|\Omega\rangle_{\mathcal{L}_1}$ summarized in table \ref{S3-A1}. 
    \begin{table}[!h]
        \centering
        \begin{tabular}{|c|c|c|}
        \hline
            Lines/cycles & $\Gamma_1$ & $\Gamma_2$ \\
            \hline
            $W_0$ & $|1,1,2,0,0,0,0,0\rangle$&$|1,1,2,0,0,0,0,0\rangle$\\
            \hline
            $W_1$ & $|1,1,2,0,0,0,0,0\rangle$&$|1,1,2,0,0,0,0,0\rangle$\\
            \hline
            $W_2$ & $|2,2,4,0,0,0,0,0\rangle$&$|2,2,4,0,0,0,0,0\rangle$\\
            \hline
            $W_3$ & $|3,-3,0,0,0,0,0,0\rangle$ & $|0,0,0,3,3,0,0,0\rangle$\\
            \hline
            $W_4$ & $|3,-3,0,0,0,0,0,0\rangle$ & $|0,0,0,3,3,0,0,0\rangle$\\
            \hline
            $W_5$ & $|2,2,-2,0,0,0,0,0\rangle$ & $|0,0,0,0,0,2,2,2\rangle$\\
            \hline
            $W_6$ & $|2,2,-2,0,0,0,0,0\rangle$ & $|0,0,0,0,0,2,2,2\rangle$\\
            \hline
            $W_7$ & $|2,2,-2,0,0,0,0,0\rangle$ & $|0,0,0,0,0,2,2,2\rangle$\\
            \hline
        \end{tabular}
        \caption{The action of line operators on the topological vacuum $|\Omega\rangle_{\mathcal{L}_1}$. Here the two vectors in each row are the results of the action of line operator $W_{\alpha}$ along $\Gamma_1$ and $\Gamma_2$ cycle on the topological vacuum $|\Omega\rangle_{\mathcal{L}_1}$, namely $W_{\alpha}[\Gamma_1]|\Omega\rangle_{\mathcal{L}_1}$ and $W_{\alpha}[\Gamma_2]|\Omega\rangle_{\mathcal{L}_1}$.}
        \label{S3-A1}
    \end{table}
    
One can see at the topological boundary we have the following identification
    \begin{equation}
        2W_0 \sim 2W_1 \sim W_2,\quad W_3 \sim W_4,\quad W_5 \sim W_6 \sim W_7.
    \end{equation}
Notice that $W_3$ has quantum dimension $3$ and $W_5$ has quantum dimension $2$, we need to determine whether they are simple or not at the boundary. Using \eqref{Hom} it is easy to find
    \begin{equation}
        \begin{split}
        \dim \textrm{Hom}_{\mathcal{L}_1}(W_{3},W_{3}) =& \langle 3 | W_{3}[\Gamma_2] |\Omega\rangle_{\mathcal{L}_1} = 3,\\
        \dim \textrm{Hom}_{\mathcal{L}_1}(W_{5},W_{5}) =& \langle 5 | W_{5}[\Gamma_2] |\Omega\rangle_{\mathcal{L}_1} = 2,
        \end{split}
    \end{equation}
which indicates $W_3$ is composed of three simple objects and $W_5$ is composed of two on the boundary. Let us denote
    \begin{equation}
        W_3 \stackrel{\mathcal{L}_1}{=} \beta_1 + \beta_2 + \beta_3,
    \end{equation}
and
    \begin{equation}
        W_5 \stackrel{\mathcal{L}_1}{=} \alpha_1 + \alpha_2.
    \end{equation}
Also, one can check the following fusion rule by acting $W_{\alpha}$ twice
        \begin{gather}
        W_3 \times_{\mathcal{L}_1} W_3  = 3 W_0 + 3W_5\\
        W_5 \times_{\mathcal{L}_1} W_5= 2W_0 + W_5\\
        W_3 \times_{\mathcal{L}_1} W_5 = W_5 \times_{\mathcal{L}_1
        } W_3 = 2 W_3.
        \end{gather}
Using the above decomposition, the second one is equivalent to
    \begin{equation}
        (\alpha_1 + \alpha_2) \times (\alpha_1 + \alpha_2) = 1 + 1 + \alpha_1 + \alpha_2,
    \end{equation}
and one can deduce that $\alpha_1,\alpha_2$ generate a $\mathbb{Z}_3$ symmetry and identify $\alpha_1 = a,\alpha_2 = a^2$ such that $a^3=1$. We also have
    \begin{equation}
        (\beta_1 + \beta_2 + \beta_3)^2 = 3 + a + a^2,\quad (\beta_1 + \beta_2 + \beta_3) \times (a+a^2) = 2\beta_1 + 2\beta_2 + 2\beta_3
    \end{equation}
and one can further deduce
    \begin{equation}
        \beta_1 = b,\quad \beta_2 = ab,\quad \beta_3 = a^2b,
    \end{equation}
where $b$ is the $\mathbb{Z}_2$ generator satisfying $b^2=1$. Therefore we can see the symmetry on the topological boundary $\mathcal{L}_1$ is $S_3$ and $W_0,W_3,W_5$ are conjugate classes of $S_3$ when restricting to the boundary. 

\subsubsection*{Topological boundary states for $\mathcal{L}_2$}
In this case, the vacuum of topological boundary state is
    \begin{equation}
        |\Omega\rangle_{\mathcal{L}_2} = |0\rangle + |1\rangle + 2|5\rangle = |1,1,0,0,0,2,0,0\rangle,
    \end{equation}
One can check the actions of line operators in table \ref{S3-A2}.
    \begin{table}[!h]
        \centering
        \begin{tabular}{|c|c|c|}
        \hline
            Lines/cycles & $\Gamma_1$ & $\Gamma_2$ \\
            \hline
            $W_0$ & $|1, 1, 0, 0, 0, 2, 0, 0\rangle$&$|1, 1, 0, 0, 0, 2, 0, 0\rangle$\\
            \hline
            $W_1$ & $|1, 1, 0, 0, 0, 2, 0, 0\rangle$&$|1, 1, 0, 0, 0, 2, 0, 0\rangle$\\
            \hline
            $W_2$ & $|2, 2, 0, 0, 0, -2, 0, 0\rangle$&$|0, 0, 2, 0, 0, 0, 2, 2\rangle$\\
            \hline
            $W_3$ & $|3, -3, 0, 0, 0, 0, 0, 0\rangle$ & $|0, 0, 0, 3, 3, 0, 0, 0\rangle$\\
            \hline
            $W_4$ & $|3, -3, 0, 0, 0, 0, 0, 0\rangle$ & $|0, 0, 0, 3, 3, 0, 0, 0\rangle$\\
            \hline
            $W_5$ & $|2, 2, 0, 0, 0, 4, 0, 0\rangle$ & $|2, 2, 0, 0, 0, 4, 0, 0\rangle$\\
            \hline
            $W_6$ & $|2, 2, 0, 0, 0, -2, 0, 0\rangle$ & $|0, 0, 2, 0, 0, 0, 2, 2\rangle$\\
            \hline
            $W_7$ & $|2, 2, 0, 0, 0, -2, 0, 0\rangle$ & $|0, 0, 2, 0, 0, 0, 2, 2\rangle$\\
            \hline
        \end{tabular}
        \caption{The action of line operators on the topological boundary state $|\Omega\rangle_{\mathcal{L}_2}$}
        \label{S3-A2}
    \end{table}
    
From the table, we should identify the following line operators on the topological boundary
    \begin{equation}
        2W_0 \sim 2W_1 \sim W_5,\quad W_3 \sim W_4,\quad W_2 \sim W_6 \sim W_7.
    \end{equation}
The discussion is the same as the $\mathcal{L}_1$ case where the role of $W_2$ is exchanged with $W_5$. One then decomposes
    \begin{equation}
        W_2 \stackrel{\mathcal{L}_2}{=} a + a^2,\quad W_3 \stackrel{\mathcal{L}_2}{=} b + ab + a^2 b,
    \end{equation}
where $a,b$ are generators of $S_3$. 
    
\subsubsection*{Topological boundary states for $\mathcal{L}_3$}
In this case, the vacuum of the topological boundary is
    \begin{equation}
        |\Omega\rangle_{\mathcal{L}_3} = |0\rangle + |2\rangle + |3\rangle =  |1,0,1,1,0,0,0,0\rangle.
    \end{equation}
One can check the actions of line operators in table \ref{S3-A3}.
    \begin{table}[!h]
        \centering
        \begin{tabular}{|c|c|c|}
        \hline
            Lines/cycles & $\Gamma_1$ & $\Gamma_2$ \\
            \hline
            $W_0$ & $|1, 0, 1, 1, 0, 0, 0, 0\rangle$&$|1, 0, 1, 1, 0, 0, 0, 0\rangle$\\
            \hline
            $W_1$ & $|1, 0, 1, -1, 0, 0, 0, 0\rangle$&$|0, 1, 1, 0, 1, 0, 0, 0\rangle$\\
            \hline
            $W_2$ & $|2, 0, 2, 0, 0, 0, 0, 0\rangle$&$|1, 1, 2, 1, 1, 0, 0, 0\rangle$\\
            \hline
            $W_3$ & $|3, 0, 0, 1, 0, 0, 0, 0\rangle$ & $|1, 0, 1, 2, 1, 1, 1, 1\rangle$\\
            \hline
            $W_4$ & $|3, 0, 0, -1, 0, 0, 0, 0\rangle$ & $|0, 1, 1, 1, 2, 1, 1, 1\rangle$\\
            \hline
            $W_5$ & $|2, 0, -1, 0, 0, 0, 0, 0\rangle$ & $|0, 0, 0, 1, 1, 1, 1, 1\rangle$\\
            \hline
            $W_6$ & $|2, 0, -1, 0, 0, 0, 0, 0\rangle$ & $|0, 0, 0, 1, 1, 1, 1, 1\rangle$\\
            \hline
            $W_7$ & $|2, 0, -1, 0, 0, 0, 0, 0\rangle$ & $|0, 0, 0, 1, 1, 1, 1, 1\rangle$\\
            \hline
        \end{tabular}
        \caption{The action of line operators on the topological boundary state $|\Omega\rangle_{\mathcal{L}_3}$}
        \label{S3-A3}
    \end{table}
    
We should identify the following line operators on the topological boundary
    \begin{equation}
        W_5 \sim W_6 \sim W_7,\quad W_2 \sim W_0+W_1,\quad W_3 \sim W_0+W_5,\quad W_4 \sim W_1+W_5,
    \end{equation}
and we can pick three of them $W_0,W_1,W_5$ as generators. From \eqref{Hom}, we can read the multiplicity
    \begin{equation}
        \begin{split}
        \dim \textrm{Hom}_{\mathcal{L}_3}(W_1,W_1) &= \langle 1 | W_1[\Gamma_2]|\Omega\rangle_{\mathcal{L}_3} = 1,\\
        \dim \textrm{Hom}_{\mathcal{L}_3}(W_5,W_5) &= \langle 5 | W_5[\Gamma_2]|\Omega\rangle_{\mathcal{L}_3} = 1,
        \end{split}
    \end{equation}
therefore all of the three line operators are simple on the boundary. We can check the fusion rule by acting them on the vacuum twice and get
    \begin{gather}
        W_1\times_{\mathcal{L}_3} W_1 = W_0\\
        W_5\times_{\mathcal{L}_3} W_5 = W_0 + W_1 + W_5\\
        W_1\times_{\mathcal{L}_3} W_5 = W_5\times_{\mathcal{L}_3} W_1 = W_5,
    \end{gather}
which are exactly the generators of Rep($S_3$): $W_0$ is the identity, $W_1$ is the $\mathbb{Z}_2$ generator and $W_5$ is the non-invertible generator.

\subsubsection*{Topological boundary states for $\mathcal{L}_4$}
In this case, the vacuum of the topological boundary is
    \begin{equation}
        |\Omega\rangle_{\mathcal{L}_4} = |0\rangle + |3\rangle + |5\rangle = |1,0,0,1,0,1,0,0\rangle
    \end{equation}
One can check the actions of line operators in the table \ref{S3-A4}.
    \begin{table}[!h]
        \centering
        \begin{tabular}{|c|c|c|}
        \hline
            Lines/cycles & $\Gamma_1$ & $\Gamma_2$ \\
            \hline
            $W_0$ & $|1, 0, 0, 1, 0, 1, 0, 0\rangle$&$|1, 0, 0, 1, 0, 1, 0, 0\rangle$\\
            \hline
            $W_1$ & $|1, 0, 0, -1, 0, 1, 0, 0\rangle$&$|0, 1, 0, 0, 1, 1, 0, 0\rangle$\\
            \hline
            $W_2$ & $|2, 0, 0, 0, 0, -1, 0, 0\rangle$&$|0, 0, 1, 1, 1, 0, 1, 1\rangle$\\
            \hline
            $W_3$ & $|3, 0, 0, 1, 0, 0, 0, 0\rangle$ & $|1, 0, 1, 2, 1, 1, 1, 1\rangle$\\
            \hline
            $W_4$ & $|3, 0, 0, -1, 0, 0, 0, 0\rangle$ & $|0, 1, 1, 1, 2, 1, 1, 1\rangle$\\
            \hline
            $W_5$ & $|2, 0, 0, 0, 0, 2, 0, 0\rangle$ & $|1, 1, 0, 1, 1, 2, 0, 0\rangle$\\
            \hline
            $W_6$ & $|2, 0, 0, 0, 0, -1, 0, 0\rangle$ & $|0, 0, 1, 1, 1, 0, 1, 1\rangle$\\
            \hline
            $W_7$ & $|2, 0, 0, 0, 0, -1, 0, 0\rangle$ & $|0, 0, 1, 1, 1, 0, 1, 1\rangle$\\
            \hline
        \end{tabular}
        \caption{The action of line operators on the topological boundary state $|\Omega\rangle_{\mathcal{L}_4}$}
        \label{S3-A4}
    \end{table}

We should identify the following line operators on the topological boundary
    \begin{equation}
        W_2 \sim W_6 \sim W_7,\quad W_5 \sim W_0+W_1,\quad W_3 \sim W_0+W_2,\quad W_4 \sim W_1+W_2,
    \end{equation}
and we can pick three generators $W_0,W_1,W_2$. The discussion is basically the same to $\mathcal{L}_1$ case where the role of $W_2$ is exchanged to $W_5$. The symmetry at the boundary is still Rep($S_3$).

\subsubsection*{Topological boundary states for $\mathcal{L}_5$}
In this case, the vacuum of the topological boundary is
    \begin{equation}
        |\Omega\rangle_{\mathcal{L}_5} = |0\rangle + |2\rangle + |4\rangle = |1,0,1,0,1,0,0,0\rangle
    \end{equation}
One can check the actions of line operators in the table \ref{S3-A5}.
    \begin{table}[!h]
        \centering
        \begin{tabular}{|c|c|c|}
        \hline
            Lines/cycles & $\Gamma_1$ & $\Gamma_2$ \\
            \hline
            $W_0$ & $|1, 0, 1, 0, 1, 0, 0, 0\rangle$&$|1, 0, 1, 0, 1, 0, 0, 0\rangle$\\
            \hline
            $W_1$ & $|1, 0, 1, 0, -1, 0, 0, 0\rangle$&$|0, 1, 1, 1, 0, 0, 0, 0\rangle$\\
            \hline
            $W_2$ & $|2, 0, 2, 0, 0, 0, 0, 0\rangle$&$|1, 1, 2, 1, 1, 0, 0, 0\rangle$\\
            \hline
            $W_3$ & $|3, 0, 0, 0, -1, 0, 0, 0\rangle$ & $|0, 1, 1, 2, 1, 1, 1, 1\rangle$\\
            \hline
            $W_4$ & $|3, 0, 0, 0, 1, 0, 0, 0\rangle$ & $|1, 0, 1, 1, 2, 1, 1, 1\rangle$\\
            \hline
            $W_5$ & $|2, 0, -1, 0, 0, 0, 0, 0\rangle$ & $|0, 0, 0, 1, 1, 1, 1, 1\rangle$\\
            \hline
            $W_6$ & $|2, 0, -1, 0, 0, 0, 0, 0\rangle$ & $|0, 0, 0, 1, 1, 1, 1, 1\rangle$\\
            \hline
            $W_7$ & $|2, 0, -1, 0, 0, 0, 0, 0\rangle$ & $|0, 0, 0, 1, 1, 1, 1, 1\rangle$\\
            \hline
        \end{tabular}
        \caption{The action of line operators on the topological boundary state $|\Omega\rangle_{\mathcal{L}_5}$}
        \label{S3-A5}
    \end{table}
First of all, the vacuum $|\Omega\rangle_{\mathcal{L}_5}$ is not invariant under the $T$ transformation and one has
    \begin{equation}
        \hat{\mathbb{T}}|\Omega\rangle_{\mathcal{L}_5} = W_1[\Gamma_1] |\Omega\rangle_{\mathcal{L}_5},
    \end{equation}
and soon we will see $W_1$ is the $\mathbb{Z}_2$ generator at the boundary. That indicates the vacuum is fermionic and the $T$-transformation switches the NS-NS sector with the NS-R sector.
    
We should identify the following line operators on the topological boundary
    \begin{equation}
        W_5 \sim W_6 \sim W_7,\quad W_2 \sim W_0+W_1,\quad W_3 \sim W_1+W_5,\quad W_4 \sim W_0+W_5,
    \end{equation}
and we can pick three generators $W_0,W_1,W_5$. According to \eqref{Hom}, we can still read
    \begin{equation}
        \begin{split}
        \dim \textrm{Hom}_{\mathcal{L}_5}(W_1,W_1) &= \langle 1 | W_1[\Gamma_2]|\Omega\rangle_{\mathcal{L}_5} = 1,\\
        \dim \textrm{Hom}_{\mathcal{L}_5}(W_5,W_5) &= \langle 5 | W_5[\Gamma_2]|\Omega\rangle_{\mathcal{L}_5} = 1. 
        \end{split}
    \end{equation}
And all of those three are simple. We can check the fusion rule by acting on the vacuum $|\Omega_5\rangle$
    \begin{gather}
        W_1\times_{\mathcal{L}_5} W_1 = W_0\\
        W_5\times_{\mathcal{L}_5} W_5 = W_0 + W_1 + W_5\\
        W_1\times_{\mathcal{L}_5} W_5 = W_5\times_{\mathcal{L}_5} W_1 = W_5,
    \end{gather}
which are exactly the generators of Rep($S_3$): $W_0$ is the identity, $W_1$ is the $\mathbb{Z}_2$ generator and $W_5$ is the non-invertible generator.

\subsubsection*{Topological boundary states for $\mathcal{L}_6$}
In this case, the vacuum of the topological boundary state is
    \begin{equation}
        |\Omega\rangle_{\mathcal{L}_6} = |0\rangle + |4\rangle + |5\rangle = |1,0,0,0,1,1,0,0\rangle.
    \end{equation}
One can check the actions of lien operators in the table \ref{S3-A6}.
    \begin{table}[!h]
        \centering
        \begin{tabular}{|c|c|c|}
        \hline
            Lines/cycles & $\Gamma_1$ & $\Gamma_2$ \\
            \hline
            $W_0$ & $|1, 0, 0, 0, 1, 1, 0, 0\rangle$&$|1, 0, 0, 0, 1, 1, 0, 0\rangle$\\
            \hline
            $W_1$ & $|1, 0, 0, 0, -1, 1, 0, 0\rangle$&$|0, 1, 0, 1, 0, 1, 0, 0\rangle$\\
            \hline
            $W_2$ & $|2, 0, 0, 0, 0, -1, 0, 0\rangle$&$|0, 0, 1, 1, 1, 0, 1, 1\rangle$\\
            \hline
            $W_3$ & $|3, 0, 0, 0, -1, 0, 0, 0\rangle$ & $|0, 1, 1, 2, 1, 1, 1, 1\rangle$\\
            \hline
            $W_4$ & $|3, 0, 0, 0, 1, 0, 0, 0\rangle$ & $|1, 0, 1, 1, 2, 1, 1, 1\rangle$\\
            \hline
            $W_5$ & $|2, 0, 0, 0, 0, 2, 0, 0\rangle$ & $|1, 1, 0, 1, 1, 2, 0, 0\rangle$\\
            \hline
            $W_6$ & $|2, 0, 0, 0, 0, -1, 0, 0\rangle$ & $|0, 0, 1, 1, 1, 0, 1, 1\rangle$\\
            \hline
            $W_7$ & $|2, 0, 0, 0, 0, -1, 0, 0\rangle$ & $|0, 0, 1, 1, 1, 0, 1, 1\rangle$\\
            \hline
        \end{tabular}
        \caption{The action of line operators on the topological boundary state $|\Omega\rangle_{\mathcal{L}_6}$}
        \label{S3-A6}
    \end{table}
Just like the $\mathcal{L}_5$ case, the vacuum $|\Omega\rangle_{\mathcal{L}_6}$ is not invariant under the $T$ transformation due to the same reason and one has
    \begin{equation}
        \hat{\mathbb{T}}|\Omega\rangle_{\mathcal{L}_6} = W_1[\Gamma_1] |\Omega\rangle_{\mathcal{L}_6},
    \end{equation}
where $W_1$ turns out to be the $\mathbb{Z}_2$ generator at the boundary. 

We should identify the following line operators on the topological boundary
    \begin{equation}
        W_2 \sim W_6 \sim W_7,\quad W_3 \sim W_1+W_2,\quad W_4 \sim W_0+W_2,\quad W_5 \sim W_0+W_1,
    \end{equation}
and we can pick three generators $W_0,W_1,W_2$. The discussion is the same as the $\mathcal{L}_5$ case where the role of $W_5$ is exchanged to $W_2$. The symmetry at the boundary is still Rep($S_3$).

\subsubsection*{Topological boundary states for $\mathcal{L}_7$ and $\mathcal{L}_8$}
Finally, let us consider the vacua of the topological boundary states for $\mathcal{L}_7$
    \begin{equation}
        |\Omega\rangle_{\mathcal{L}_7} = |0\rangle + |1\rangle + 2|6\rangle = |1,1,0,0,0,0,2,0\rangle
    \end{equation}
 and $\mathcal{L}_8$
    \begin{equation}
        |\Omega\rangle_{\mathcal{L}_8} = |0\rangle + |1\rangle + 2|7\rangle = |1,1,0,0,0,0,0,2\rangle.
    \end{equation}
and they exchange with each other under the S-transformation
    \begin{equation}
        \hat{\mathbb{S}}|\Omega\rangle_{\mathcal{L}_7} = |\Omega\rangle_{\mathcal{L}_8},\quad \hat{\mathbb{S}}|\Omega\rangle_{\mathcal{L}_8} = |\Omega\rangle_{\mathcal{L}_7}.
    \end{equation}
The actions of line operators on the topological boundary states are collected in table \ref{S3-L7} and table \ref{S3-L8} for $\mathcal{L}_7$ and $\mathcal{L}_8$. 
    \begin{table}[!h]
        \centering
        \begin{tabular}{|c|c|c|}
        \hline
            Lines/cycles & $\Gamma_1$ & $\Gamma_2$ \\
            \hline
            $W_0$ & $|1, 1, 0, 0, 0, 0, 2, 0\rangle$&$|1, 1, 0, 0, 0, 0, 2, 0\rangle$\\
            \hline
            $W_1$ & $|1, 1, 0, 0, 0, 0, 2, 0\rangle$&$|1, 1, 0, 0, 0, 0, 2, 0\rangle$\\
            \hline
            $W_2$ & $|2, 2, 0, 0, 0, 0, -2, 0\rangle$&$|0, 0, 2, 0, 0, 2, 0, 2\rangle$\\
            \hline
            $W_3$ & $|3, -3, 0, 0, 0, 0, 0, 0\rangle$ & $|0, 0, 0, 3, 3, 0, 0, 0\rangle$\\
            \hline
            $W_4$ & $|3, -3, 0, 0, 0, 0, 0, 0\rangle$ & $|0, 0, 0, 3, 3, 0, 0, 0\rangle$\\
            \hline
            $W_5$ & $|2, 2, 0, 0, 0, 0, -2, 0\rangle$ & $|0, 0, 2, 0, 0, 2, 0, 2\rangle$\\
            \hline
            $W_6$ & $|2, 2, 0, 0, 0, 0, -2, 0\rangle$ & $|2, 2, 0, 0, 0, 0, 4, 0\rangle$\\
            \hline
            $W_7$ & $|2, 2, 0, 0, 0, 0, 4, 0\rangle$ & $|0, 0, 2, 0, 0, 2, 0, 2\rangle$\\
            \hline
        \end{tabular}
        \caption{The action of line operators on the topological boundary state $|\Omega\rangle_{\mathcal{L}_7}$}
        \label{S3-L7}
    \end{table}

    \begin{table}[!h]
        \centering
        \begin{tabular}{|c|c|c|}
        \hline
            Lines/cycles & $\Gamma_1$ & $\Gamma_2$ \\
            \hline
            $W_0$ & $|1, 1, 0, 0, 0, 0, 0, 2\rangle$&$|1, 1, 0, 0, 0, 0, 0, 2\rangle$\\
            \hline
            $W_1$ & $|1, 1, 0, 0, 0, 0, 0, 2\rangle$&$|1, 1, 0, 0, 0, 0, 0, 2\rangle$\\
            \hline
            $W_2$ & $|2, 2, 0, 0, 0, 0, 0, -2\rangle$&$|0, 0, 2, 0, 0, 2, 2, 0\rangle$\\
            \hline
            $W_3$ & $|3, -3, 0, 0, 0, 0, 0, 0\rangle$ & $|0, 0, 0, 3, 3, 0, 0, 0\rangle$\\
            \hline
            $W_4$ & $|3, -3, 0, 0, 0, 0, 0, 0\rangle$ & $|0, 0, 0, 3, 3, 0, 0, 0\rangle$\\
            \hline
            $W_5$ & $|2, 2, 0, 0, 0, 0, 0, -2\rangle$ & $|0, 0, 2, 0, 0, 2, 2, 0\rangle$\\
            \hline
            $W_6$ & $|2, 2, 0, 0, 0, 0, 0, 4\rangle$ & $|0, 0, 2, 0, 0, 2, 2, 0\rangle$\\
            \hline
            $W_7$ & $|2, 2, 0, 0, 0, 0, 0, -2\rangle$ & $|2, 2, 0, 0, 0, 0, 0, 4\rangle$\\
            \hline
        \end{tabular}
        \caption{The action of line operators on the topological boundary state $|\Omega\rangle_{\mathcal{L}_8}$}
        \label{S3-L8}
    \end{table}

Similar to the $\mathcal{L}_{pf}$ in $\mathbb{Z}_N$ case, the identification of line operators along $\Gamma_1$ and $\Gamma_2$ are different. For example, upon $|\Omega\rangle_{\mathcal{L}_7}$ one has
\begin{equation}
    2W_0[\Gamma_1] \sim 2W_1[\Gamma_1] \sim W_7[\Gamma_1],\quad W_2[\Gamma_1] \sim W_5[\Gamma_1] \sim W_6[\Gamma_1],\quad W_3[\Gamma_1] \sim W_4[\Gamma_1]
\end{equation}
along $\Gamma_1$ cycle and
\begin{equation}
    2W_0[\Gamma_2] \sim 2W_1[\Gamma_2] \sim W_6[\Gamma_2],\quad W_2[\Gamma_2] \sim W_5[\Gamma_2] \sim W_7[\Gamma_2],\quad W_3[\Gamma_2] \sim W_4[\Gamma_2]
\end{equation}
along $\Gamma_2$ cycle. Similar to $|\Omega\rangle_{\mathcal{L}_8}$ with the role of $W_6,W_7$ exchanged. Nevertheless, for both $\Gamma_1$ and $\Gamma_2$ one can check the symmetries on the topological boundary $|\Omega\rangle_{\mathcal{L}_7}$ and $|\Omega\rangle_{\mathcal{L}_8}$ are still given by $S_3$.

\subsubsection*{Dualities in terms of topological boundary states}
For any theory $\mathfrak{T}_{S_3}$ with a non-anomalous $S_3$ symmetry, we can first build a dynamical boundary state $|\chi\rangle_{\mathfrak{T}_{S_3}}$ based on the topological boundary states of $\mathcal{L}_1$. Recall that we have identified $W_3$ as the conjugate class of $b$ and $W_5$ as the conjugate class of $a$, the dynamical boundary state is defined as
    \begin{align}
        |\chi\rangle_{\mathfrak{T}_{S_3}} =& Z_{\mathfrak{T}_{S_3}}
\,\begin{gathered}
\begin{tikzpicture}[scale=.5]
\node at (0, 1.45) {};
\draw[line] (-.15,-.2)--(-.25, -.2) -- (-.25, 1.4)-- (-.15,1.4);
\draw[very thick] (0,0) rectangle (1.2,1.2);
\draw[line] (1.35, -.2)--(1.45, -.2) -- (1.45, 1.4)--(1.35, 1.4);
\end{tikzpicture}
\end{gathered}\,
|\Omega\rangle_{\mathcal{L}_1} + Z_{\mathfrak{T}_{S_3}}
\,\begin{gathered}
\begin{tikzpicture}[scale=.5]
\node at (0, 1.45) {};
\draw[line] (-.15,-.2)--(-.25, -.2) -- (-.25, 1.4)-- (-.15,1.4);
\draw[very thick] (0,0) rectangle (1.2,1.2);
\draw[line, thick] (0,.6)--(1.2,.6);
\draw[line] (1.35, -.2)--(1.45, -.2) -- (1.45, 1.4)--(1.35, 1.4);
\end{tikzpicture}
\end{gathered}\,
W_3[\Gamma_1]|\Omega\rangle_{\mathcal{L}_1} + Z_{\mathfrak{T}_{S_3}}
\,\begin{gathered}
\begin{tikzpicture}[scale=.5]
\node at (0, 1.45) {};
\draw[line] (-.15,-.2)--(-.25, -.2) -- (-.25, 1.4)-- (-.15,1.4);
\draw[very thick] (0,0) rectangle (1.2,1.2);
\draw[line, thick] (.6,0)--(.6,1.2);
\draw[line] (1.35, -.2)--(1.45, -.2) -- (1.45, 1.4)--(1.35, 1.4);
\end{tikzpicture}
\end{gathered}\,
W_3[\Gamma_2]|\Omega\rangle_{\mathcal{L}_1} \nonumber\\
       +&  Z_{\mathfrak{T}_{S_3}}
\,\begin{gathered}
\begin{tikzpicture}[scale=.5]
\node at (0, 1.45) {};
\draw[line] (-.15,-.2)--(-.25, -.2) -- (-.25, 1.4)-- (-.15,1.4);
\draw[very thick] (0,0) rectangle (1.2,1.2);
\draw  (.6,0) arc (0:90:.6); 
\draw (.6,1.2) arc (-180:-90:.6);
\draw[line] (1.35, -.2)--(1.45, -.2) -- (1.45, 1.4)--(1.35, 1.4);
\end{tikzpicture}
\end{gathered}\,
\hat{\mathbb{T}} W_3[\Gamma_2]|\Omega\rangle_{\mathcal{L}_1} + Z_{\mathfrak{T}_{S_3}}
\,\begin{gathered}
\begin{tikzpicture}[scale=.5]
\node at (0, 1.45) {};
\draw[line] (-.15,-.2)--(-.25, -.2) -- (-.25, 1.4)-- (-.15,1.4);
\draw[very thick] (0,0) rectangle (1.2,1.2);
\draw[line, densely dotted, thick] (0,.6)--(1.2,.6);
\draw[line] (1.35, -.2)--(1.45, -.2) -- (1.45, 1.4)--(1.35, 1.4);
\end{tikzpicture}
\end{gathered}\,
W_5[\Gamma_1]|\Omega\rangle_{\mathcal{L}_1} + Z_{\mathfrak{T}_{S_3}}
\,\begin{gathered}
\begin{tikzpicture}[scale=.5]
\node at (0, 1.45) {};
\draw[line] (-.15,-.2)--(-.25, -.2) -- (-.25, 1.4)-- (-.15,1.4);
\draw[very thick] (0,0) rectangle (1.2,1.2);
\draw[line, densely dotted, thick] (.6,0)--(.6,1.2);
\draw[line] (1.35, -.2)--(1.45, -.2) -- (1.45, 1.4)--(1.35, 1.4);
\end{tikzpicture}
\end{gathered}\,
W_5[\Gamma_2]|\Omega\rangle_{\mathcal{L}_1}\nonumber\\
       +& Z_{\mathfrak{T}_{S_3}}
\,\begin{gathered}
\begin{tikzpicture}[scale=.5]
\node at (0, 1.45) {};
\draw[line] (-.15,-.2)--(-.25, -.2) -- (-.25, 1.4)-- (-.15,1.4);
\draw[very thick] (0,0) rectangle (1.2,1.2);
\draw[densely dotted,thick]  (.6,0) arc (0:90:.6); 
\draw[densely dotted,thick] (.6,1.2) arc (-180:-90:.6);
\draw[line] (1.35, -.2)--(1.45, -.2) -- (1.45, 1.4)--(1.35, 1.4);
\end{tikzpicture}
\end{gathered}\,
\hat{\mathbb{T}} W_5[\Gamma_2]|\Omega\rangle_{\mathcal{L}_1} + Z_{\mathfrak{T}_{S_3}}
\,\begin{gathered}
\begin{tikzpicture}[scale=.5]
\node at (0, 1.45) {};
\draw[line] (-.15,-.2)--(-.25, -.2) -- (-.25, 1.4)-- (-.15,1.4);
\draw[very thick] (0,0) rectangle (1.2,1.2);
\draw[densely dotted,thick]  (.6,0) .. controls (.6,.4) and (.0,.4).. (0,.4);
\draw[densely dotted,thick]  (.6,1.2) .. controls (.6,.8) and (1.2,.8) .. (1.2,.8);
\draw[densely dotted,thick] (0,.8) .. controls (.6,.8) and (.6,.4) .. (1.2,.4);
\draw[line] (1.35, -.2)--(1.45, -.2) -- (1.45, 1.4)--(1.35, 1.4);
\end{tikzpicture}
\end{gathered}\,
\hat{\mathbb{T}}^2 W_5[\Gamma_2]|\Omega\rangle_{\mathcal{L}_1}
    \end{align}
where the action of line operators are collected in the table \ref{S3-A1} and the $T$-matrix is \eqref{S3-T}. The coefficients are partition functions of $\mathfrak{T}_{S_3}$ for each sector, where the solid line represents the $\mathbb{Z}_2$ generator $b$ and the dashed line represents the $\mathbb{Z}_3$ generator $a$. For later convenience, we will also write down the dynamical boundary states in terms of the explicit state vector
    \begin{align}
        |\chi\rangle_{\mathfrak{T}_{S_3}} =& Z_{\mathfrak{T}_{S_3}}\,\begin{gathered}
\begin{tikzpicture}[scale=.5]
\node at (0, 1.45) {};
\draw[line] (-.15,-.2)--(-.25, -.2) -- (-.25, 1.4)-- (-.15,1.4);
\draw[very thick] (0,0) rectangle (1.2,1.2);
\draw[line] (1.35, -.2)--(1.45, -.2) -- (1.45, 1.4)--(1.35, 1.4);
\end{tikzpicture}
\end{gathered}\,|1,1,2,0,0,0,0,0\rangle + Z_{\mathfrak{T}_{S_3}}
\,\begin{gathered}
\begin{tikzpicture}[scale=.5]
\node at (0, 1.45) {};
\draw[line] (-.15,-.2)--(-.25, -.2) -- (-.25, 1.4)-- (-.15,1.4);
\draw[very thick] (0,0) rectangle (1.2,1.2);
\draw[line, thick] (0,.6)--(1.2,.6);
\draw[line] (1.35, -.2)--(1.45, -.2) -- (1.45, 1.4)--(1.35, 1.4);
\end{tikzpicture}
\end{gathered}\,
|3,-3,0,0,0,0,0,0\rangle \nonumber\\
        +& Z_{\mathfrak{T}_{S_3}}
        \,\begin{gathered}
\begin{tikzpicture}[scale=.5]
\node at (0, 1.45) {};
\draw[line] (-.15,-.2)--(-.25, -.2) -- (-.25, 1.4)-- (-.15,1.4);
\draw[very thick] (0,0) rectangle (1.2,1.2);
\draw[line, thick] (.6,0)--(.6,1.2);
\draw[line] (1.35, -.2)--(1.45, -.2) -- (1.45, 1.4)--(1.35, 1.4);
\end{tikzpicture}
\end{gathered}\,
|0,0,0,3,3,0,0,0\rangle + Z_{\mathfrak{T}_{S_3}}
\,\begin{gathered}
\begin{tikzpicture}[scale=.5]
\node at (0, 1.45) {};
\draw[line] (-.15,-.2)--(-.25, -.2) -- (-.25, 1.4)-- (-.15,1.4);
\draw[very thick] (0,0) rectangle (1.2,1.2);
\draw  (.6,0) arc (0:90:.6); 
\draw (.6,1.2) arc (-180:-90:.6);
\draw[line] (1.35, -.2)--(1.45, -.2) -- (1.45, 1.4)--(1.35, 1.4);
\end{tikzpicture}
\end{gathered}\,
|0,0,0,3,-3,0,0,0\rangle\nonumber\\
        +& Z_{\mathfrak{T}_{S_3}}
        \,\begin{gathered}
\begin{tikzpicture}[scale=.5]
\node at (0, 1.45) {};
\draw[line] (-.15,-.2)--(-.25, -.2) -- (-.25, 1.4)-- (-.15,1.4);
\draw[very thick] (0,0) rectangle (1.2,1.2);
\draw[line, densely dotted, thick] (0,.6)--(1.2,.6);
\draw[line] (1.35, -.2)--(1.45, -.2) -- (1.45, 1.4)--(1.35, 1.4);
\end{tikzpicture}
\end{gathered}\,
|2,2,-2,0,0,0,0,0\rangle + Z_{\mathfrak{T}_{S_3}}
\,\begin{gathered}
\begin{tikzpicture}[scale=.5]
\node at (0, 1.45) {};
\draw[line] (-.15,-.2)--(-.25, -.2) -- (-.25, 1.4)-- (-.15,1.4);
\draw[very thick] (0,0) rectangle (1.2,1.2);
\draw[line, densely dotted, thick] (.6,0)--(.6,1.2);
\draw[line] (1.35, -.2)--(1.45, -.2) -- (1.45, 1.4)--(1.35, 1.4);
\end{tikzpicture}
\end{gathered}\,
|0,0,0,0,0,2,2,2\rangle\nonumber\\
       +& Z_{\mathfrak{T}_{S_3}}
       \,\begin{gathered}
\begin{tikzpicture}[scale=.5]
\node at (0, 1.45) {};
\draw[line] (-.15,-.2)--(-.25, -.2) -- (-.25, 1.4)-- (-.15,1.4);
\draw[very thick] (0,0) rectangle (1.2,1.2);
\draw[densely dotted,thick]  (.6,0) arc (0:90:.6); 
\draw[densely dotted,thick] (.6,1.2) arc (-180:-90:.6);
\draw[line] (1.35, -.2)--(1.45, -.2) -- (1.45, 1.4)--(1.35, 1.4);
\end{tikzpicture}
\end{gathered}\,
|0,0,0,0,0,2,2\omega,2\omega^2\rangle + Z_{\mathfrak{T}_{S_3}}
\,\begin{gathered}
\begin{tikzpicture}[scale=.5]
\node at (0, 1.45) {};
\draw[line] (-.15,-.2)--(-.25, -.2) -- (-.25, 1.4)-- (-.15,1.4);
\draw[very thick] (0,0) rectangle (1.2,1.2);
\draw[densely dotted,thick]  (.6,0) .. controls (.6,.4) and (.0,.4).. (0,.4);
\draw[densely dotted,thick]  (.6,1.2) .. controls (.6,.8) and (1.2,.8) .. (1.2,.8);
\draw[densely dotted,thick] (0,.8) .. controls (.6,.8) and (.6,.4) .. (1.2,.4);
\draw[line] (1.35, -.2)--(1.45, -.2) -- (1.45, 1.4)--(1.35, 1.4);
\end{tikzpicture}
\end{gathered}\,
|0,0,0,0,0,2,2\omega^2,2\omega\rangle\ 
    \end{align}
where $\omega$ is the third root of unity. One may wonder why we do not consider the partition functions of other sectors, for example, the partition function $Z_{\mathfrak{T}_{S_3}}\,\begin{gathered}
\begin{tikzpicture}[scale=.5]
\node at (0, 1.45) {};
\draw[line] (-.15,-.2)--(-.25, -.2) -- (-.25, 1.4)-- (-.15,1.4);
\draw[very thick] (0,0) rectangle (1.2,1.2);
\draw[line, densely dotted, thick] (.3,0)--(.3,1.2);
\draw[line, densely dotted, thick] (.9,0)--(.9,1.2);
\draw[line] (1.35, -.2)--(1.45, -.2) -- (1.45, 1.4)--(1.35, 1.4);
\end{tikzpicture}
\end{gathered}$ twisted by $a^2$. It is because $a^2$ is conjugated to $a$ by $a^2 = b a b^{-1}$ and therefore we simply have $Z_{\mathfrak{T}_{S_3}}\,\begin{gathered}
\begin{tikzpicture}[scale=.5]
\node at (0, 1.45) {};
\draw[line] (-.15,-.2)--(-.25, -.2) -- (-.25, 1.4)-- (-.15,1.4);
\draw[very thick] (0,0) rectangle (1.2,1.2);
\draw[line, densely dotted, thick] (.3,0)--(.3,1.2);
\draw[line, densely dotted, thick] (.9,0)--(.9,1.2);
\draw[line] (1.35, -.2)--(1.45, -.2) -- (1.45, 1.4)--(1.35, 1.4);
\end{tikzpicture}
\end{gathered} =  Z_{\mathfrak{T}_{S_3}}\,\begin{gathered}
\begin{tikzpicture}[scale=.5]
\node at (0, 1.45) {};
\draw[line] (-.15,-.2)--(-.25, -.2) -- (-.25, 1.4)-- (-.15,1.4);
\draw[very thick] (0,0) rectangle (1.2,1.2);
\draw[line, thick] (.3,0)--(.3,1.2);
\draw[line, densely dotted, thick] (.6,0)--(.6,1.2);
\draw[line, thick] (.9,0)--(.9,1.2);
\draw[line] (1.35, -.2)--(1.45, -.2) -- (1.45, 1.4)--(1.35, 1.4);
\end{tikzpicture}
\end{gathered} = Z_{\mathfrak{T}_{S_3}}\,\begin{gathered}
\begin{tikzpicture}[scale=.5]
\node at (0, 1.45) {};
\draw[line] (-.15,-.2)--(-.25, -.2) -- (-.25, 1.4)-- (-.15,1.4);
\draw[very thick] (0,0) rectangle (1.2,1.2);
\draw[line, densely dotted, thick] (.6,0)--(.6,1.2);
\draw[line] (1.35, -.2)--(1.45, -.2) -- (1.45, 1.4)--(1.35, 1.4);
\end{tikzpicture}
\end{gathered}$, where in the middle we can move one of the solid $b$-line to fuse another $b$-line along the torus. Indeed, if we denote the partition function twisted by $g,h$ along $\Gamma_1$ and $\Gamma_2$ direction as $Z[g,h]$, where $gh=hg$. Then one has $Z[kgk^{-1},khk^{-1}] = Z[g,h]$ and the partition function is only dependent on the conjugacy class\cite{Dijkgraaf:1989hb}. Actually, one can write
    \begin{equation}
        Z_{\mathfrak{T}_{S_3}}\,\begin{gathered}
\begin{tikzpicture}[scale=.5]
\node at (0, 1.45) {};
\draw[line] (-.15,-.2)--(-.25, -.2) -- (-.25, 1.4)-- (-.15,1.4);
\draw[very thick] (0,0) rectangle (1.2,1.2);
\draw[line, densely dotted, thick] (.6,0)--(.6,1.2);
\draw[line] (1.35, -.2)--(1.45, -.2) -- (1.45, 1.4)--(1.35, 1.4);
\end{tikzpicture}
\end{gathered}\,
|0,0,0,0,0,2,2,2\rangle = Z_{\mathfrak{T}_{S_3}}\,\begin{gathered}
\begin{tikzpicture}[scale=.5]
\node at (0, 1.45) {};
\draw[line] (-.15,-.2)--(-.25, -.2) -- (-.25, 1.4)-- (-.15,1.4);
\draw[very thick] (0,0) rectangle (1.2,1.2);
\draw[line, densely dotted, thick] (.6,0)--(.6,1.2);
\draw[line] (1.35, -.2)--(1.45, -.2) -- (1.45, 1.4)--(1.35, 1.4);
\end{tikzpicture}
\end{gathered}\,
|0,0,0,0,0,1,1,1\rangle + Z_{\mathfrak{T}_{S_3}}\,\begin{gathered}
\begin{tikzpicture}[scale=.5]
\node at (0, 1.45) {};
\draw[line] (-.15,-.2)--(-.25, -.2) -- (-.25, 1.4)-- (-.15,1.4);
\draw[very thick] (0,0) rectangle (1.2,1.2);
\draw[line, densely dotted, thick] (.3,0)--(.3,1.2);
\draw[line, densely dotted, thick] (.9,0)--(.9,1.2);
\draw[line] (1.35, -.2)--(1.45, -.2) -- (1.45, 1.4)--(1.35, 1.4);
\end{tikzpicture}
\end{gathered}\,
|0,0,0,0,0,1,1,1\rangle
    \end{equation}
if one wants to take account of both sectors in the conjugacy class $[a]$.

The dualities in $\mathfrak{T}_{S_3}$ can be described by changing the topological boundary state while fixing the dynamical boundary state $|\chi\rangle$. In general, we have
    \begin{equation}
        \widetilde{Z}_{\mathfrak{T}_{S_3}} = \frac{1}{6} \langle \textrm{top} | \chi\rangle_{\mathfrak{T}_{S_3}},
    \end{equation}
where the normalization factor can also be absorbed into the definition of the vacuum state. We will present the projection onto the vacua $|\Omega\rangle_{\mathcal{L}}$ for each $\mathcal{L}$ in the following. For the vacuum $|\Omega\rangle_{\mathcal{L}_1}$ and other twisted sectors, we simply get the partition functions of the original theory.
We will illustrate the procedure using the topological boundary states for $\mathcal{L}_2$ in detail. Project the dynamical boundary state $|\chi\rangle_{\mathfrak{T}_{S_3}}$ onto the vacuum $|\Omega\rangle_{\mathcal{L}_2}=|1,1,0,0,0,2,0,0\rangle$ we obtain the dual partition function
    \begin{align}
        Z_{\mathfrak{T}_{S_3}/\mathbb{Z}_3} 
        \,\begin{gathered}
\begin{tikzpicture}[scale=.5]
\node at (0, 1.45) {};
\draw[line] (-.15,-.2)--(-.25, -.2) -- (-.25, 1.4)-- (-.15,1.4);
\draw[very thick] (0,0) rectangle (1.2,1.2);
\draw[line] (1.35, -.2)--(1.45, -.2) -- (1.45, 1.4)--(1.35, 1.4);
\end{tikzpicture}
\end{gathered}\,
= \frac{1}{3}Z_{\mathfrak{T}_{S_3}}
\,\begin{gathered}
\begin{tikzpicture}[scale=.5]
\node at (0, 1.45) {};
\draw[line] (-.15,-.2)--(-.25, -.2) -- (-.25, 1.4)-- (-.15,1.4);
\draw[very thick] (0,0) rectangle (1.2,1.2);
\draw[line] (1.35, -.2)--(1.45, -.2) -- (1.45, 1.4)--(1.35, 1.4);
\end{tikzpicture}
\end{gathered}\,
+ \frac{2}{3}Z_{\mathfrak{T}_{S_3}}
\,\begin{gathered}
\begin{tikzpicture}[scale=.5]
\node at (0, 1.45) {};
\draw[line] (-.15,-.2)--(-.25, -.2) -- (-.25, 1.4)-- (-.15,1.4);
\draw[very thick] (0,0) rectangle (1.2,1.2);
\draw[line, densely dotted, thick] (0,.6)--(1.2,.6);
\draw[line] (1.35, -.2)--(1.45, -.2) -- (1.45, 1.4)--(1.35, 1.4);
\end{tikzpicture}
\end{gathered}\,
+ \frac{2}{3}Z_{\mathfrak{T}_{S_3}}
\,\begin{gathered}
\begin{tikzpicture}[scale=.5]
\node at (0, 1.45) {};
\draw[line] (-.15,-.2)--(-.25, -.2) -- (-.25, 1.4)-- (-.15,1.4);
\draw[very thick] (0,0) rectangle (1.2,1.2);
\draw[line, densely dotted, thick] (.6,0)--(.6,1.2);
\draw[line] (1.35, -.2)--(1.45, -.2) -- (1.45, 1.4)--(1.35, 1.4);
\end{tikzpicture}
\end{gathered}\,
+ \frac{2}{3}Z_{\mathfrak{T}_{S_3}}
\,\begin{gathered}
\begin{tikzpicture}[scale=.5]
\node at (0, 1.45) {};
\draw[line] (-.15,-.2)--(-.25, -.2) -- (-.25, 1.4)-- (-.15,1.4);
\draw[very thick] (0,0) rectangle (1.2,1.2);
\draw[densely dotted,thick]  (.6,0) arc (0:90:.6); 
\draw[densely dotted,thick] (.6,1.2) arc (-180:-90:.6);
\draw[line] (1.35, -.2)--(1.45, -.2) -- (1.45, 1.4)--(1.35, 1.4);
\end{tikzpicture}
\end{gathered}\,
+ \frac{2}{3}Z_{\mathfrak{T}_{S_3}}
\,\begin{gathered}
\begin{tikzpicture}[scale=.5]
\node at (0, 1.45) {};
\draw[line] (-.15,-.2)--(-.25, -.2) -- (-.25, 1.4)-- (-.15,1.4);
\draw[very thick] (0,0) rectangle (1.2,1.2);
\draw[densely dotted,thick]  (.6,0) .. controls (.6,.4) and (.0,.4).. (0,.4);
\draw[densely dotted,thick]  (.6,1.2) .. controls (.6,.8) and (1.2,.8) .. (1.2,.8);
\draw[densely dotted,thick] (0,.8) .. controls (.6,.8) and (.6,.4) .. (1.2,.4);
\draw[line] (1.35, -.2)--(1.45, -.2) -- (1.45, 1.4)--(1.35, 1.4);
\end{tikzpicture}
\end{gathered}\,
.
    \end{align}
The factor $2$ on the numerators reflects the contributions of both $a$ and $a^2$ in the same conjugacy class $[a]$. We can see it corresponds to gauging the $\mathbb{Z}_3$ symmetry. Recall that on $\mathcal{L}_2$ boundary the conjugate class $[a]$ and $[b]$ are represented by $W_2$ and $W_3$. Project $|\chi\rangle_{\mathfrak{T}_{S_3}}$ onto the states $W_2[\Gamma_1]|\Omega\rangle_{\mathcal{L}_2} = |2,2,0,0,0,-2,0,0\rangle$ and $W_2[\Gamma_2]|\Omega\rangle_{\mathcal{L}_2} = |0,0,2,0,0,0,2,2\rangle$ we get separately
    \begin{align}
        Z_{\mathfrak{T}_{S_3}/\mathbb{Z}_3}
        \,\begin{gathered}
\begin{tikzpicture}[scale=.5]
\node at (0, 1.45) {};
\draw[line] (-.15,-.2)--(-.25, -.2) -- (-.25, 1.4)-- (-.15,1.4);
\draw[very thick] (0,0) rectangle (1.2,1.2);
\draw[line, densely dotted, thick] (0,.6)--(1.2,.6);
\draw[line] (1.35, -.2)--(1.45, -.2) -- (1.45, 1.4)--(1.35, 1.4);
\end{tikzpicture}
\end{gathered}\,
= \frac{1}{3}Z_{\mathfrak{T}_{S_3}}
\,\begin{gathered}
\begin{tikzpicture}[scale=.5]
\node at (0, 1.45) {};
\draw[line] (-.15,-.2)--(-.25, -.2) -- (-.25, 1.4)-- (-.15,1.4);
\draw[very thick] (0,0) rectangle (1.2,1.2);
\draw[line] (1.35, -.2)--(1.45, -.2) -- (1.45, 1.4)--(1.35, 1.4);
\end{tikzpicture}
\end{gathered}\,
+ \frac{2}{3}Z_{\mathfrak{T}_{S_3}}
\,\begin{gathered}
\begin{tikzpicture}[scale=.5]
\node at (0, 1.45) {};
\draw[line] (-.15,-.2)--(-.25, -.2) -- (-.25, 1.4)-- (-.15,1.4);
\draw[very thick] (0,0) rectangle (1.2,1.2);
\draw[line, densely dotted, thick] (0,.6)--(1.2,.6);
\draw[line] (1.35, -.2)--(1.45, -.2) -- (1.45, 1.4)--(1.35, 1.4);
\end{tikzpicture}
\end{gathered}\,
- \frac{1}{3}Z_{\mathfrak{T}_{S_3}}
\,\begin{gathered}
\begin{tikzpicture}[scale=.5]
\node at (0, 1.45) {};
\draw[line] (-.15,-.2)--(-.25, -.2) -- (-.25, 1.4)-- (-.15,1.4);
\draw[very thick] (0,0) rectangle (1.2,1.2);
\draw[line, densely dotted, thick] (.6,0)--(.6,1.2);
\draw[line] (1.35, -.2)--(1.45, -.2) -- (1.45, 1.4)--(1.35, 1.4);
\end{tikzpicture}
\end{gathered}\,
- \frac{1}{3}Z_{\mathfrak{T}_{S_3}}
\,\begin{gathered}
\begin{tikzpicture}[scale=.5]
\node at (0, 1.45) {};
\draw[line] (-.15,-.2)--(-.25, -.2) -- (-.25, 1.4)-- (-.15,1.4);
\draw[very thick] (0,0) rectangle (1.2,1.2);
\draw[densely dotted,thick]  (.6,0) arc (0:90:.6); 
\draw[densely dotted,thick] (.6,1.2) arc (-180:-90:.6);
\draw[line] (1.35, -.2)--(1.45, -.2) -- (1.45, 1.4)--(1.35, 1.4);
\end{tikzpicture}
\end{gathered}\,
- \frac{1}{3}Z_{\mathfrak{T}_{S_3}}
\,\begin{gathered}
\begin{tikzpicture}[scale=.5]
\node at (0, 1.45) {};
\draw[line] (-.15,-.2)--(-.25, -.2) -- (-.25, 1.4)-- (-.15,1.4);
\draw[very thick] (0,0) rectangle (1.2,1.2);
\draw[densely dotted,thick]  (.6,0) .. controls (.6,.4) and (.0,.4).. (0,.4);
\draw[densely dotted,thick]  (.6,1.2) .. controls (.6,.8) and (1.2,.8) .. (1.2,.8);
\draw[densely dotted,thick] (0,.8) .. controls (.6,.8) and (.6,.4) .. (1.2,.4);
\draw[line] (1.35, -.2)--(1.45, -.2) -- (1.45, 1.4)--(1.35, 1.4);
\end{tikzpicture}
\end{gathered}\,
.
    \end{align}    
and 
    \begin{align}
        Z_{\mathfrak{T}_{S_3}/\mathbb{Z}_3}
        \,\begin{gathered}
\begin{tikzpicture}[scale=.5]
\node at (0, 1.45) {};
\draw[line] (-.15,-.2)--(-.25, -.2) -- (-.25, 1.4)-- (-.15,1.4);
\draw[very thick] (0,0) rectangle (1.2,1.2);
\draw[line, densely dotted, thick] (.6,0)--(.6,1.2);
\draw[line] (1.35, -.2)--(1.45, -.2) -- (1.45, 1.4)--(1.35, 1.4);
\end{tikzpicture}
\end{gathered}\,
=  \frac{1}{3}Z_{\mathfrak{T}_{S_3}}
\,\begin{gathered}
\begin{tikzpicture}[scale=.5]
\node at (0, 1.45) {};
\draw[line] (-.15,-.2)--(-.25, -.2) -- (-.25, 1.4)-- (-.15,1.4);
\draw[very thick] (0,0) rectangle (1.2,1.2);
\draw[line] (1.35, -.2)--(1.45, -.2) -- (1.45, 1.4)--(1.35, 1.4);
\end{tikzpicture}
\end{gathered}\,
- \frac{1}{3}Z_{\mathfrak{T}_{S_3}}
\,\begin{gathered}
\begin{tikzpicture}[scale=.5]
\node at (0, 1.45) {};
\draw[line] (-.15,-.2)--(-.25, -.2) -- (-.25, 1.4)-- (-.15,1.4);
\draw[very thick] (0,0) rectangle (1.2,1.2);
\draw[line, densely dotted, thick] (0,.6)--(1.2,.6);
\draw[line] (1.35, -.2)--(1.45, -.2) -- (1.45, 1.4)--(1.35, 1.4);
\end{tikzpicture}
\end{gathered}\,
+ \frac{2}{3}Z_{\mathfrak{T}_{S_3}}
\,\begin{gathered}
\begin{tikzpicture}[scale=.5]
\node at (0, 1.45) {};
\draw[line] (-.15,-.2)--(-.25, -.2) -- (-.25, 1.4)-- (-.15,1.4);
\draw[very thick] (0,0) rectangle (1.2,1.2);
\draw[line, densely dotted, thick] (.6,0)--(.6,1.2);
\draw[line] (1.35, -.2)--(1.45, -.2) -- (1.45, 1.4)--(1.35, 1.4);
\end{tikzpicture}
\end{gathered}\,
- \frac{1}{3}Z_{\mathfrak{T}_{S_3}}
\,\begin{gathered}
\begin{tikzpicture}[scale=.5]
\node at (0, 1.45) {};
\draw[line] (-.15,-.2)--(-.25, -.2) -- (-.25, 1.4)-- (-.15,1.4);
\draw[very thick] (0,0) rectangle (1.2,1.2);
\draw[densely dotted,thick]  (.6,0) arc (0:90:.6); 
\draw[densely dotted,thick] (.6,1.2) arc (-180:-90:.6);
\draw[line] (1.35, -.2)--(1.45, -.2) -- (1.45, 1.4)--(1.35, 1.4);
\end{tikzpicture}
\end{gathered}\,
- \frac{1}{3}Z_{\mathfrak{T}_{S_3}}
\,\begin{gathered}
\begin{tikzpicture}[scale=.5]
\node at (0, 1.45) {};
\draw[line] (-.15,-.2)--(-.25, -.2) -- (-.25, 1.4)-- (-.15,1.4);
\draw[very thick] (0,0) rectangle (1.2,1.2);
\draw[densely dotted,thick]  (.6,0) .. controls (.6,.4) and (.0,.4).. (0,.4);
\draw[densely dotted,thick]  (.6,1.2) .. controls (.6,.8) and (1.2,.8) .. (1.2,.8);
\draw[densely dotted,thick] (0,.8) .. controls (.6,.8) and (.6,.4) .. (1.2,.4);
\draw[line] (1.35, -.2)--(1.45, -.2) -- (1.45, 1.4)--(1.35, 1.4);
\end{tikzpicture}
\end{gathered}\,
,
    \end{align}
where we need the identity $1+\omega + \omega^2=0$ to recover the definition of $\mathbb{Z}_3$ gauging. Further project $|\chi\rangle_{\mathfrak{T}_{S_3}}$ onto the states  $\hat{\mathbb{T}}W_2[\Gamma_2]|\Omega\rangle_{\mathcal{L}_2} = |0,0,2,0,0,0,2\omega,2\omega^2\rangle$ and $\hat{\mathbb{T}}^2 W_2[\Gamma_2]|\Omega\rangle_{\mathcal{L}_2} = |0,0,2,0,0,0,2\omega^2,2\omega\rangle$ we get
    \begin{align}
        Z_{\mathfrak{T}_{S_3}/\mathbb{Z}_3} 
        \,\begin{gathered}
\begin{tikzpicture}[scale=.5]
\node at (0, 1.45) {};
\draw[line] (-.15,-.2)--(-.25, -.2) -- (-.25, 1.4)-- (-.15,1.4);
\draw[very thick] (0,0) rectangle (1.2,1.2);
\draw[densely dotted,thick]  (.6,0) arc (0:90:.6); 
\draw[densely dotted,thick] (.6,1.2) arc (-180:-90:.6);
\draw[line] (1.35, -.2)--(1.45, -.2) -- (1.45, 1.4)--(1.35, 1.4);
\end{tikzpicture}
\end{gathered}\,
=  \frac{1}{3}Z_{\mathfrak{T}_{S_3}}
\,\begin{gathered}
\begin{tikzpicture}[scale=.5]
\node at (0, 1.45) {};
\draw[line] (-.15,-.2)--(-.25, -.2) -- (-.25, 1.4)-- (-.15,1.4);
\draw[very thick] (0,0) rectangle (1.2,1.2);
\draw[line] (1.35, -.2)--(1.45, -.2) -- (1.45, 1.4)--(1.35, 1.4);
\end{tikzpicture}
\end{gathered}\,
- \frac{1}{3}Z_{\mathfrak{T}_{S_3}}
\,\begin{gathered}
\begin{tikzpicture}[scale=.5]
\node at (0, 1.45) {};
\draw[line] (-.15,-.2)--(-.25, -.2) -- (-.25, 1.4)-- (-.15,1.4);
\draw[very thick] (0,0) rectangle (1.2,1.2);
\draw[line, densely dotted, thick] (0,.6)--(1.2,.6);
\draw[line] (1.35, -.2)--(1.45, -.2) -- (1.45, 1.4)--(1.35, 1.4);
\end{tikzpicture}
\end{gathered}\,
- \frac{1}{3}Z_{\mathfrak{T}_{S_3}}
\,\begin{gathered}
\begin{tikzpicture}[scale=.5]
\node at (0, 1.45) {};
\draw[line] (-.15,-.2)--(-.25, -.2) -- (-.25, 1.4)-- (-.15,1.4);
\draw[very thick] (0,0) rectangle (1.2,1.2);
\draw[line, densely dotted, thick] (.6,0)--(.6,1.2);
\draw[line] (1.35, -.2)--(1.45, -.2) -- (1.45, 1.4)--(1.35, 1.4);
\end{tikzpicture}
\end{gathered}\,
+ \frac{2}{3}Z_{\mathfrak{T}_{S_3}}
\,\begin{gathered}
\begin{tikzpicture}[scale=.5]
\node at (0, 1.45) {};
\draw[line] (-.15,-.2)--(-.25, -.2) -- (-.25, 1.4)-- (-.15,1.4);
\draw[very thick] (0,0) rectangle (1.2,1.2);
\draw[densely dotted,thick]  (.6,0) arc (0:90:.6); 
\draw[densely dotted,thick] (.6,1.2) arc (-180:-90:.6);
\draw[line] (1.35, -.2)--(1.45, -.2) -- (1.45, 1.4)--(1.35, 1.4);
\end{tikzpicture}
\end{gathered}\,
- \frac{1}{3}Z_{\mathfrak{T}_{S_3}}
\,\begin{gathered}
\begin{tikzpicture}[scale=.5]
\node at (0, 1.45) {};
\draw[line] (-.15,-.2)--(-.25, -.2) -- (-.25, 1.4)-- (-.15,1.4);
\draw[very thick] (0,0) rectangle (1.2,1.2);
\draw[densely dotted,thick]  (.6,0) .. controls (.6,.4) and (.0,.4).. (0,.4);
\draw[densely dotted,thick]  (.6,1.2) .. controls (.6,.8) and (1.2,.8) .. (1.2,.8);
\draw[densely dotted,thick] (0,.8) .. controls (.6,.8) and (.6,.4) .. (1.2,.4);
\draw[line] (1.35, -.2)--(1.45, -.2) -- (1.45, 1.4)--(1.35, 1.4);
\end{tikzpicture}
\end{gathered}\,
,
    \end{align}
and also
    \begin{align}
        Z_{\mathfrak{T}_{S_3}/\mathbb{Z}_3}
        \,\begin{gathered}
\begin{tikzpicture}[scale=.5]
\node at (0, 1.45) {};
\draw[line] (-.15,-.2)--(-.25, -.2) -- (-.25, 1.4)-- (-.15,1.4);
\draw[very thick] (0,0) rectangle (1.2,1.2);
\draw[densely dotted,thick]  (.6,0) .. controls (.6,.4) and (.0,.4).. (0,.4);
\draw[densely dotted,thick]  (.6,1.2) .. controls (.6,.8) and (1.2,.8) .. (1.2,.8);
\draw[densely dotted,thick] (0,.8) .. controls (.6,.8) and (.6,.4) .. (1.2,.4);
\draw[line] (1.35, -.2)--(1.45, -.2) -- (1.45, 1.4)--(1.35, 1.4);
\end{tikzpicture}
\end{gathered}\,
=  \frac{1}{3}Z_{\mathfrak{T}_{S_3}}
\,\begin{gathered}
\begin{tikzpicture}[scale=.5]
\node at (0, 1.45) {};
\draw[line] (-.15,-.2)--(-.25, -.2) -- (-.25, 1.4)-- (-.15,1.4);
\draw[very thick] (0,0) rectangle (1.2,1.2);
\draw[line] (1.35, -.2)--(1.45, -.2) -- (1.45, 1.4)--(1.35, 1.4);
\end{tikzpicture}
\end{gathered}\,
- \frac{1}{3}Z_{\mathfrak{T}_{S_3}}
\,\begin{gathered}
\begin{tikzpicture}[scale=.5]
\node at (0, 1.45) {};
\draw[line] (-.15,-.2)--(-.25, -.2) -- (-.25, 1.4)-- (-.15,1.4);
\draw[very thick] (0,0) rectangle (1.2,1.2);
\draw[line, densely dotted, thick] (0,.6)--(1.2,.6);
\draw[line] (1.35, -.2)--(1.45, -.2) -- (1.45, 1.4)--(1.35, 1.4);
\end{tikzpicture}
\end{gathered}\,
- \frac{1}{3}Z_{\mathfrak{T}_{S_3}}
\,\begin{gathered}
\begin{tikzpicture}[scale=.5]
\node at (0, 1.45) {};
\draw[line] (-.15,-.2)--(-.25, -.2) -- (-.25, 1.4)-- (-.15,1.4);
\draw[very thick] (0,0) rectangle (1.2,1.2);
\draw[line, densely dotted, thick] (.6,0)--(.6,1.2);
\draw[line] (1.35, -.2)--(1.45, -.2) -- (1.45, 1.4)--(1.35, 1.4);
\end{tikzpicture}
\end{gathered}\,
- \frac{1}{3}Z_{\mathfrak{T}_{S_3}}
\,\begin{gathered}
\begin{tikzpicture}[scale=.5]
\node at (0, 1.45) {};
\draw[line] (-.15,-.2)--(-.25, -.2) -- (-.25, 1.4)-- (-.15,1.4);
\draw[very thick] (0,0) rectangle (1.2,1.2);
\draw[densely dotted,thick]  (.6,0) arc (0:90:.6); 
\draw[densely dotted,thick] (.6,1.2) arc (-180:-90:.6);
\draw[line] (1.35, -.2)--(1.45, -.2) -- (1.45, 1.4)--(1.35, 1.4);
\end{tikzpicture}
\end{gathered}\,
+ \frac{2}{3}Z_{\mathfrak{T}_{S_3}}
\,\begin{gathered}
\begin{tikzpicture}[scale=.5]
\node at (0, 1.45) {};
\draw[line] (-.15,-.2)--(-.25, -.2) -- (-.25, 1.4)-- (-.15,1.4);
\draw[very thick] (0,0) rectangle (1.2,1.2);
\draw[densely dotted,thick]  (.6,0) .. controls (.6,.4) and (.0,.4).. (0,.4);
\draw[densely dotted,thick]  (.6,1.2) .. controls (.6,.8) and (1.2,.8) .. (1.2,.8);
\draw[densely dotted,thick] (0,.8) .. controls (.6,.8) and (.6,.4) .. (1.2,.4);
\draw[line] (1.35, -.2)--(1.45, -.2) -- (1.45, 1.4)--(1.35, 1.4);
\end{tikzpicture}
\end{gathered}\,
.
    \end{align}    
For the conjugate class $[b]$ we can project $|\chi\rangle_{\mathfrak{T}_{S_3}}$ onto the state $W_3[\Gamma_1]|\Omega\rangle_{\mathcal{L}_2} = |3,-3,0,0,0,0,0,0\rangle$,  $W_3[\Gamma_2]|\Omega\rangle_{\mathcal{L}_2} = |0,0,0,3,3,0,0,0\rangle$ and also $\hat{\mathbb{T}}W_3[\Gamma_2]|\Omega\rangle_{\mathcal{L}_2} = |0,0,0,3,-3,0,0,0\rangle$. We simply get
\begin{equation}
    Z_{\mathfrak{T}_{S_3}/\mathbb{Z}_3}\,\begin{gathered}
\begin{tikzpicture}[scale=.5]
\node at (0, 1.45) {};
\draw[line] (-.15,-.2)--(-.25, -.2) -- (-.25, 1.4)-- (-.15,1.4);
\draw[very thick] (0,0) rectangle (1.2,1.2);
\draw[line, thick] (0,.6)--(1.2,.6);
\draw[line] (1.35, -.2)--(1.45, -.2) -- (1.45, 1.4)--(1.35, 1.4);
\end{tikzpicture}
\end{gathered}\, = Z_{\mathfrak{T}_{S_3}}\,\begin{gathered}
\begin{tikzpicture}[scale=.5]
\node at (0, 1.45) {};
\draw[line] (-.15,-.2)--(-.25, -.2) -- (-.25, 1.4)-- (-.15,1.4);
\draw[very thick] (0,0) rectangle (1.2,1.2);
\draw[line, thick] (0,.6)--(1.2,.6);
\draw[line] (1.35, -.2)--(1.45, -.2) -- (1.45, 1.4)--(1.35, 1.4);
\end{tikzpicture}
\end{gathered}\,,\quad Z_{\mathfrak{T}_{S_3}/\mathbb{Z}_3}\,\begin{gathered}
\begin{tikzpicture}[scale=.5]
\node at (0, 1.45) {};
\draw[line] (-.15,-.2)--(-.25, -.2) -- (-.25, 1.4)-- (-.15,1.4);
\draw[very thick] (0,0) rectangle (1.2,1.2);
\draw[line, thick] (.6,0)--(.6,1.2);
\draw[line] (1.35, -.2)--(1.45, -.2) -- (1.45, 1.4)--(1.35, 1.4);
\end{tikzpicture}
\end{gathered}\, = Z_{\mathfrak{T}_{S_3}}\,\begin{gathered}
\begin{tikzpicture}[scale=.5]
\node at (0, 1.45) {};
\draw[line] (-.15,-.2)--(-.25, -.2) -- (-.25, 1.4)-- (-.15,1.4);
\draw[very thick] (0,0) rectangle (1.2,1.2);
\draw[line, thick] (.6,0)--(.6,1.2);
\draw[line] (1.35, -.2)--(1.45, -.2) -- (1.45, 1.4)--(1.35, 1.4);
\end{tikzpicture}
\end{gathered}\,,\quad Z_{\mathfrak{T}_{S_3}/\mathbb{Z}_3},\begin{gathered}
\begin{tikzpicture}[scale=.5]
\node at (0, 1.45) {};
\draw[line] (-.15,-.2)--(-.25, -.2) -- (-.25, 1.4)-- (-.15,1.4);
\draw[very thick] (0,0) rectangle (1.2,1.2);
\draw  (.6,0) arc (0:90:.6); 
\draw (.6,1.2) arc (-180:-90:.6);
\draw[line] (1.35, -.2)--(1.45, -.2) -- (1.45, 1.4)--(1.35, 1.4);
\end{tikzpicture}
\end{gathered}\, = Z_{\mathfrak{T}_{S_3}},\begin{gathered}
\begin{tikzpicture}[scale=.5]
\node at (0, 1.45) {};
\draw[line] (-.15,-.2)--(-.25, -.2) -- (-.25, 1.4)-- (-.15,1.4);
\draw[very thick] (0,0) rectangle (1.2,1.2);
\draw  (.6,0) arc (0:90:.6); 
\draw (.6,1.2) arc (-180:-90:.6);
\draw[line] (1.35, -.2)--(1.45, -.2) -- (1.45, 1.4)--(1.35, 1.4);
\end{tikzpicture}
\end{gathered}\,,
\end{equation}
which means we do not touch $\mathbb{Z}_2$ part in $S_3$.

For other topological boundary states the method is the same and we will only focus on the vacuum sector. For $\mathcal{L}_3$ the vacuum is $|\Omega\rangle_{\mathcal{L}_3} = |1,0,1,1,0,0,0,0\rangle$ and we have
    \begin{equation}
        \frac{1}{6}\langle \Omega |_{\mathcal{L}_3} | \chi\rangle_{\mathfrak{T}_{S_3}} = \frac{1}{2} Z_{\mathfrak{T}_{S_3}}
        \,\begin{gathered}
\begin{tikzpicture}[scale=.5]
\node at (0, 1.45) {};
\draw[line] (-.15,-.2)--(-.25, -.2) -- (-.25, 1.4)-- (-.15,1.4);
\draw[very thick] (0,0) rectangle (1.2,1.2);
\draw[line] (1.35, -.2)--(1.45, -.2) -- (1.45, 1.4)--(1.35, 1.4);
\end{tikzpicture}
\end{gathered}\,
+ \frac{1}{2}Z_{\mathfrak{T}_{S_3}}
\,\begin{gathered}
\begin{tikzpicture}[scale=.5]
\node at (0, 1.45) {};
\draw[line] (-.15,-.2)--(-.25, -.2) -- (-.25, 1.4)-- (-.15,1.4);
\draw[very thick] (0,0) rectangle (1.2,1.2);
\draw[line, thick] (0,.6)--(1.2,.6);
\draw[line] (1.35, -.2)--(1.45, -.2) -- (1.45, 1.4)--(1.35, 1.4);
\end{tikzpicture}
\end{gathered}\,
+ \frac{1}{2}Z_{\mathfrak{T}_{S_3}}
\,\begin{gathered}
\begin{tikzpicture}[scale=.5]
\node at (0, 1.45) {};
\draw[line] (-.15,-.2)--(-.25, -.2) -- (-.25, 1.4)-- (-.15,1.4);
\draw[very thick] (0,0) rectangle (1.2,1.2);
\draw[line, thick] (.6,0)--(.6,1.2);
\draw[line] (1.35, -.2)--(1.45, -.2) -- (1.45, 1.4)--(1.35, 1.4);
\end{tikzpicture}
\end{gathered}\,
+ \frac{1}{2}Z_{\mathfrak{T}_{S_3}}
\,\begin{gathered}
\begin{tikzpicture}[scale=.5]
\node at (0, 1.45) {};
\draw[line] (-.15,-.2)--(-.25, -.2) -- (-.25, 1.4)-- (-.15,1.4);
\draw[very thick] (0,0) rectangle (1.2,1.2);
\draw  (.6,0) arc (0:90:.6); 
\draw (.6,1.2) arc (-180:-90:.6);
\draw[line] (1.35, -.2)--(1.45, -.2) -- (1.45, 1.4)--(1.35, 1.4);
\end{tikzpicture}
\end{gathered}\,
    \end{equation}
which corresponds to gauging $\mathbb{Z}_2 \in S_3$ symmetry. For $\mathcal{L}_4$ the vacuum is $|\Omega\rangle_{\mathcal{L}_4} = |1,0,0,1,0,1,0,0\rangle$ and we have
    \begin{align}
        \frac{1}{6}\langle \Omega |_{\mathcal{L}_4} | \chi\rangle_{\mathfrak{T}_{S_3}} =& \frac{1}{6} Z_{\mathfrak{T}_{S_3}}
        \,\begin{gathered}
\begin{tikzpicture}[scale=.5]
\node at (0, 1.45) {};
\draw[line] (-.15,-.2)--(-.25, -.2) -- (-.25, 1.4)-- (-.15,1.4);
\draw[very thick] (0,0) rectangle (1.2,1.2);
\draw[line] (1.35, -.2)--(1.45, -.2) -- (1.45, 1.4)--(1.35, 1.4);
\end{tikzpicture}
\end{gathered}\,
+ \frac{3}{6}Z_{\mathfrak{T}_{S_3}}
\,\begin{gathered}
\begin{tikzpicture}[scale=.5]
\node at (0, 1.45) {};
\draw[line] (-.15,-.2)--(-.25, -.2) -- (-.25, 1.4)-- (-.15,1.4);
\draw[very thick] (0,0) rectangle (1.2,1.2);
\draw[line, thick] (0,.6)--(1.2,.6);
\draw[line] (1.35, -.2)--(1.45, -.2) -- (1.45, 1.4)--(1.35, 1.4);
\end{tikzpicture}
\end{gathered}\,
+ \frac{3}{6}Z_{\mathfrak{T}_{S_3}}
\,\begin{gathered}
\begin{tikzpicture}[scale=.5]
\node at (0, 1.45) {};
\draw[line] (-.15,-.2)--(-.25, -.2) -- (-.25, 1.4)-- (-.15,1.4);
\draw[very thick] (0,0) rectangle (1.2,1.2);
\draw[line, thick] (.6,0)--(.6,1.2);
\draw[line] (1.35, -.2)--(1.45, -.2) -- (1.45, 1.4)--(1.35, 1.4);
\end{tikzpicture}
\end{gathered}\,
+ \frac{3}{6}Z_{\mathfrak{T}_{S_3}}
\,\begin{gathered}
\begin{tikzpicture}[scale=.5]
\node at (0, 1.45) {};
\draw[line] (-.15,-.2)--(-.25, -.2) -- (-.25, 1.4)-- (-.15,1.4);
\draw[very thick] (0,0) rectangle (1.2,1.2);
\draw  (.6,0) arc (0:90:.6); 
\draw (.6,1.2) arc (-180:-90:.6);
\draw[line] (1.35, -.2)--(1.45, -.2) -- (1.45, 1.4)--(1.35, 1.4);
\end{tikzpicture}
\end{gathered}\,
\nonumber\\
        +&\frac{2}{6} Z_{\mathfrak{T}_{S_3}}
        \,\begin{gathered}
\begin{tikzpicture}[scale=.5]
\node at (0, 1.45) {};
\draw[line] (-.15,-.2)--(-.25, -.2) -- (-.25, 1.4)-- (-.15,1.4);
\draw[very thick] (0,0) rectangle (1.2,1.2);
\draw[line, densely dotted, thick] (0,.6)--(1.2,.6);
\draw[line] (1.35, -.2)--(1.45, -.2) -- (1.45, 1.4)--(1.35, 1.4);
\end{tikzpicture}
\end{gathered}\,
+\frac{2}{6} Z_{\mathfrak{T}_{S_3}}
\,\begin{gathered}
\begin{tikzpicture}[scale=.5]
\node at (0, 1.45) {};
\draw[line] (-.15,-.2)--(-.25, -.2) -- (-.25, 1.4)-- (-.15,1.4);
\draw[very thick] (0,0) rectangle (1.2,1.2);
\draw[line, densely dotted, thick] (.6,0)--(.6,1.2);
\draw[line] (1.35, -.2)--(1.45, -.2) -- (1.45, 1.4)--(1.35, 1.4);
\end{tikzpicture}
\end{gathered}\,
+\frac{2}{6} Z_{\mathfrak{T}_{S_3}}
\,\begin{gathered}
\begin{tikzpicture}[scale=.5]
\node at (0, 1.45) {};
\draw[line] (-.15,-.2)--(-.25, -.2) -- (-.25, 1.4)-- (-.15,1.4);
\draw[very thick] (0,0) rectangle (1.2,1.2);
\draw[densely dotted,thick]  (.6,0) arc (0:90:.6); 
\draw[densely dotted,thick] (.6,1.2) arc (-180:-90:.6);
\draw[line] (1.35, -.2)--(1.45, -.2) -- (1.45, 1.4)--(1.35, 1.4);
\end{tikzpicture}
\end{gathered}\,
+\frac{2}{6} Z_{\mathfrak{T}_{S_3}}
\,\begin{gathered}
\begin{tikzpicture}[scale=.5]
\node at (0, 1.45) {};
\draw[line] (-.15,-.2)--(-.25, -.2) -- (-.25, 1.4)-- (-.15,1.4);
\draw[very thick] (0,0) rectangle (1.2,1.2);
\draw[densely dotted,thick]  (.6,0) .. controls (.6,.4) and (.0,.4).. (0,.4);
\draw[densely dotted,thick]  (.6,1.2) .. controls (.6,.8) and (1.2,.8) .. (1.2,.8);
\draw[densely dotted,thick] (0,.8) .. controls (.6,.8) and (.6,.4) .. (1.2,.4);
\draw[line] (1.35, -.2)--(1.45, -.2) -- (1.45, 1.4)--(1.35, 1.4);
\end{tikzpicture}
\end{gathered}\,,
    \end{align}
which corresponds to gauging the whole $S_3$ symmetry. For the fermionic Lagrangian algebra $\mathcal{L}_5$ the vacuum is $|\Omega\rangle_{\mathcal{L}_5} = |1,0,1,0,1,0,0,0\rangle$ and we have
    \begin{equation}
        \frac{1}{6}\langle \Omega |_{\mathcal{L}_5} | \chi\rangle_{\mathfrak{T}_{S_3}} = \frac{1}{2} Z_{\mathfrak{T}_{S_3}}
        \,\begin{gathered}
\begin{tikzpicture}[scale=.5]
\node at (0, 1.45) {};
\draw[line] (-.15,-.2)--(-.25, -.2) -- (-.25, 1.4)-- (-.15,1.4);
\draw[very thick] (0,0) rectangle (1.2,1.2);
\draw[line] (1.35, -.2)--(1.45, -.2) -- (1.45, 1.4)--(1.35, 1.4);
\end{tikzpicture}
\end{gathered}\,
+ \frac{1}{2}Z_{\mathfrak{T}_{S_3}}
\,\begin{gathered}
\begin{tikzpicture}[scale=.5]
\node at (0, 1.45) {};
\draw[line] (-.15,-.2)--(-.25, -.2) -- (-.25, 1.4)-- (-.15,1.4);
\draw[very thick] (0,0) rectangle (1.2,1.2);
\draw[line, thick] (0,.6)--(1.2,.6);
\draw[line] (1.35, -.2)--(1.45, -.2) -- (1.45, 1.4)--(1.35, 1.4);
\end{tikzpicture}
\end{gathered}\,
+ \frac{1}{2}Z_{\mathfrak{T}_{S_3}}
\,\begin{gathered}
\begin{tikzpicture}[scale=.5]
\node at (0, 1.45) {};
\draw[line] (-.15,-.2)--(-.25, -.2) -- (-.25, 1.4)-- (-.15,1.4);
\draw[very thick] (0,0) rectangle (1.2,1.2);
\draw[line, thick] (.6,0)--(.6,1.2);
\draw[line] (1.35, -.2)--(1.45, -.2) -- (1.45, 1.4)--(1.35, 1.4);
\end{tikzpicture}
\end{gathered}\,
-\frac{1}{2}Z_{\mathfrak{T}_{S_3}}
\,\begin{gathered}
\begin{tikzpicture}[scale=.5]
\node at (0, 1.45) {};
\draw[line] (-.15,-.2)--(-.25, -.2) -- (-.25, 1.4)-- (-.15,1.4);
\draw[very thick] (0,0) rectangle (1.2,1.2);
\draw  (.6,0) arc (0:90:.6); 
\draw (.6,1.2) arc (-180:-90:.6);
\draw[line] (1.35, -.2)--(1.45, -.2) -- (1.45, 1.4)--(1.35, 1.4);
\end{tikzpicture}
\end{gathered}\,.
    \end{equation}
Compared to the $\mathcal{L}_3$ case, the only difference is the sign of the last sector is flipped, which corresponds to the fermionization. Finally, for $\mathcal{L}_6$ the vacuum is $|\Omega\rangle_{\mathcal{L}_6} = |1,0,0,0,1,1,0,0\rangle$ and we have
    \begin{align}
        \frac{1}{6}\langle \Omega |_{\mathcal{L}_6} | \chi\rangle_{\mathfrak{T}_{S_3}} =& \frac{1}{6} Z_{\mathfrak{T}_{S_3}}
        \,\begin{gathered}
\begin{tikzpicture}[scale=.5]
\node at (0, 1.45) {};
\draw[line] (-.15,-.2)--(-.25, -.2) -- (-.25, 1.4)-- (-.15,1.4);
\draw[very thick] (0,0) rectangle (1.2,1.2);
\draw[line] (1.35, -.2)--(1.45, -.2) -- (1.45, 1.4)--(1.35, 1.4);
\end{tikzpicture}
\end{gathered}\,
+ \frac{3}{6}Z_{\mathfrak{T}_{S_3}}
\,\begin{gathered}
\begin{tikzpicture}[scale=.5]
\node at (0, 1.45) {};
\draw[line] (-.15,-.2)--(-.25, -.2) -- (-.25, 1.4)-- (-.15,1.4);
\draw[very thick] (0,0) rectangle (1.2,1.2);
\draw[line, thick] (0,.6)--(1.2,.6);
\draw[line] (1.35, -.2)--(1.45, -.2) -- (1.45, 1.4)--(1.35, 1.4);
\end{tikzpicture}
\end{gathered}\,
+ \frac{3}{6}Z_{\mathfrak{T}_{S_3}}
\,\begin{gathered}
\begin{tikzpicture}[scale=.5]
\node at (0, 1.45) {};
\draw[line] (-.15,-.2)--(-.25, -.2) -- (-.25, 1.4)-- (-.15,1.4);
\draw[very thick] (0,0) rectangle (1.2,1.2);
\draw[line, thick] (.6,0)--(.6,1.2);
\draw[line] (1.35, -.2)--(1.45, -.2) -- (1.45, 1.4)--(1.35, 1.4);
\end{tikzpicture}
\end{gathered}\,
- \frac{3}{6}Z_{\mathfrak{T}_{S_3}}
\,\begin{gathered}
\begin{tikzpicture}[scale=.5]
\node at (0, 1.45) {};
\draw[line] (-.15,-.2)--(-.25, -.2) -- (-.25, 1.4)-- (-.15,1.4);
\draw[very thick] (0,0) rectangle (1.2,1.2);
\draw  (.6,0) arc (0:90:.6); 
\draw (.6,1.2) arc (-180:-90:.6);
\draw[line] (1.35, -.2)--(1.45, -.2) -- (1.45, 1.4)--(1.35, 1.4);
\end{tikzpicture}
\end{gathered}\,
\nonumber\\
        +&\frac{2}{6} Z_{\mathfrak{T}_{S_3}}
        \,\begin{gathered}
\begin{tikzpicture}[scale=.5]
\node at (0, 1.45) {};
\draw[line] (-.15,-.2)--(-.25, -.2) -- (-.25, 1.4)-- (-.15,1.4);
\draw[very thick] (0,0) rectangle (1.2,1.2);
\draw[line, densely dotted, thick] (0,.6)--(1.2,.6);
\draw[line] (1.35, -.2)--(1.45, -.2) -- (1.45, 1.4)--(1.35, 1.4);
\end{tikzpicture}
\end{gathered}\,
+\frac{2}{6} Z_{\mathfrak{T}_{S_3}}
\,\begin{gathered}
\begin{tikzpicture}[scale=.5]
\node at (0, 1.45) {};
\draw[line] (-.15,-.2)--(-.25, -.2) -- (-.25, 1.4)-- (-.15,1.4);
\draw[very thick] (0,0) rectangle (1.2,1.2);
\draw[line, densely dotted, thick] (.6,0)--(.6,1.2);
\draw[line] (1.35, -.2)--(1.45, -.2) -- (1.45, 1.4)--(1.35, 1.4);
\end{tikzpicture}
\end{gathered}\,
+\frac{2}{6} Z_{\mathfrak{T}_{S_3}}
\,\begin{gathered}
\begin{tikzpicture}[scale=.5]
\node at (0, 1.45) {};
\draw[line] (-.15,-.2)--(-.25, -.2) -- (-.25, 1.4)-- (-.15,1.4);
\draw[very thick] (0,0) rectangle (1.2,1.2);
\draw[densely dotted,thick]  (.6,0) arc (0:90:.6); 
\draw[densely dotted,thick] (.6,1.2) arc (-180:-90:.6);
\draw[line] (1.35, -.2)--(1.45, -.2) -- (1.45, 1.4)--(1.35, 1.4);
\end{tikzpicture}
\end{gathered}\,
+\frac{2}{6} Z_{\mathfrak{T}_{S_3}}
\,\begin{gathered}
\begin{tikzpicture}[scale=.5]
\node at (0, 1.45) {};
\draw[line] (-.15,-.2)--(-.25, -.2) -- (-.25, 1.4)-- (-.15,1.4);
\draw[very thick] (0,0) rectangle (1.2,1.2);
\draw[densely dotted,thick]  (.6,0) .. controls (.6,.4) and (.0,.4).. (0,.4);
\draw[densely dotted,thick]  (.6,1.2) .. controls (.6,.8) and (1.2,.8) .. (1.2,.8);
\draw[densely dotted,thick] (0,.8) .. controls (.6,.8) and (.6,.4) .. (1.2,.4);
\draw[line] (1.35, -.2)--(1.45, -.2) -- (1.45, 1.4)--(1.35, 1.4);
\end{tikzpicture}
\end{gathered}\,.
    \end{align}
where the sign of the $\begin{gathered}
\begin{tikzpicture}[scale=.5]
\node at (0, 1.25) {};
\draw[very thick] (0,0) rectangle (1.2,1.2);
\draw  (.6,0) arc (0:90:.6); 
\draw (.6,1.2) arc (-180:-90:.6);
\end{tikzpicture}
\end{gathered}$ -sector is flipped compared to $\mathcal{L}_4$ case. The partition function corresponds to gauging the $\mathbb{Z}_3$ symmetry first and then doing a fermionization.

Finally, let us consider the parafermionic algebra $\mathcal{L}_7$ and $\mathcal{L}_8$. The vacua are
    \begin{equation}
        |\Omega\rangle_{\mathcal{L}_7} = |1,1,0,0,0,0,2,0\rangle,\quad |\Omega\rangle_{\mathcal{L}_8} = |1,1,0,0,0,0,0,2\rangle,
    \end{equation}
and we get the dual partition functions
    \begin{align}
        \frac{1}{6}\langle \Omega |_{\mathcal{L}_7} | \chi\rangle_{\mathfrak{T}_{S_3}} 
= \frac{1}{3}Z_{\mathfrak{T}_{S_3}}
\,\begin{gathered}
\begin{tikzpicture}[scale=.5]
\node at (0, 1.45) {};
\draw[line] (-.15,-.2)--(-.25, -.2) -- (-.25, 1.4)-- (-.15,1.4);
\draw[very thick] (0,0) rectangle (1.2,1.2);
\draw[line] (1.35, -.2)--(1.45, -.2) -- (1.45, 1.4)--(1.35, 1.4);
\end{tikzpicture}
\end{gathered}\,
+ \frac{2}{3}Z_{\mathfrak{T}_{S_3}}
\,\begin{gathered}
\begin{tikzpicture}[scale=.5]
\node at (0, 1.45) {};
\draw[line] (-.15,-.2)--(-.25, -.2) -- (-.25, 1.4)-- (-.15,1.4);
\draw[very thick] (0,0) rectangle (1.2,1.2);
\draw[line, densely dotted, thick] (0,.6)--(1.2,.6);
\draw[line] (1.35, -.2)--(1.45, -.2) -- (1.45, 1.4)--(1.35, 1.4);
\end{tikzpicture}
\end{gathered}\,
+ \frac{2}{3}Z_{\mathfrak{T}_{S_3}}
\,\begin{gathered}
\begin{tikzpicture}[scale=.5]
\node at (0, 1.45) {};
\draw[line] (-.15,-.2)--(-.25, -.2) -- (-.25, 1.4)-- (-.15,1.4);
\draw[very thick] (0,0) rectangle (1.2,1.2);
\draw[line, densely dotted, thick] (.6,0)--(.6,1.2);
\draw[line] (1.35, -.2)--(1.45, -.2) -- (1.45, 1.4)--(1.35, 1.4);
\end{tikzpicture}
\end{gathered}\,
+ \frac{2\omega}{3}Z_{\mathfrak{T}_{S_3}}
\,\begin{gathered}
\begin{tikzpicture}[scale=.5]
\node at (0, 1.45) {};
\draw[line] (-.15,-.2)--(-.25, -.2) -- (-.25, 1.4)-- (-.15,1.4);
\draw[very thick] (0,0) rectangle (1.2,1.2);
\draw[densely dotted,thick]  (.6,0) arc (0:90:.6); 
\draw[densely dotted,thick] (.6,1.2) arc (-180:-90:.6);
\draw[line] (1.35, -.2)--(1.45, -.2) -- (1.45, 1.4)--(1.35, 1.4);
\end{tikzpicture}
\end{gathered}\,
+ \frac{2 \omega^2}{3}Z_{\mathfrak{T}_{S_3}}
\,\begin{gathered}
\begin{tikzpicture}[scale=.5]
\node at (0, 1.45) {};
\draw[line] (-.15,-.2)--(-.25, -.2) -- (-.25, 1.4)-- (-.15,1.4);
\draw[very thick] (0,0) rectangle (1.2,1.2);
\draw[densely dotted,thick]  (.6,0) .. controls (.6,.4) and (.0,.4).. (0,.4);
\draw[densely dotted,thick]  (.6,1.2) .. controls (.6,.8) and (1.2,.8) .. (1.2,.8);
\draw[densely dotted,thick] (0,.8) .. controls (.6,.8) and (.6,.4) .. (1.2,.4);
\draw[line] (1.35, -.2)--(1.45, -.2) -- (1.45, 1.4)--(1.35, 1.4);
\end{tikzpicture}
\end{gathered}\,
,
    \end{align}
and
    \begin{align}
        \frac{1}{6}\langle \Omega |_{\mathcal{L}_8} | \chi\rangle_{\mathfrak{T}_{S_3}} 
= \frac{1}{3}Z_{\mathfrak{T}_{S_3}}
\,\begin{gathered}
\begin{tikzpicture}[scale=.5]
\node at (0, 1.45) {};
\draw[line] (-.15,-.2)--(-.25, -.2) -- (-.25, 1.4)-- (-.15,1.4);
\draw[very thick] (0,0) rectangle (1.2,1.2);
\draw[line] (1.35, -.2)--(1.45, -.2) -- (1.45, 1.4)--(1.35, 1.4);
\end{tikzpicture}
\end{gathered}\,
+ \frac{2}{3}Z_{\mathfrak{T}_{S_3}}
\,\begin{gathered}
\begin{tikzpicture}[scale=.5]
\node at (0, 1.45) {};
\draw[line] (-.15,-.2)--(-.25, -.2) -- (-.25, 1.4)-- (-.15,1.4);
\draw[very thick] (0,0) rectangle (1.2,1.2);
\draw[line, densely dotted, thick] (0,.6)--(1.2,.6);
\draw[line] (1.35, -.2)--(1.45, -.2) -- (1.45, 1.4)--(1.35, 1.4);
\end{tikzpicture}
\end{gathered}\,
+ \frac{2}{3}Z_{\mathfrak{T}_{S_3}}
\,\begin{gathered}
\begin{tikzpicture}[scale=.5]
\node at (0, 1.45) {};
\draw[line] (-.15,-.2)--(-.25, -.2) -- (-.25, 1.4)-- (-.15,1.4);
\draw[very thick] (0,0) rectangle (1.2,1.2);
\draw[line, densely dotted, thick] (.6,0)--(.6,1.2);
\draw[line] (1.35, -.2)--(1.45, -.2) -- (1.45, 1.4)--(1.35, 1.4);
\end{tikzpicture}
\end{gathered}\,
+ \frac{2 \omega^2}{3}Z_{\mathfrak{T}_{S_3}}
\,\begin{gathered}
\begin{tikzpicture}[scale=.5]
\node at (0, 1.45) {};
\draw[line] (-.15,-.2)--(-.25, -.2) -- (-.25, 1.4)-- (-.15,1.4);
\draw[very thick] (0,0) rectangle (1.2,1.2);
\draw[densely dotted,thick]  (.6,0) arc (0:90:.6); 
\draw[densely dotted,thick] (.6,1.2) arc (-180:-90:.6);
\draw[line] (1.35, -.2)--(1.45, -.2) -- (1.45, 1.4)--(1.35, 1.4);
\end{tikzpicture}
\end{gathered}\,
+ \frac{2 \omega}{3}Z_{\mathfrak{T}_{S_3}}
\,\begin{gathered}
\begin{tikzpicture}[scale=.5]
\node at (0, 1.45) {};
\draw[line] (-.15,-.2)--(-.25, -.2) -- (-.25, 1.4)-- (-.15,1.4);
\draw[very thick] (0,0) rectangle (1.2,1.2);
\draw[densely dotted,thick]  (.6,0) .. controls (.6,.4) and (.0,.4).. (0,.4);
\draw[densely dotted,thick]  (.6,1.2) .. controls (.6,.8) and (1.2,.8) .. (1.2,.8);
\draw[densely dotted,thick] (0,.8) .. controls (.6,.8) and (.6,.4) .. (1.2,.4);
\draw[line] (1.35, -.2)--(1.45, -.2) -- (1.45, 1.4)--(1.35, 1.4);
\end{tikzpicture}
\end{gathered}\,
,
    \end{align}
where $\omega = e^{\frac{2\pi i}{3}}$ is the 3rd root of unity. Therefore we see the dual theories can be understood as $\mathbb{Z}_3$ parafermionization of the original theory where the dual partition functions are already presented in the previous sections.

\subsection{Ising Category}
When the symmetry category is a modular tensor category (MTC) denoted as $\mathcal{C}$, the Drinfeld center $\mathcal{Z}(\mathcal{C})$ of the SymTFT is simply obtained as
    \begin{equation}
        \mathcal{Z}(\mathcal{C}) = \bar{\mathcal{C}} \times \mathcal{C}
    \end{equation}
and the $S$ and $T$ matrices are given by the direct product
    \begin{equation}
        S = S^*_{\mathcal{C}} \otimes S_{\mathcal{C}},\quad  T = T^*_{\mathcal{C}} \otimes T_{\mathcal{C}},
    \end{equation}
where $S_{\mathcal{C}}$ and $T_{\mathcal{C}}$ are the $S$ and $T$ matrices in $\mathcal{C}$ and $S^*_{\mathcal{C}}, T^*_{\mathcal{C}}$ are complex conjugate of $S_{\mathcal{C}}, T_{\mathcal{C}}$. In this paper, we will mainly consider the Ising category and the Lee-Yang category as two examples.

The $S$ and $T$ matrices for Ising SymTFT are presented as
    \begin{equation}
        S = \frac{1}{4}\left(\begin{array}{ccccccccc}
            1&1&1&1&\sqrt{2}&\sqrt{2}&\sqrt{2}&\sqrt{2}&2\\
            1&1&1&1&-\sqrt{2}&-\sqrt{2}&-\sqrt{2}&-\sqrt{2}&2\\            1&1&1&1&\sqrt{2}&\sqrt{2}&-\sqrt{2}&-\sqrt{2}&-2\\
            1&1&1&1&-\sqrt{2}&-\sqrt{2}&\sqrt{2}&\sqrt{2}&-2\\
            \sqrt{2}&-\sqrt{2}&\sqrt{2}&-\sqrt{2}&0&0&2&-2&0\\
            \sqrt{2}&-\sqrt{2}&\sqrt{2}&-\sqrt{2}&0&0&-2&2&0\\
            \sqrt{2}&-\sqrt{2}&-\sqrt{2}&\sqrt{2}&2&-2&0&0&0\\
            \sqrt{2}&-\sqrt{2}&-\sqrt{2}&\sqrt{2}&-2&2&0&0&0\\
            2&2&-2&-2&0&0&0&0&0
        \end{array} \right)
    \end{equation}
and
    \begin{equation}
        T = \textrm{Diag}\left(1,1,-1,-1,e^{\frac{2\pi i}{16}},-e^{\frac{2\pi i}{16}},e^{-\frac{2\pi i}{16}},-e^{-\frac{2\pi i}{16}},1 \right)
    \end{equation}
up to relabeling of line operators. The fusion table is presented in table \ref{fusion-ising}.
\begin{table}[!h]
    \footnotesize
    \centering
    \begin{tabular}{|c|c|c|c|c|c|c|c|c|c|}
    \hline
    $\otimes$& $W_0$ & $W_1$ &$W_2$&$W_3$&$W_4$&$W_5$&$W_6$&$W_7$&$W_8$\\
    \hline
    $W_0$& $W_0$& $W_1$ & $W_2$ & $W_3$ & $W_4$ & $W_5$ & $W_6$ & $W_7$ & $W_8$\\
    \hline
    $W_1$& $W_1$& $W_0$ & $W_3$& $W_2$& $W_5$&$W_4$&$W_7$&$W_6$& $W_8$\\
    \hline
    $W_2$&$W_2$&$W_3$&$W_0$&$W_1$&$W_5$&$W_4$&$W_6$& $W_7$&$W_8$\\
    \hline
    $W_3$&$W_3$&$W_2$&$W_1$&$W_0$&$W_4$&$W_5$&$W_7$&$W_6$&$W_8$\\
    \hline
    $W_4$&$W_4$&$W_5$&$W_5$&$W_4$&$W_0 \oplus W_3$&$W_1 \oplus W_2$& $W_8$ & $W_8$& $W_6 \oplus W_7$\\
    \hline
    $W_5$&$W_5$&$W_4$&$W_4$&$W_5$&$W_1 \oplus W_2$& $W_0 \oplus W_3$& $W_8$&$W_8$&$W_6 \oplus W_7$\\
    \hline
    $W_6$&$W_6$&$W_7$&$W_6$&$W_7$&$W_8$&$W_8$&$W_0 \oplus W_2$& $W_1 \oplus W_3$& $W_4 \oplus W_5$\\
    \hline
    $W_7$&$W_7$&$W_6$&$W_7$&$W_6$&$W_8$&$W_8$&$W_1 \oplus W_3$&$W_0 \oplus W_2$& $W_4 \oplus W_5$\\
    \hline
    $W_8$&$W_8$&$W_8$&$W_8$&$W_8$&$W_6 \oplus W_7$&$W_6 \oplus W_7$&$W_4 \oplus W_5$&$W_4 \oplus W_5$&$W_0\oplus W_1 \oplus W_2 \oplus W_3$\\
    \hline
    \end{tabular}
    \caption{The fusion rule of Ising SymTFT}
    \label{fusion-ising}
\end{table}
And there exist two kinds of Lagrangian algebras. The bosonic one is
    \begin{equation}
        \mathcal{L}_1 = W_0 \oplus W_1 \oplus W_8,
    \end{equation}
and the fermionic one is
    \begin{equation}
        \mathcal{L}_2 = W_0 \oplus W_1 \oplus W_2 \oplus W_3.
    \end{equation}

\subsubsection*{Topological boundary states for $\mathcal{L}_1$}
The vacuum of the topological boundary state is
    \begin{equation}
        |\Omega\rangle_{\mathcal{L}_1} = |0\rangle + |1\rangle + |8\rangle = |1,1,0,0,0,0,0,0,1\rangle,
    \end{equation}
which is invariant under both $S$- and $T$- transformations. The actions of line operators upon the topological boundary state are given in table \ref{Ising-L1}.
        \begin{table}[!h]
        \centering
        \begin{tabular}{|c|c|c|}
        \hline
            Lines/cycles & $\Gamma_1$ & $\Gamma_2$ \\
            \hline
            $W_0$&$|1,1,0,0,0,0,0,0,1\rangle$ & $|1,1,0,0,0,0,0,0,1\rangle$\\
            \hline
            $W_1$ & $|1,1,0,0,0,0,0,0,1\rangle$ & $|1,1,0,0,0,0,0,0,1\rangle$\\
            \hline
            $W_2$ & $|1,1,0,0,0,0,0,0,-1\rangle$&$|0,0,1,1,0,0,0,0,1\rangle$\\
            \hline
            $W_3$ & $|1,1,0,0,0,0,0,0,-1\rangle$&$|0,0,1,1,0,0,0,0,1\rangle$\\
            \hline
            $W_4$ & $|\sqrt{2},-\sqrt{2},0,0,0,0,0,0,0\rangle$ & $|0,0,0,0,1,1,1,1,0\rangle$\\
            \hline
            $W_5$ & $|\sqrt{2},-\sqrt{2},0,0,0,0,0,0,0\rangle$ & $|0,0,0,0,1,1,1,1,0\rangle$\\
            \hline
            $W_6$ & $|\sqrt{2},-\sqrt{2},0,0,0,0,0,0,0\rangle$ & $|0,0,0,0,1,1,1,1,0\rangle$\\
            \hline
            $W_7$& $|\sqrt{2},-\sqrt{2},0,0,0,0,0,0,0\rangle$ & $|0,0,0,0,1,1,1,1,0\rangle$\\
            \hline
            $W_8$ & $|2,2,0,0,0,0,0,0,0\rangle$ & $|1,1,1,1,0,0,0,0,2\rangle$\\
            \hline
        \end{tabular}
        \caption{The action of line operators on the topological boundary state $|\Omega\rangle_{\mathcal{L}_1}$}
        \label{Ising-L1}
        \end{table}
Here one can check the identification
    \begin{equation}
        W_0\sim W_1,\quad W_2 \sim W_3,\quad W_4\sim W_5\sim W_6 \sim W_7,\quad W_8 \sim W_0 + W_2.
    \end{equation}
on the topological boundary. Let us choose $W_0,W_3,W_4$ as three representatives, We can further examine the fusion rule
    \begin{equation}
        W_3 \times_{\mathcal{L}_1} W_3 = W_0,\quad W_3 \times_{\mathcal{L}_1} W_4 = W_4 \times_{\mathcal{L}_1} W_3 = W_4,\quad  W_4 \times_{\mathcal{L}_1} W_4 = W_0 + W_3,
    \end{equation}
which gives the Ising fusion category. We can identify
    \begin{equation}
        W_0 = I,\quad W_3 = \eta,\quad W_4 = \mathcal{N},
    \end{equation}
where $(I,\eta,\mathcal{N})$ are the identity, $\mathbb{Z}_2$ and non-invertible TDLs in Ising category.

\subsubsection*{Topological boundary states for $\mathcal{L}_2$}

The vacuum of the topological boundary state is
    \begin{equation}
        |\Omega\rangle_{\mathcal{L}_2} = |0\rangle + |1\rangle + |2\rangle + |3\rangle = |1,1,1,1,0,0,0,0,0\rangle,
    \end{equation}
which is not invariant under T-transformation due to its fermionic nature.
        \begin{table}[!h]
        \centering
        \begin{tabular}{|c|c|c|}
        \hline
            Lines/cycles & $\Gamma_1$ & $\Gamma_2$ \\
            \hline
            $W_0$&$|1,1,1,1,0,0,0,0,0\rangle$ & $|1,1,1,1,0,0,0,0,0\rangle$\\
            \hline
            $W_1$ &$|1,1,1,1,0,0,0,0,0\rangle$ & $|1,1,1,1,0,0,0,0,0\rangle$\\
            \hline
            $W_2$ & $|1,1,1,1,0,0,0,0,0\rangle$ & $|1,1,1,1,0,0,0,0,0\rangle$\\
            \hline
            $W_3$&$|1,1,1,1,0,0,0,0,0\rangle$ & $|1,1,1,1,0,0,0,0,0\rangle$\\
            \hline
            $W_4$ & $|\sqrt{2},-\sqrt{2},\sqrt{2},-\sqrt{2},0,0,0,0,0\rangle$ & $|0,0,0,0,2,2,0,0,0\rangle$\\
            \hline
            $W_5$ & $|\sqrt{2},-\sqrt{2},\sqrt{2},-\sqrt{2},0,0,0,0,0\rangle$ & $|0,0,0,0,2,2,0,0,0\rangle$\\
            \hline
            $W_6$ & $|\sqrt{2},-\sqrt{2},-\sqrt{2},\sqrt{2},0,0,0,0,0\rangle$ & $|0,0,0,0,0,0,2,2,0\rangle$\\
            \hline
            $W_7$& $|\sqrt{2},-\sqrt{2},-\sqrt{2},\sqrt{2},0,0,0,0,0\rangle$ & $|0,0,0,0,0,0,2,2,0\rangle$\\
            \hline
            $W_8$ & $|2,2,-2,-2,0,0,0,0,0\rangle$ & $|0,0,0,0,0,0,0,0,4\rangle$\\
            \hline
        \end{tabular}
        \caption{The action of line operators on the topological boundary state $|\Omega\rangle_{\mathcal{L}_2}$}
        \label{Ising-L2}
        \end{table}
One can check the identification following the table
        \begin{equation}
            W_0 \sim W_1 \sim W_2 \sim W_3,\quad W_4\sim W_5,\quad W_6\sim W_7,\quad 
        \end{equation}
and we can choose $W_0,W_4,W_6,W_8$ as three representatives. The fusion rules are given by
    \begin{equation}
        W_4\times_{\mathcal{L}_2} W_4 = 2 W_0,\quad W_6 \times_{\mathcal{L}_2} W_6 = 2W_0,\quad W_8 \times_{\mathcal{L}_2} W_8 =4 W_0,
    \end{equation}
    and also
    \begin{equation}
        W_4 \times_{\mathcal{L}_2} W_6 = W_6 \times_{\mathcal{L}_2} W_4 = W_8,\quad W_4 \times_{\mathcal{L}_2} W_8 = W_8 \times_{\mathcal{L}_2} W_4 = W_6,\quad W_6 \times_{\mathcal{L}_2} W_8 = W_8 \times_{\mathcal{L}_2} W_6 = W_4.
    \end{equation}
Therefore one can identify
        \begin{equation}
            W_0 = I,\quad W_4 = (-1)^{F_L},\quad W_6 = (-1)^{F_R},\quad W_8 = 2 (-1)^F.
        \end{equation}
\subsubsection{$F$-moves, half-braidings, and self-duality in Ising Model}

In addition to the fusion rule, we can also extract the F-move and half-braiding coefficients from the analysis of the topological boundary states as we will show next. Here we will give a very brief review of the $F$-move and half-braiding structure.

Consider a 2d TDL network obeying the fusion rules given by the LHS in the following equation, where $L_1,\cdots,L_5$ are all simple TDLs. The three TDLs attached to each junction must obey the fusion rule. For example for the $L_1,L_2,L_5$ junction, $L_5\in L_1 \times L_2$. We will also assume all the fusion coefficients are zero or one for simplicity. We can consider moving the endpoint of $L_2$ line and let it end on $L_3$ as shown in the RHS, such a topological deformation is called an F-move. The general relations of the two diagrams are related by the following equations, where on the RHS we need to sum over the simple line $L_6$ obeying the fusion rule with the complex $F$-coefficients $F^{L_1 L_2 L_3}_{L_4}(L_5,L_6)$
\begin{equation}\label{F-coefficient}
\begin{gathered}
\begin{tikzpicture}[scale=.5]
\node at (0, 1.2) {};
\draw[thick] (0,-0.6) -- (0,1.2);
\draw[thick] (0,1.2) -- (-2.4,3.6);
\draw[thick] (0,1.2) -- (2.4,3.6);
\draw[thick] (-1.2,2.4) -- (0,3.6);
\node at (0,-1.2) {$L_4$};
\node at (-2.4,4.2) {$L_1$};
\node at (0,4.2) {$L_2$};
\node at (2.4,4.2) {$L_3$};
\node at (-1.2,1.2) {$L_5$};
\end{tikzpicture}
\end{gathered} \quad = \quad \sum_{L_6} 
\begin{gathered}
\begin{tikzpicture}[scale=.5]
\node at (0, 1.2) {};
\draw[thick] (0,-0.6) -- (0,1.2);
\draw[thick] (0,1.2) -- (-2.4,3.6);
\draw[thick] (0,1.2) -- (2.4,3.6);
\draw[thick] (1.2,2.4) -- (0,3.6);
\node at (0,-1.2) {$L_4$};
\node at (-2.4,4.2) {$L_1$};
\node at (0,4.2) {$L_2$};
\node at (2.4,4.2) {$L_3$};
\node at (1.2,1.2) {$L_6$};
\end{tikzpicture}
\end{gathered} \quad \times \quad F^{L_1 L_2 L_3}_{L_4} (L_5,L_6).
\end{equation}
The $F$-coefficients $F^{L_1 L_2 L_3}_{L_4}(L_5,L_6)$ are constrained by the pentagon equations
\begin{equation}
    \sum_{L_{10}} F^{L_2 L_3 L_4}_{L_9} (L_{10},L_{8}) F^{L_1 L_{10} L_{4}}_{L_5}(L_7,L_9) F^{L_1 L_2 L_3}_{L_7}(L_6,L_{10}) = F^{L_1 L_2 L_8}_{L_5} (L_6,L_9) F^{L_6 L_3 L_4}_{L_5} (L_7,L_8)
\end{equation}
for any given simple TDL $L_1,\cdots,L_9$. For example, when the symmetry category $\mathcal{S}$ is a group $G$, the simple TDLs are identified as the generators of the group, and the solutions of pentagon equations are classified by the 3rd group cohomology $H^3(G,U(1))$, which characterizes the 't Hooft anomaly of $G$ in 2d.

In some cases, the simple TDLs are objects of an MTC and satisfy additional constraints. To define an MTC, we need to lift 2d TDLs into 3d topological line operators so they can braid with other line operators. The braiding structure is characterized by another set of coefficients defined below. Suppose we have obtained a solution of the pentagon equation, we can further consider the structure of half-braiding defined in the following diagram
\begin{equation}\label{R-coefficient}
    \begin{gathered}
        \begin{tikzpicture}
            \draw[line,thick] (0,0)--(1.2,1.2);
            \draw[line,thick] (1.2,0)--(0.7,0.5);
            \draw[line,thick] (0.5,0.7)--(0,1.2);
            \node at (0,-0.3) {$L_2$};
            \node at (1.3,-0.3) {$L_1$};
        \end{tikzpicture}
    \end{gathered}\quad = \quad \sum_{L_3} \sqrt{\frac{d_{L_3}}{d_{L_1} d_{L_2}}}\quad 
    \begin{gathered}
        \begin{tikzpicture}
            \draw[line,thick] (0,0)--(0.6,0.6);
            \draw[line,thick] (1.2,0)--(0.6,0.6);
            \draw[line,thick] (0.6,0.6) -- (0.6,1.2);
            \draw[line,thick] (0.6,1.2) -- (0,1.8);
            \draw[line,thick] (0.6,1.2) -- (1.2,1.8);
            \node at (0.9,0.9) {$L_3$};
            \node at (0,-0.3) {$L_2$};
            \node at (1.3,-0.3) {$L_1$};
            \node at (0.0,2.1) {$L_1$};
            \node at (1.3,2.1) {$L_2$};
        \end{tikzpicture}     
    \end{gathered}\quad \times \quad R_{L_3}^{L_1 L_2}
\end{equation}
where $L_2$ is on the top of $L_1$ in the LHS and we still assume the fusion coefficients are zero or one. The R-coefficients are complex numbers that satisfy two sets of hexagon equations given by
    \begin{equation}
        \sum_{L_7} F^{L_1 L_2 L_3}_{L_4} (L_7,L_6) R_{L_4}^{L_7 L_3} R^{L_3 L_1 L_2}_{L_4} (L_5,L_7) = R^{L_2 L_3}_{L_6} F^{L_1 L_3 L_2}_{L_4}(L_5,L_6)R^{L_1 L_3}_{L_5}
    \end{equation}
for each $L_1,\cdots,L_6$ and another one by replacing $R^{AB}_C \rightarrow (R^{BA}_C)^{-1}$. 

In general, the solution of hexagon equations might not exist. But if we can find a solution of hexagon equations for a given set of solutions of $\{F^{ABC}_D(E,F)\}$ for pentagon equations, then roughly speaking we can assign the modular structure to the underlying 2D system. For example, the topological spins for each TDL ($T$-matrix) and $S$-matrix are given by
    \begin{equation}
        \theta_{L_1} = \sum_{L_2} \frac{d_{L_2}}{d_{L_1}} R^{L_1 L_1}_{L_2},\quad S_{L_1 L_2} = D^{-1} \sum_{L_3} N^{L_3}_{\bar{L}_1 L_2} \frac{\theta_{L_3}}{\theta_{L_1}\theta_{L_2}} d_{L_3},
    \end{equation}
In summary, begin with a collection of TDLs satisfying some fusion rules, there might exist three layers of structures upon that: $F$-coefficients constrained by the pentagon equations, $R$-coefficients constrained by the F-move solutions and hexagon equations, and the modular matrices $S/T$. Each layer of structure is determined by the previous one.

So far we built the topological boundary only using the $S/T$-matrices and did not talk about the F-move and half-braiding of the underlying 2D theory. It is interesting to see whether we can recover those data from the picture of SymTFT. In this section, we will compute the $F$-move and half-braiding coefficients of the Ising category from the SymTFT picture.

\noindent
\textbf{$F$-moves}\\
First, by acting $W_3$ and $W_4$ on the topological vacuum $|\Omega\rangle_{\mathcal{L}_1}$ and applying $T$-transformation we can obtain the following basis vectors
\begin{align}
\begin{gathered}
\begin{tikzpicture}[scale=.5]
\node at (0, 1.2) {};
\draw[very thick] (0,0) rectangle (1.2,1.2);
\end{tikzpicture}
\end{gathered}
:\quad &\, |0\rangle_{\mathcal{L}_1}\equiv |\Omega\rangle_{\mathcal L_1}=|1,1,0,0,0,0,0,0,1\rangle\notag\\
\begin{gathered}
\begin{tikzpicture}[scale=.5]
\node at (0, 1.2) {};
\draw[very thick] (0,0) rectangle (1.2,1.2);
\draw [line,densely dotted,thick] (0,.60) -- (1.2,.60); 
\end{tikzpicture}
\end{gathered}
:\quad &\, |1\rangle_{\mathcal{L}_1}\equiv W_{3}[\Gamma_1]|\Omega\rangle_{\mathcal L_1}=|1,1,0,0,0,0,0,0,-1\rangle\notag\\
\begin{gathered}
\begin{tikzpicture}[scale=.5]
\node at (0, 1.2) {};
\draw[very thick] (0,0) rectangle (1.2,1.2);
\draw [line,densely dotted,thick] (.60,0) -- (.60,1.20); 
\end{tikzpicture}
\end{gathered}
:\quad &\, |2\rangle_{\mathcal{L}_1}\equiv W_{3}[\Gamma_2]|\Omega\rangle_{\mathcal L_1}=|0,0,1,1,0,0,0,0,1\rangle\notag\\
\begin{gathered}
\begin{tikzpicture}[scale=.5]
\node at (0, 1.2) {};
\draw[very thick] (0,0) rectangle (1.2,1.2);
\draw [line, thick] (0,.60) -- (1.2,.60); 
\end{tikzpicture}
\end{gathered}
:\quad &\, |3\rangle_{\mathcal{L}_1}\equiv W_{4}[\Gamma_1]|\Omega\rangle_{\mathcal L_1}=|\sqrt{2},-\sqrt{2},0,0,0,0,0,0,0\rangle\notag\\
\begin{gathered}
\begin{tikzpicture}[scale=.5]
\node at (0, 1.2) {};
\draw[very thick] (0,0) rectangle (1.2,1.2);
\draw [line, thick] (.60,0) -- (.60,1.20); 
\end{tikzpicture}
\end{gathered}
:\quad &\, |4\rangle_{\mathcal{L}_1}\equiv W_{4}[\Gamma_2]|\Omega\rangle_{\mathcal L_1}=|0,0,0,0,1,1,1,1,0\rangle\notag\\
\begin{gathered}
\begin{tikzpicture}[scale=.5]
\node at (0, 1.2) {};
\draw[very thick] (0,0) rectangle (1.2,1.2);
\draw[thick]  (.6,0) arc (0:90:.6); 
\draw[thick] (.6,1.2) arc (-180:-90:.6);
\end{tikzpicture}
\end{gathered}
:\quad &\, |5\rangle_{\mathcal{L}_1}\equiv\hat{\mathbb T}\cdot W_{4}[\Gamma_2]|\Omega\rangle_{\mathcal L_1}=|0,0,0,0,\omega,-\omega,\omega^{-1},-\omega^{-1},0\rangle\notag\\
\begin{gathered}
\begin{tikzpicture}[scale=0.5]
\node at (0, 1.2) {};
\draw[very thick] (0,0) rectangle (1.2,1.2);
\draw[thick]  (.6,0) arc (-180:-270:.6); 
\draw[thick] (.6,1.2) arc (0:-90:.6);
\end{tikzpicture}
\end{gathered}:\quad &\, |6\rangle_{\mathcal{L}_1} \equiv \hat{\mathbb T}^{-1} \cdot W_{4}[\Gamma_2]|\Omega\rangle_{\mathcal L_1}=|0,0,0,0,\omega^{-1},-\omega^{-1},\omega,-\omega,0\rangle,
\end{align}
where $\omega=e^{\frac{2\pi i}{16}}$. The dashed line is the $\mathbb{Z}_2$ generator $\eta$ and the solid line is the non-invertible TDL $\mathcal{N}$. Besides, one also have
    \begin{equation}
\begin{gathered}
\begin{tikzpicture}[scale=.5]
\node at (0, 1.2) {};
\draw[very thick] (0,0) rectangle (1.2,1.2);
\draw[densely dotted, thick]  (.6,0) arc (0:90:.6); 
\draw[densely dotted, thick] (.6,1.2) arc (-180:-90:.6);
\end{tikzpicture}
\end{gathered}
:\quad \, \hat{\mathbb T}\cdot W_{3}[\Gamma_2]|\Omega\rangle_{\mathcal L_1}=|0,0,-1,-1,0,0,0,0,1\rangle.       
    \end{equation}
However, one can check it is not independent and satisfies
    \begin{equation}\label{Ising-Z2-relation}
        \begin{gathered}
\begin{tikzpicture}[scale=.5]
\node at (0, 1.2) {};
\draw[very thick] (0,0) rectangle (1.2,1.2);
\draw[densely dotted, thick]  (.6,0) arc (0:90:.6); 
\draw[densely dotted, thick] (.6,1.2) arc (-180:-90:.6);
\end{tikzpicture}
\end{gathered} = \begin{gathered}
\begin{tikzpicture}[scale=.5]
\node at (0, 1.2) {};
\draw[very thick] (0,0) rectangle (1.2,1.2);
\end{tikzpicture}
\end{gathered} - \begin{gathered}
\begin{tikzpicture}[scale=.5]
\node at (0, 1.2) {};
\draw[very thick] (0,0) rectangle (1.2,1.2);
\draw [line,densely dotted,thick] (0,.60) -- (1.2,.60); 
\end{tikzpicture}
\end{gathered} - \begin{gathered}
\begin{tikzpicture}[scale=.5]
\node at (0, 1.2) {};
\draw[very thick] (0,0) rectangle (1.2,1.2);
\draw [line,densely dotted,thick] (.60,0) -- (.60,1.20); 
\end{tikzpicture}
\end{gathered}
    \end{equation}
which is different compared to the $\mathbb{Z}_2$ sector in the previous examples. This is related to the fact that the boundary theory is self-dual under $\mathbb{Z}_2$ gauging as we will show later.

To derive the $F$-move of the TDLs, we first notice that on the topological boundary one has
\begin{align}
\begin{gathered}
\begin{tikzpicture}[scale=1]
\node at (0, 1.2) {};
\draw[very thick] (0,0) rectangle (1.2,1.2);
\draw [line, thick] (.4,0) -- (.40,1.20); 
\draw [line, thick] (.8,0) -- (.80,1.20); 
\end{tikzpicture}
\end{gathered}
\quad=\quad
\begin{gathered}
\begin{tikzpicture}[scale=1]
\node at (0, 1.2) {};
\draw[very thick] (0,0) rectangle (1.2,1.2);
\end{tikzpicture}
\end{gathered}
\quad+\quad
\begin{gathered}
\begin{tikzpicture}[scale=1]
\node at (0, 1.2) {};
\draw[very thick] (0,0) rectangle (1.2,1.2);
\draw [line,densely dotted,thick] (.6,0) -- (.60,1.20); 
\end{tikzpicture}
\end{gathered}
\,
\end{align}
by the fusion rule $\mathcal{N}^2 = I + \eta$. On the other hand, F-move of the two $\mathcal{N}$-lines implies
\begin{align}
\begin{gathered}
\begin{tikzpicture}[scale=1]
\node at (0, 1.2) {};
\draw[very thick] (0,0) rectangle (1.2,1.2);
\draw [line, thick] (.4,0) -- (.40,1.20); 
\draw [line, thick] (.8,0) -- (.80,1.20); 
\end{tikzpicture}
\end{gathered}
\quad=\quad \alpha\quad
\begin{gathered}
\begin{tikzpicture}[scale=1]
\node at (0, 1.2) {};
\draw[very thick] (0,0) rectangle (1.2,1.2);
\draw[thick] (.9,1.2) arc (0:-180:.3); 
\draw[thick] (.3,0) arc (180:0:.3); 
\end{tikzpicture}
\end{gathered}
\quad+\quad \beta\quad
\begin{gathered}
\begin{tikzpicture}[scale=1]
\node at (0, 1.2) {};
\draw[very thick] (0,0) rectangle (1.2,1.2);
\draw[thick] (.9,1.2) arc (0:-180:.3); 
\draw[thick] (.3,0) arc (180:0:.3); 
\draw[line,densely dotted,thick] (.6, .3) -- (.6, .9);
\end{tikzpicture}
\end{gathered}
\end{align}
where $\alpha,\beta$ are undetermined coefficients. We will assume the vevs of the $\mathcal{N}$-bubbles give the quantum dimension of the $\mathcal{N}$-line on the boundary,
\begin{align}
\begin{gathered}
\begin{tikzpicture}[scale=1]
\node at (0, 1.2) {};
\draw[very thick] (0,0) rectangle (1.2,1.2);
\draw[thick] (.6,.6) circle (.25); 
\end{tikzpicture}
\end{gathered}&
\quad=\quad d_{\mathcal{N}} \quad
\begin{gathered}
\begin{tikzpicture}[scale=1]
\node at (0, 1.2) {};
\draw[very thick] (0,0) rectangle (1.2,1.2);
\end{tikzpicture}
\end{gathered}
\, \notag\\
\begin{gathered}
\begin{tikzpicture}[scale=1]
\node at (0, 1.2) {};
\draw[very thick] (0,0) rectangle (1.2,1.2);
\draw[thick] (.6,.6) circle (.25); 
\draw[line,densely dotted, thick] (.6,0) -- (.6, .35);
\draw[line,densely dotted, thick] (.6,1.2) -- (.6, .85);
\end{tikzpicture}
\end{gathered}&
\quad=\quad d_{\mathcal{N}} \quad
\begin{gathered}
\begin{tikzpicture}[scale=1]
\node at (0, 1.2) {};
\draw[very thick] (0,0) rectangle (1.2,1.2);
\draw [line,densely dotted,thick] (.60,0) -- (.60,1.20); 
\end{tikzpicture}
\end{gathered}
\end{align}
where $d_{\mathcal{N}}=\sqrt{2}$. It thus implies that
\begin{align}
\alpha=\beta=d_{\mathcal{N}}^{-1} = \frac{1}{\sqrt{2}} \, 
\end{align}
Therefore one has the following relations
\begin{align}
    &\begin{gathered}
\begin{tikzpicture}[scale=1]
\node at (0, 1.2) {};
\draw[very thick] (0,0) rectangle (1.2,1.2);
\draw[thick]  (.6,0) arc (0:90:.6); 
\draw[thick] (.6,1.2) arc (-180:-90:.6);
\end{tikzpicture}
\end{gathered} \quad = \quad \frac{1}{\sqrt{2}}\quad \begin{gathered}
\begin{tikzpicture}[scale=1]
\node at (0, 1.2) {};
\draw[very thick] (0,0) rectangle (1.2,1.2);
\draw[thick]  (.6,0) arc (-180:-270:.6); 
\draw[thick] (.6,1.2) arc (0:-90:.6);
\end{tikzpicture}
\end{gathered} \quad + \quad \frac{1}{\sqrt{2}}\quad\begin{gathered}
\begin{tikzpicture}[scale=1]
\node at (0, 1.2) {};
\draw[very thick] (0,0) rectangle (1.2,1.2);
\draw[thick]  (.6,0) arc (180:90:.6); 
\draw[thick] (.6,1.2) arc (0:-90:.6);
\draw[line, thick,densely dotted] (.776,.424)--(.424,.776);
\end{tikzpicture}
\end{gathered} \notag\\
&\begin{gathered}
\begin{tikzpicture}[scale=1]
\node at (0, 1.2) {};
\draw[very thick] (0,0) rectangle (1.2,1.2);
\draw[thick]  (.6,0) arc (-180:-270:.6); 
\draw[thick] (.6,1.2) arc (0:-90:.6);
\end{tikzpicture}
\end{gathered} \quad = \quad \frac{1}{\sqrt{2}}\quad \begin{gathered}
\begin{tikzpicture}[scale=1]
\node at (0, 1.2) {};
\draw[very thick] (0,0) rectangle (1.2,1.2);
\draw[thick]  (.6,0) arc (0:90:.6); 
\draw[thick] (.6,1.2) arc (-180:-90:.6);
\end{tikzpicture}
\end{gathered} \quad + \quad \frac{1}{\sqrt{2}}\quad\begin{gathered}
\begin{tikzpicture}[scale=1]
\node at (0, 1.2) {};
\draw[very thick] (0,0) rectangle (1.2,1.2);
\draw[thick]  (.6,0) arc (0:90:.6); 
\draw[thick] (.6,1.2) arc (-180:-90:.6);
\draw[line, thick,densely dotted] (.424,.424)--(.776,.776);
\end{tikzpicture}
\end{gathered}
\end{align}
from where one can solve the states
\begin{equation}
    \begin{gathered}
\begin{tikzpicture}[scale=0.5]
\node at (0, 1.2) {};
\draw[very thick] (0,0) rectangle (1.2,1.2);
\draw[thick]  (.6,0) arc (180:90:.6); 
\draw[thick] (.6,1.2) arc (0:-90:.6);
\draw[line, thick,densely dotted] (.776,.424)--(.424,.776);
\end{tikzpicture}
\end{gathered}:\quad |0,0,0,0,i\omega^{-1},-i\omega^{-1},-i\omega,i\omega,0\rangle,\quad \begin{gathered}
\begin{tikzpicture}[scale=0.5]
\node at (0, 1.2) {};
\draw[very thick] (0,0) rectangle (1.2,1.2);
\draw[thick]  (.6,0) arc (0:90:.6); 
\draw[thick] (.6,1.2) arc (-180:-90:.6);
\draw[line, thick,densely dotted] (.424,.424)--(.776,.776);
\end{tikzpicture}
\end{gathered}:\quad |0,0,0,0,-i\omega,i\omega,i\omega^{-1},-i\omega^{-1},0\rangle,
\end{equation}
and verify the $F$-move\footnote{Notice that we only need the properties of TDLs to obtain this F-move coefficients. We thank Zhihao Duan for pointing this out.}
\begin{align}
\begin{pmatrix}
\begin{gathered}
\begin{tikzpicture}[scale=.5]
\node at (0, 1.2) {};
\draw[very thick] (0,0) rectangle (1.2,1.2);
\draw[thick]  (.6,0) arc (-180:-270:.6); 
\draw[thick] (.6,1.2) arc (0:-90:.6);
\end{tikzpicture}
\end{gathered}
\,\, {}\\
\begin{gathered}
\begin{tikzpicture}[scale=.5]
\node at (0, 1.2) {};
\draw[very thick] (0,0) rectangle (1.2,1.2);
\draw[thick]  (.6,0) arc (180:90:.6); 
\draw[thick] (.6,1.2) arc (0:-90:.6);
\draw[line, thick, densely dotted] (.776,.424)--(.424,.776);
\end{tikzpicture}
\end{gathered}
\,\,{}
\end{pmatrix}
\quad=\quad
\begin{pmatrix}
\frac{1}{\sqrt{2}} & \frac{1}{\sqrt{2}}
\,\,{}\\
\frac{1}{\sqrt{2}} & -\frac{1}{\sqrt{2}}\,\,{}
\end{pmatrix}
\cdot
\begin{pmatrix}
\begin{gathered}
\begin{tikzpicture}[scale=.5]
\node at (0, 1.2) {};
\draw[very thick] (0,0) rectangle (1.2,1.2);
\draw[thick]  (.6,0) arc (0:90:.6); 
\draw[thick] (.6,1.2) arc (-180:-90:.6);
\end{tikzpicture}
\end{gathered}
\,\,{}\\
\begin{gathered}
\begin{tikzpicture}[scale=.5]
\node at (0, 1.2) {};
\draw[very thick] (0,0) rectangle (1.2,1.2);
\draw[thick]  (.6,0) arc (0:90:.6); 
\draw[thick] (.6,1.2) arc (-180:-90:.6);
\draw[line, thick,densely dotted] (.424,.424)--(.776,.776);
\end{tikzpicture}
\end{gathered}
\,\,{}
\end{pmatrix}
\,.
\end{align}
Compared to \eqref{F-coefficient}, we can read
    \begin{equation}
        F^{\mathcal{N}\mathcal{N}\mathcal{N}}_{\mathcal{N}}(I,\mathcal{N}) = F^{\mathcal{N}\mathcal{N}\mathcal{N}}_{\mathcal{N}}(I,\eta)=F^{\mathcal{N}\mathcal{N}\mathcal{N}}_{\mathcal{N}}(\mathcal{N},I) = \frac{1}{\sqrt{2}},\quad F^{\mathcal{N}\mathcal{N}\mathcal{N}}_{\mathcal{N}}(\mathcal{N},\mathcal{N}) = -\frac{1}{\sqrt{2}}.
    \end{equation}
Further, notice that we have the following relations under $T$-transformation
    \begin{equation}
        \begin{gathered}
\begin{tikzpicture}[scale=1]
\node at (0, 1.2) {};
\draw[very thick] (0,0) rectangle (1.2,1.2);
\draw[thick]  (.6,0) arc (180:90:.6); 
\draw[thick] (.6,1.2) arc (0:-90:.6);
\draw[line, thick,densely dotted] (.776,.424)--(.424,.776);
\end{tikzpicture}
\end{gathered}\quad = \quad \hat{\mathbb{T}}^{-1} \cdot \begin{gathered}
\begin{tikzpicture}[scale=1]
\node at (0, 1.2) {};
\draw[very thick] (0,0) rectangle (1.2,1.2);
\draw[line,thick,densely dotted] (0,.60) -- (.60,.30);
\draw[line,thick,densely dotted] (.60,.90) -- (1.20,.60);
\draw [line, thick] (.60,0) -- (.60,1.20); 
\end{tikzpicture}
\end{gathered} \quad ,\qquad \begin{gathered}
\begin{tikzpicture}[scale=1]
\node at (0, 1.2) {};
\draw[very thick] (0,0) rectangle (1.2,1.2);
\draw[thick]  (.6,0) arc (0:90:.6); 
\draw[thick] (.6,1.2) arc (-180:-90:.6);
\draw[line, thick,densely dotted] (.424,.424)--(.776,.776);
\end{tikzpicture}
\end{gathered} \quad = \quad \hat{\mathbb{T}} \cdot \begin{gathered}
\begin{tikzpicture}[scale=1]
\node at (0, 1.2) {};
\draw[very thick] (0,0) rectangle (1.2,1.2);
\draw[line,thick,densely dotted] (0,.60) -- (.60,.90);
\draw[line,thick,densely dotted] (.60,.30) -- (1.20,.60);
\draw [line, thick] (.60,0) -- (.60,1.20); 
\end{tikzpicture}
\end{gathered} ,
    \end{equation}
and we can obtain the other two basis as
\begin{align}
    \begin{gathered}
\begin{tikzpicture}[scale=0.5]
\node at (0, 1.2) {};
\draw[very thick] (0,0) rectangle (1.2,1.2);
\draw[line,thick,densely dotted] (0,.60) -- (.60,.30);
\draw[line,thick,densely dotted] (.60,.90) -- (1.20,.60);
\draw [line, thick] (.60,0) -- (.60,1.20); 
\end{tikzpicture}
\end{gathered}:&\quad \,|7\rangle_{\mathcal{L}_1} = |0,0,0,0,i,i,-i,-i,0\rangle\notag\\
\begin{gathered}
\begin{tikzpicture}[scale=.5]
\node at (0, 1.2) {};
\draw[very thick] (0,0) rectangle (1.2,1.2);
\draw [line, thick] (0,.60) -- (1.2,.60); 
\draw [line, thick,densely dotted] (0.6,1.20) -- (0.30,.60); 
\draw [line, thick,densely dotted] (0.9,.60) -- (0.60,0); 
\end{tikzpicture}
\end{gathered}:&\quad \,|8\rangle_{\mathcal{L}_1} = \hat{\mathbb{S}}\cdot |7\rangle_{\mathcal{L}_1} = |0,0,\sqrt{2} i,-\sqrt{2} i,0,0,0,0,0\rangle
\end{align}
and also
\begin{equation}
    \begin{gathered}
\begin{tikzpicture}[scale=0.5]
\node at (0, 1.2) {};
\draw[very thick] (0,0) rectangle (1.2,1.2);
\draw[line,thick,densely dotted] (0,.60) -- (.60,.90);
\draw[line,thick,densely dotted] (.60,.30) -- (1.20,.60);
\draw [line, thick] (.60,0) -- (.60,1.20); 
\end{tikzpicture}
\end{gathered}: \quad \,|7'\rangle_{\mathcal{L}_1} = |0,0,0,0,-i,-i,i,i,0\rangle.
\end{equation}
Compared $|7\rangle_{\mathcal{L}_1}$ and $|7'\rangle_{\mathcal{L}_1}$ we can recover the other non-trivial $F$-move relation
    \begin{equation}
        \begin{gathered}
\begin{tikzpicture}[scale=1]
\node at (0, 1.2) {};
\draw[very thick] (0,0) rectangle (1.2,1.2);
\draw[line,thick,densely dotted] (0,.60) -- (.60,.30);
\draw[line,thick,densely dotted] (.60,.90) -- (1.20,.60);
\draw [line, thick] (.60,0) -- (.60,1.20); 
\end{tikzpicture}
\end{gathered}\quad =\quad -\quad \begin{gathered}
\begin{tikzpicture}[scale=1]
\node at (0, 1.2) {};
\draw[very thick] (0,0) rectangle (1.2,1.2);
\draw[line,thick,densely dotted] (0,.60) -- (.60,.90);
\draw[line,thick,densely dotted] (.60,.30) -- (1.20,.60);
\draw [line, thick] (.60,0) -- (.60,1.20); 
\end{tikzpicture}
\end{gathered}\quad,
    \end{equation}
or in terms of $F$-coefficient
    \begin{equation}
        F^{\eta \mathcal{N}\eta}_{\mathcal{N}}(\mathcal{N},\mathcal{N}) = -1.
    \end{equation}
\noindent
\textbf{Half-braidings}\\
Another important property of the category of the symmetry defects on the boundary is the half-braiding, which, roughly speaking, connects a fusion of two defects $L_1
\otimes L_2$ to that of $L_2\otimes L_1$

We can also derive the half-braiding on the topological boundary by turning on a symmetry defect $S$ along $\Gamma_2$ and putting the line operators $W$ onto the topological boundary. 
\begin{align}
    \begin{gathered}
    \begin{tikzpicture}[scale=1]
\draw[very thick] (0,0)--(0.6,2.4);
\draw[very thick] (0.6,2.4)--(3.0,2.4);
\draw[very thick] (3.0,2.4)--(2.4,0);
\draw[very thick] (2.4,0)--(0,0);
\draw[thick,red] (1.2,0.8)--(1.8,3.2);
\draw[dashed] (1.2,0)--(1.2,0.8);
\draw[dashed] (1.8,2.4)--(1.8,3.2);
\draw[dashed] (1.2,0)--(1.8,2.4);
\draw[thick,blue] (0.3,1.2)--(2.7,1.2);
\node at (0,1.2) {$S$};
\node at (0.9,0.9) {$W$};
\draw[->] (3.3,1.2)--(5.4,1.2);
\draw[very thick] (6,0)--(6.6,2.4);
\draw[very thick] (6.6,2.4)--(9.0,2.4);
\draw[very thick] (9.0,2.4)--(8.4,0);
\draw[very thick] (8.4,0)--(6,0);
\draw[thick,red] (7.2,0)--(7.8,2.4);
\draw[thick,blue] (6.3,1.2)--(8.7,1.2);
\node at (7.2,-0.3) {$W$};
\node at (6,1.2) {$S$};
\end{tikzpicture}\end{gathered}\nonumber
\end{align}
When $S$ is the identity line the half-braiding is trivial. We will consider $S=(\eta,\mathcal{N})$ separately for the $\mathbb{Z}_2$ and non-invertible TDL. When $W=W_0$ and $W_1$, we have
\begin{equation}
\begin{gathered}
\begin{tikzpicture}[scale=1]
\node at (0, 1.8) {};
\draw[very thick] (0,0) rectangle (1.2,1.2);
\draw [line, thick,red] (.60,0) -- (.60,1.20); 
\draw [line, thick,densely dotted] (0,.60)--(1.20,0.60);
\node at (-0.3,0.6) {$\eta$};
\node at (0.6,-0.3) {$W_0$};
\end{tikzpicture}
\end{gathered}\quad \ = \quad \begin{gathered}
\begin{tikzpicture}[scale=1]
\node at (0, 1.2) {};
\draw[very thick] (0,0) rectangle (1.2,1.2);
\draw [line, thick,densely dotted] (0,.60)--(1.20,0.60);
\end{tikzpicture}
\end{gathered} \quad , \qquad \begin{gathered}
\begin{tikzpicture}[scale=1]
\node at (0, 1.8) {};
\draw[very thick] (0,0) rectangle (1.2,1.2);
\draw [line, thick,red] (.60,0) -- (.60,1.20); 
\draw [line, thick] (0,.60)--(1.20,0.60);
\node at (-0.3,0.6) {$\mathcal{N}$};
\node at (0.6,-0.3) {$W_0$};
\end{tikzpicture}
\end{gathered}\quad \ = \quad \begin{gathered}
\begin{tikzpicture}[scale=1]
\node at (0, 1.2) {};
\draw[very thick] (0,0) rectangle (1.2,1.2);
\draw [line, thick] (0,.60)--(1.20,0.60);
\end{tikzpicture}
\end{gathered}
\end{equation}
for $W_0$ and
\begin{equation}
\begin{gathered}
\begin{tikzpicture}[scale=1]
\node at (0, 1.8) {};
\draw[very thick] (0,0) rectangle (1.2,1.2);
\draw [line, thick,red] (.60,0) -- (.60,1.20); 
\draw [line, thick,densely dotted] (0,.60)--(1.20,0.60);
\node at (-0.3,0.6) {$\eta$};
\node at (0.6,-0.3) {$W_1$};
\end{tikzpicture}
\end{gathered}\quad \ = \quad \begin{gathered}
\begin{tikzpicture}[scale=1]
\node at (0, 1.2) {};
\draw[very thick] (0,0) rectangle (1.2,1.2);
\draw [line, thick,densely dotted] (0,.60)--(1.20,0.60);
\end{tikzpicture}
\end{gathered} \quad , \qquad \begin{gathered}
\begin{tikzpicture}[scale=1]
\node at (0, 1.8) {};
\draw[very thick] (0,0) rectangle (1.2,1.2);
\draw [line, thick,red] (.60,0) -- (.60,1.20); 
\draw [line, thick] (0,.60)--(1.20,0.60);
\node at (-0.3,0.6) {$\mathcal{N}$};
\node at (0.6,-0.3) {$W_1$};
\end{tikzpicture}
\end{gathered}\quad \ = \quad -\quad \begin{gathered}
\begin{tikzpicture}[scale=1]
\node at (0, 1.2) {};
\draw[very thick] (0,0) rectangle (1.2,1.2);
\draw [line, thick] (0,.60)--(1.20,0.60);
\end{tikzpicture}
\end{gathered}
\end{equation}
for $W_1$. Therefore $W_0$ and $W_1$ give the untwist sector with $S=(\eta,\mathcal{N})$ acting as $(1,\pm 1)$. When $W=W_2$ and $W_3$, we have
\begin{equation}
\begin{gathered}
\begin{tikzpicture}[scale=1]
\node at (0, 1.8) {};
\draw[very thick] (0,0) rectangle (1.2,1.2);
\draw [line, thick,red] (.60,0) -- (.60,1.20); 
\draw [line, thick,densely dotted] (0,.60)--(1.20,0.60);
\node at (-0.3,0.6) {$\eta$};
\node at (0.6,-0.3) {$W_2$};
\end{tikzpicture}
\end{gathered}\quad \ = \quad - \quad \begin{gathered}
\begin{tikzpicture}[scale=1]
\node at (0, 1.2) {};
\draw[very thick] (0,0) rectangle (1.2,1.2);
\draw[densely dotted, thick]  (.6,0) arc (0:90:.6); 
\draw[densely dotted, thick] (.6,1.2) arc (-180:-90:.6);
\end{tikzpicture}
\end{gathered} \quad , \qquad \begin{gathered}
\begin{tikzpicture}[scale=1]
\node at (0, 1.8) {};
\draw[very thick] (0,0) rectangle (1.2,1.2);
\draw [line, thick,red] (.60,0) -- (.60,1.20); 
\draw [line, thick] (0,.60)--(1.20,0.60);
\node at (-0.3,0.6) {$\mathcal{N}$};
\node at (0.6,-0.3) {$W_2$};
\end{tikzpicture}
\end{gathered}\quad \ = \quad i \quad \begin{gathered}
\begin{tikzpicture}[scale=1]
\node at (0, 1.2) {};
\draw[very thick] (0,0) rectangle (1.2,1.2);
\draw [line, thick] (0,.60) -- (1.2,.60); 
\draw [line, thick,densely dotted] (0.6,1.20) -- (0.90,.60); 
\draw [line, thick,densely dotted] (0.3,.60) -- (0.60,0); 
\end{tikzpicture}
\end{gathered}
\end{equation}
for $W_2$ and
\begin{equation}
\begin{gathered}
\begin{tikzpicture}[scale=1]
\node at (0, 1.8) {};
\draw[very thick] (0,0) rectangle (1.2,1.2);
\draw [line, thick,red] (.60,0) -- (.60,1.20); 
\draw [line, thick,densely dotted] (0,.60)--(1.20,0.60);
\node at (-0.3,0.6) {$\eta$};
\node at (0.6,-0.3) {$W_3$};
\end{tikzpicture}
\end{gathered}\quad \ = \quad - \quad \begin{gathered}
\begin{tikzpicture}[scale=1]
\node at (0, 1.2) {};
\draw[very thick] (0,0) rectangle (1.2,1.2);
\draw[densely dotted, thick]  (.6,0) arc (0:90:.6); 
\draw[densely dotted, thick] (.6,1.2) arc (-180:-90:.6);
\end{tikzpicture}
\end{gathered} \quad , \qquad \begin{gathered}
\begin{tikzpicture}[scale=1]
\node at (0, 1.8) {};
\draw[very thick] (0,0) rectangle (1.2,1.2);
\draw [line, thick,red] (.60,0) -- (.60,1.20); 
\draw [line, thick] (0,.60)--(1.20,0.60);
\node at (-0.3,0.6) {$\mathcal{N}$};
\node at (0.6,-0.3) {$W_3$};
\end{tikzpicture}
\end{gathered}\quad \ = \quad -i \quad \begin{gathered}
\begin{tikzpicture}[scale=1]
\node at (0, 1.2) {};
\draw[very thick] (0,0) rectangle (1.2,1.2);
\draw [line, thick] (0,.60) -- (1.2,.60); 
\draw [line, thick,densely dotted] (0.6,1.20) -- (0.90,.60); 
\draw [line, thick,densely dotted] (0.3,.60) -- (0.60,0); 
\end{tikzpicture}
\end{gathered}
\end{equation}
for $W_3$. Here we can see $W_2$ and $W_3$ give the $\eta$-twist sector with $S=(\eta,\mathcal{N})$ acting as $(-1,\pm i)$. For $W=W_4,W_5,W_6,W_7$ we have
\begin{align}\label{W4-W5-half-braiding}
\begin{gathered}
    \begin{tikzpicture}[scale=1]
\node at (0, 1.8) {};
\draw[very thick] (0,0) rectangle (1.2,1.2);
\draw [line, thick,red] (.60,0) -- (.60,1.20); 
\draw [line, thick,densely dotted] (0,.60)--(1.20,0.60);
\node at (-0.3,0.6) {$\eta$};
\node at (0.6,-0.3) {$W_4$};
\end{tikzpicture}
\end{gathered}&\quad = \quad -i\quad    \begin{gathered}
\begin{tikzpicture}[scale=1]
\node at (0, 1.2) {};
\draw[very thick] (0,0) rectangle (1.2,1.2);
\draw[line,thick,densely dotted] (0,.60) -- (.60,.30);
\draw[line,thick,densely dotted] (.60,.90) -- (1.20,.60);
\draw [line, thick] (.60,0) -- (.60,1.20); 
\end{tikzpicture}
\end{gathered}\quad , \qquad \begin{gathered}
\begin{tikzpicture}[scale=1]
\node at (0, 1.8) {};
\draw[very thick] (0,0) rectangle (1.2,1.2);
\draw [line, thick,red] (.60,0) -- (.60,1.20); 
\draw [line, thick] (0,.60)--(1.20,0.60);
\node at (-0.3,0.6) {$\mathcal{N}$};
\node at (0.6,-0.3) {$W_4$};
\end{tikzpicture}
\end{gathered}\quad = \quad \frac{\omega^{-1}}{\sqrt{2}}\quad \begin{gathered}
\begin{tikzpicture}[scale=1]
\node at (0, 1.2) {};
\draw[very thick] (0,0) rectangle (1.2,1.2);
\draw[thick]  (.6,0) arc (0:90:.6); 
\draw[thick] (.6,1.2) arc (-180:-90:.6);
\end{tikzpicture}
\end{gathered} \quad + \quad \frac{i \omega^{-1}}{\sqrt{2}}\quad\begin{gathered}
\begin{tikzpicture}[scale=1]
\node at (0, 1.2) {};
\draw[very thick] (0,0) rectangle (1.2,1.2);
\draw[thick]  (.6,0) arc (0:90:.6); 
\draw[thick] (.6,1.2) arc (-180:-90:.6);
\draw[line, thick,densely dotted] (.424,.424)--(.776,.776);
\end{tikzpicture}
\end{gathered}\notag\\
\begin{gathered}
    \begin{tikzpicture}[scale=1]
\node at (0, 1.8) {};
\draw[very thick] (0,0) rectangle (1.2,1.2);
\draw [line, thick,red] (.60,0) -- (.60,1.20); 
\draw [line, thick,densely dotted] (0,.60)--(1.20,0.60);
\node at (-0.3,0.6) {$\eta$};
\node at (0.6,-0.3) {$W_5$};
\end{tikzpicture}
\end{gathered}&\quad = \quad -i\quad    \begin{gathered}
\begin{tikzpicture}[scale=1]
\node at (0, 1.2) {};
\draw[very thick] (0,0) rectangle (1.2,1.2);
\draw[line,thick,densely dotted] (0,.60) -- (.60,.30);
\draw[line,thick,densely dotted] (.60,.90) -- (1.20,.60);
\draw [line, thick] (.60,0) -- (.60,1.20); 
\end{tikzpicture}
\end{gathered}\quad , \qquad \begin{gathered}
\begin{tikzpicture}[scale=1]
\node at (0, 1.8) {};
\draw[very thick] (0,0) rectangle (1.2,1.2);
\draw [line, thick,red] (.60,0) -- (.60,1.20); 
\draw [line, thick] (0,.60)--(1.20,0.60);
\node at (-0.3,0.6) {$\mathcal{N}$};
\node at (0.6,-0.3) {$W_5$};
\end{tikzpicture}
\end{gathered}\quad = \quad \frac{\omega^{7}}{\sqrt{2}}\quad \begin{gathered}
\begin{tikzpicture}[scale=1]
\node at (0, 1.2) {};
\draw[very thick] (0,0) rectangle (1.2,1.2);
\draw[thick]  (.6,0) arc (0:90:.6); 
\draw[thick] (.6,1.2) arc (-180:-90:.6);
\end{tikzpicture}
\end{gathered} \quad + \quad \frac{i \omega^{7}}{\sqrt{2}}\quad\begin{gathered}
\begin{tikzpicture}[scale=1]
\node at (0, 1.2) {};
\draw[very thick] (0,0) rectangle (1.2,1.2);
\draw[thick]  (.6,0) arc (0:90:.6); 
\draw[thick] (.6,1.2) arc (-180:-90:.6);
\draw[line, thick,densely dotted] (.424,.424)--(.776,.776);
\end{tikzpicture}
\end{gathered}
\end{align}
for $W_4,W_5$ and
\begin{align}
\begin{gathered}
    \begin{tikzpicture}[scale=1]
\node at (0, 1.8) {};
\draw[very thick] (0,0) rectangle (1.2,1.2);
\draw [line, thick,red] (.60,0) -- (.60,1.20); 
\draw [line, thick,densely dotted] (0,.60)--(1.20,0.60);
\node at (-0.3,0.6) {$\eta$};
\node at (0.6,-0.3) {$W_6$};
\end{tikzpicture}
\end{gathered}&\quad = \quad i\quad    \begin{gathered}
\begin{tikzpicture}[scale=1]
\node at (0, 1.2) {};
\draw[very thick] (0,0) rectangle (1.2,1.2);
\draw[line,thick,densely dotted] (0,.60) -- (.60,.30);
\draw[line,thick,densely dotted] (.60,.90) -- (1.20,.60);
\draw [line, thick] (.60,0) -- (.60,1.20); 
\end{tikzpicture}
\end{gathered}\quad , \qquad \begin{gathered}
\begin{tikzpicture}[scale=1]
\node at (0, 1.8) {};
\draw[very thick] (0,0) rectangle (1.2,1.2);
\draw [line, thick,red] (.60,0) -- (.60,1.20); 
\draw [line, thick] (0,.60)--(1.20,0.60);
\node at (-0.3,0.6) {$\mathcal{N}$};
\node at (0.6,-0.3) {$W_6$};
\end{tikzpicture}
\end{gathered}\quad = \quad \frac{\omega}{\sqrt{2}}\quad \begin{gathered}
\begin{tikzpicture}[scale=1]
\node at (0, 1.2) {};
\draw[very thick] (0,0) rectangle (1.2,1.2);
\draw[thick]  (.6,0) arc (0:90:.6); 
\draw[thick] (.6,1.2) arc (-180:-90:.6);
\end{tikzpicture}
\end{gathered} \quad + \quad \frac{-i \omega}{\sqrt{2}}\quad\begin{gathered}
\begin{tikzpicture}[scale=1]
\node at (0, 1.2) {};
\draw[very thick] (0,0) rectangle (1.2,1.2);
\draw[thick]  (.6,0) arc (0:90:.6); 
\draw[thick] (.6,1.2) arc (-180:-90:.6);
\draw[line, thick,densely dotted] (.424,.424)--(.776,.776);
\end{tikzpicture}
\end{gathered}\notag\\
\begin{gathered}
    \begin{tikzpicture}[scale=1]
\node at (0, 1.8) {};
\draw[very thick] (0,0) rectangle (1.2,1.2);
\draw [line, thick,red] (.60,0) -- (.60,1.20); 
\draw [line, thick,densely dotted] (0,.60)--(1.20,0.60);
\node at (-0.3,0.6) {$\eta$};
\node at (0.6,-0.3) {$W_7$};
\end{tikzpicture}
\end{gathered}&\quad = \quad i\quad    \begin{gathered}
\begin{tikzpicture}[scale=1]
\node at (0, 1.2) {};
\draw[very thick] (0,0) rectangle (1.2,1.2);
\draw[line,thick,densely dotted] (0,.60) -- (.60,.30);
\draw[line,thick,densely dotted] (.60,.90) -- (1.20,.60);
\draw [line, thick] (.60,0) -- (.60,1.20); 
\end{tikzpicture}
\end{gathered}\quad , \qquad \begin{gathered}
\begin{tikzpicture}[scale=1]
\node at (0, 1.8) {};
\draw[very thick] (0,0) rectangle (1.2,1.2);
\draw [line, thick,red] (.60,0) -- (.60,1.20); 
\draw [line, thick] (0,.60)--(1.20,0.60);
\node at (-0.3,0.6) {$\mathcal{N}$};
\node at (0.6,-0.3) {$W_7$};
\end{tikzpicture}
\end{gathered}\quad = \quad \frac{\omega^{-7}}{\sqrt{2}}\quad \begin{gathered}
\begin{tikzpicture}[scale=1]
\node at (0, 1.2) {};
\draw[very thick] (0,0) rectangle (1.2,1.2);
\draw[thick]  (.6,0) arc (0:90:.6); 
\draw[thick] (.6,1.2) arc (-180:-90:.6);
\end{tikzpicture}
\end{gathered} \quad + \quad \frac{-i \omega^{-7}}{\sqrt{2}}\quad\begin{gathered}
\begin{tikzpicture}[scale=1]
\node at (0, 1.2) {};
\draw[very thick] (0,0) rectangle (1.2,1.2);
\draw[thick]  (.6,0) arc (0:90:.6); 
\draw[thick] (.6,1.2) arc (-180:-90:.6);
\draw[line, thick,densely dotted] (.424,.424)--(.776,.776);
\end{tikzpicture}
\end{gathered}
\end{align}
for $W_6,W_7$, where $\omega=e^{\frac{2\pi i}{16}}$. We can read those four line operators give the $\mathcal{N}$-twist sector with $\eta$ acting as $\mp i$ and $\mathcal{N}$ acting on a 2-dim vector space with diagonal matrix elements $\pm(\omega^{-1},i\omega^{-1})$ and $\pm(\omega,-i\omega)$. Here the missing factor $\frac{1}{\sqrt{2}}$ is recognized as the factor $\sqrt{\frac{d_{I}}{d_{\mathcal{N}}d_{\mathcal{N}}}}$ or $\sqrt{\frac{d_{\eta}}{d_{\mathcal{N}}d_{\mathcal{N}}}}$ in \eqref{R-coefficient}. Finally, for $W_8$ one can examine
    \begin{equation}
        \begin{gathered}
\begin{tikzpicture}[scale=1]
\node at (0, 1.8) {};
\draw[very thick] (0,0) rectangle (1.2,1.2);
\draw [line, thick,red] (.60,0) -- (.60,1.20); 
\draw [line, thick,densely dotted] (0,.60)--(1.20,0.60);
\node at (-0.3,0.6) {$\eta$};
\node at (0.6,-0.3) {$W_8$};
\end{tikzpicture}
\end{gathered}\quad = \quad -\quad \begin{gathered}
\begin{tikzpicture}[scale=1]
\node at (0, 1.2) {};
\draw[very thick] (0,0) rectangle (1.2,1.2);
\draw [line,densely dotted,thick] (0,.60) -- (1.2,.60); 
\end{tikzpicture}
\end{gathered}\quad + \quad \begin{gathered}
\begin{tikzpicture}[scale=1]
\node at (0, 1.2) {};
\draw[very thick] (0,0) rectangle (1.2,1.2);
\draw[densely dotted, thick]  (.6,0) arc (0:90:.6); 
\draw[densely dotted, thick] (.6,1.2) arc (-180:-90:.6);
\end{tikzpicture}
\end{gathered}\quad ,
    \end{equation}
such that $W_8$ compose a multiplet of both untwist sector and $\mathbb{Z}_2$-twist sector, and $\eta$ act on them as $(-1,1)$. Moreover, one has
    \begin{equation}
        \begin{gathered}
\begin{tikzpicture}[scale=1]
\node at (0, 1.8) {};
\draw[very thick] (0,0) rectangle (1.2,1.2);
\draw [line, thick,red] (.60,0) -- (.60,1.20); 
\draw [line, thick] (0,.60)--(1.20,0.60);
\node at (-0.3,0.6) {$\mathcal{N}$};
\node at (0.6,-0.3) {$W_8$};
\end{tikzpicture}
\end{gathered}\quad = \quad 0,
    \end{equation}
and the reason is that $\mathcal{N}$ will exchange the twist/untwist sector in $W_8$ so that the partition function is zero since the sector is not allowed on the torus. The results we obtained here agree with those in \cite{Bhardwaj:2023ayw}. 

We can obtain the $R$-coefficients of the Ising category by focusing on the cases where $W$ equals to $W_0,W_3$ and $W_4$, or namely $I,\eta$ and $\mathcal{N}$. For example, choose $L_1 = \mathcal{N}$ and $L_2 = \mathcal{N}\equiv W_4$ in \eqref{R-coefficient} we have
\begin{equation}
    \begin{gathered}
        \begin{tikzpicture}
            \draw[line,thick,red] (0,0)--(1.2,1.2);
            \draw[line,thick] (1.2,0)--(0.7,0.5);
            \draw[line,thick] (0.5,0.7)--(0,1.2);
            \node at (0,-0.3) {$W_4$};
            \node at (1.3,-0.3) {$\mathcal{N}$};
        \end{tikzpicture}
    \end{gathered}\quad = \quad \frac{1}{\sqrt{2}}\, R_{I}^{\mathcal{N} \mathcal{N}} \quad 
    \begin{gathered}
        \begin{tikzpicture}
            \draw[line,thick] (0,0)--(0.6,0.6);
            \draw[line,thick] (1.2,0)--(0.6,0.6);
            \draw[line,thick] (0.6,1.2) -- (0,1.8);
            \draw[line,thick] (0.6,1.2) -- (1.2,1.8);
            \node at (0.9,0.9) {$I$};
            \node at (0,-0.3) {$\mathcal{N}$};
            \node at (1.3,-0.3) {$\mathcal{N}$};
            \node at (0.0,2.1) {$\mathcal{N}$};
            \node at (1.3,2.1) {$\mathcal{N}$};
        \end{tikzpicture}     
    \end{gathered}\quad + \quad \frac{1}{\sqrt{2}}\, R_{\eta}^{\mathcal{N}\mathcal{N}} \quad 
    \begin{gathered}
        \begin{tikzpicture}
            \draw[line,thick] (0,0)--(0.6,0.6);
            \draw[line,thick] (1.2,0)--(0.6,0.6);
            \draw[line, densely dotted, thick] (0.6,0.6) -- (0.6,1.2);
            \draw[line,thick] (0.6,1.2) -- (0,1.8);
            \draw[line,thick] (0.6,1.2) -- (1.2,1.8);
            \node at (0.9,0.9) {$\eta$};
            \node at (0,-0.3) {$\mathcal{N}$};
            \node at (1.3,-0.3) {$\mathcal{N}$};
            \node at (0.0,2.1) {$\mathcal{N}$};
            \node at (1.3,2.1) {$\mathcal{N}$};
        \end{tikzpicture}     
    \end{gathered}
\end{equation}
and we can read
    \begin{equation}
        R^{\mathcal{N} \mathcal{N}}_{I} = \omega^{-1},\quad R^{\mathcal{N} \mathcal{N}}_{\eta} = i \omega^{-1},
    \end{equation}
from \eqref{W4-W5-half-braiding}. Similarly, we can obtain the other two non-trivial half-braiding coefficients
    \begin{equation}
        R^{\eta \eta}_I = -1,\quad R^{\eta \mathcal{N}}_{\mathcal{N}} = R^{ \mathcal{N}\eta}_{\mathcal{N}} = -i.
    \end{equation}
One can check they are indeed the solution of hexagon equations providing the $F$-move coefficients discussed above.

\noindent
\textbf{Dualities}\\
For Ising CFT, the dynamical boundary can be constructed as
    \begin{align}
        |\chi\rangle_{\textrm{Ising}} =& |\chi_0|^2 |0\rangle + |\chi_{\frac{1}{2}}|^2 |1\rangle +  \chi_0 \bar{\chi}_{\frac{1}{2}}|2\rangle + \chi_{\frac{1}{2}}\bar{\chi}_0|3\rangle + \chi_{\frac{1}{16}} \bar{\chi}_0 |4\rangle + \chi_{\frac{1}{16}} \bar{\chi}_{\frac{1}{2}} |5\rangle \nonumber\\
        &+ \chi_0 \bar{\chi}_{\frac{1}{16}} |6\rangle + \chi_{\frac{1}{2}}\bar{\chi}_{\frac{1}{16}}|7\rangle + |\chi_{\frac{1}{16}}|^2 |8\rangle,
    \end{align}
where the coefficients are the characters of the corresponding line operators. The partition function of Ising CFT can be easily obtained by projecting $|\chi\rangle_{\textrm{Ising}}$ onto different sectors $|i\rangle_{\mathcal{L}_1}$. For other theory $\mathfrak{T}$ whose symmetry is also characterized by the Ising category, one can construct the dynamical boundary state using the partition functions as coefficients
    \begin{equation}
        |\chi\rangle_{\mathfrak{T}} = \sum_{i,j=0,\cdots,8} Z_{\mathfrak{T}}[i] g^{ij} 
  |j\rangle_{\mathcal{L}_1}
    \end{equation}
where $g^{ij}$ is the inverse of the metric $g_{ij} \equiv \,_{\mathcal{L}_1}\langle i | j\rangle_{\mathcal{L}_1}$.

The identity\eqref{Ising-Z2-relation} translates to the relation of the partition function
\begin{equation}
    Z_{\mathfrak{T}}
\,\begin{gathered}
\begin{tikzpicture}[scale=.5]
\node at (0, 1.45) {};
\draw[line] (-.15,-.2)--(-.25, -.2) -- (-.25, 1.4)-- (-.15,1.4);
\draw[very thick] (0,0) rectangle (1.2,1.2);
\draw[densely dotted,thick]  (.6,0) arc (0:90:.6); 
\draw[densely dotted,thick] (.6,1.2) arc (-180:-90:.6);
\draw[line] (1.35, -.2)--(1.45, -.2) -- (1.45, 1.4)--(1.35, 1.4);
\end{tikzpicture}
\end{gathered}\,=\,Z_{\mathfrak{T}} 
        \,\begin{gathered}
\begin{tikzpicture}[scale=.5]
\node at (0, 1.45) {};
\draw[line] (-.15,-.2)--(-.25, -.2) -- (-.25, 1.4)-- (-.15,1.4);
\draw[very thick] (0,0) rectangle (1.2,1.2);
\draw[line] (1.35, -.2)--(1.45, -.2) -- (1.45, 1.4)--(1.35, 1.4);
\end{tikzpicture}
\end{gathered}\,-Z_{\mathfrak{T}}
\,\begin{gathered}
\begin{tikzpicture}[scale=.5]
\node at (0, 1.45) {};
\draw[line] (-.15,-.2)--(-.25, -.2) -- (-.25, 1.4)-- (-.15,1.4);
\draw[very thick] (0,0) rectangle (1.2,1.2);
\draw[line, densely dotted, thick] (0,.6)--(1.2,.6);
\draw[line] (1.35, -.2)--(1.45, -.2) -- (1.45, 1.4)--(1.35, 1.4);
\end{tikzpicture}
\end{gathered}\,
- Z_{\mathfrak{T}}
\,\begin{gathered}
\begin{tikzpicture}[scale=.5]
\node at (0, 1.45) {};
\draw[line] (-.15,-.2)--(-.25, -.2) -- (-.25, 1.4)-- (-.15,1.4);
\draw[very thick] (0,0) rectangle (1.2,1.2);
\draw[line, densely dotted, thick] (.6,0)--(.6,1.2);
\draw[line] (1.35, -.2)--(1.45, -.2) -- (1.45, 1.4)--(1.35, 1.4);
\end{tikzpicture}
\end{gathered}
\end{equation}
which can be rearranged into
\begin{equation}
    Z_{\mathfrak{T}} 
        \,\begin{gathered}
\begin{tikzpicture}[scale=.5]
\node at (0, 1.45) {};
\draw[line] (-.15,-.2)--(-.25, -.2) -- (-.25, 1.4)-- (-.15,1.4);
\draw[very thick] (0,0) rectangle (1.2,1.2);
\draw[line] (1.35, -.2)--(1.45, -.2) -- (1.45, 1.4)--(1.35, 1.4);
\end{tikzpicture}
\end{gathered}\,=\,\frac{1}{2} \left(Z_{\mathfrak{T}} 
        \,\begin{gathered}
\begin{tikzpicture}[scale=.5]
\node at (0, 1.45) {};
\draw[line] (-.15,-.2)--(-.25, -.2) -- (-.25, 1.4)-- (-.15,1.4);
\draw[very thick] (0,0) rectangle (1.2,1.2);
\draw[line] (1.35, -.2)--(1.45, -.2) -- (1.45, 1.4)--(1.35, 1.4);
\end{tikzpicture}
\end{gathered}\, + Z_{\mathfrak{T}}
\,\begin{gathered}
\begin{tikzpicture}[scale=.5]
\node at (0, 1.45) {};
\draw[line] (-.15,-.2)--(-.25, -.2) -- (-.25, 1.4)-- (-.15,1.4);
\draw[very thick] (0,0) rectangle (1.2,1.2);
\draw[densely dotted,thick]  (.6,0) arc (0:90:.6); 
\draw[densely dotted,thick] (.6,1.2) arc (-180:-90:.6);
\draw[line] (1.35, -.2)--(1.45, -.2) -- (1.45, 1.4)--(1.35, 1.4);
\end{tikzpicture}
\end{gathered}\,+Z_{\mathfrak{T}}
\,\begin{gathered}
\begin{tikzpicture}[scale=.5]
\node at (0, 1.45) {};
\draw[line] (-.15,-.2)--(-.25, -.2) -- (-.25, 1.4)-- (-.15,1.4);
\draw[very thick] (0,0) rectangle (1.2,1.2);
\draw[line, densely dotted, thick] (0,.6)--(1.2,.6);
\draw[line] (1.35, -.2)--(1.45, -.2) -- (1.45, 1.4)--(1.35, 1.4);
\end{tikzpicture}
\end{gathered}\,+Z_{\mathfrak{T}}
\,\begin{gathered}
\begin{tikzpicture}[scale=.5]
\node at (0, 1.45) {};
\draw[line] (-.15,-.2)--(-.25, -.2) -- (-.25, 1.4)-- (-.15,1.4);
\draw[very thick] (0,0) rectangle (1.2,1.2);
\draw[line, densely dotted, thick] (.6,0)--(.6,1.2);
\draw[line] (1.35, -.2)--(1.45, -.2) -- (1.45, 1.4)--(1.35, 1.4);
\end{tikzpicture}
\end{gathered}\right)\, \equiv \,Z_{\mathfrak{T}/\mathbb{Z}_2} 
        \,\begin{gathered}
\begin{tikzpicture}[scale=.5]
\node at (0, 1.45) {};
\draw[line] (-.15,-.2)--(-.25, -.2) -- (-.25, 1.4)-- (-.15,1.4);
\draw[very thick] (0,0) rectangle (1.2,1.2);
\draw[line] (1.35, -.2)--(1.45, -.2) -- (1.45, 1.4)--(1.35, 1.4);
\end{tikzpicture}
\end{gathered}
\end{equation}
which indicates the Ising category is self-dual under the $\mathbb{Z}_2$-gauging. Projecting the dynamical boundary state onto the fermionic topological boundary states $|\Omega\rangle_{\mathcal{L}_2}$ we obtain
    \begin{equation}
        \langle\Omega|_{\mathcal{L}_2}|\chi\rangle_{\mathfrak{T}}\,=\,Z_{\mathfrak{T}}\,\begin{gathered}
\begin{tikzpicture}[scale=.5]
\node at (0, 1.45) {};
\draw[line] (-.15,-.2)--(-.25, -.2) -- (-.25, 1.4)-- (-.15,1.4);
\draw[very thick] (0,0) rectangle (1.2,1.2);
\draw[line, densely dotted, thick] (0,.6)--(1.2,.6);
\draw[line] (1.35, -.2)--(1.45, -.2) -- (1.45, 1.4)--(1.35, 1.4);
\end{tikzpicture}
\end{gathered}\,+Z_{\mathfrak{T}}
\,\begin{gathered}
\begin{tikzpicture}[scale=.5]
\node at (0, 1.45) {};
\draw[line] (-.15,-.2)--(-.25, -.2) -- (-.25, 1.4)-- (-.15,1.4);
\draw[very thick] (0,0) rectangle (1.2,1.2);
\draw[line, densely dotted, thick] (.6,0)--(.6,1.2);
\draw[line] (1.35, -.2)--(1.45, -.2) -- (1.45, 1.4)--(1.35, 1.4);
\end{tikzpicture}
\end{gathered}\,=\,\frac{1}{2} \left(Z_{\mathfrak{T}} 
        \,\begin{gathered}
\begin{tikzpicture}[scale=.5]
\node at (0, 1.45) {};
\draw[line] (-.15,-.2)--(-.25, -.2) -- (-.25, 1.4)-- (-.15,1.4);
\draw[very thick] (0,0) rectangle (1.2,1.2);
\draw[line] (1.35, -.2)--(1.45, -.2) -- (1.45, 1.4)--(1.35, 1.4);
\end{tikzpicture}
\end{gathered}\, + Z_{\mathfrak{T}}
\,\begin{gathered}
\begin{tikzpicture}[scale=.5]
\node at (0, 1.45) {};
\draw[line] (-.15,-.2)--(-.25, -.2) -- (-.25, 1.4)-- (-.15,1.4);
\draw[very thick] (0,0) rectangle (1.2,1.2);
\draw[densely dotted,thick]  (.6,0) arc (0:90:.6); 
\draw[densely dotted,thick] (.6,1.2) arc (-180:-90:.6);
\draw[line] (1.35, -.2)--(1.45, -.2) -- (1.45, 1.4)--(1.35, 1.4);
\end{tikzpicture}
\end{gathered}\,+Z_{\mathfrak{T}}
\,\begin{gathered}
\begin{tikzpicture}[scale=.5]
\node at (0, 1.45) {};
\draw[line] (-.15,-.2)--(-.25, -.2) -- (-.25, 1.4)-- (-.15,1.4);
\draw[very thick] (0,0) rectangle (1.2,1.2);
\draw[line, densely dotted, thick] (0,.6)--(1.2,.6);
\draw[line] (1.35, -.2)--(1.45, -.2) -- (1.45, 1.4)--(1.35, 1.4);
\end{tikzpicture}
\end{gathered}\,-Z_{\mathfrak{T}}
\,\begin{gathered}
\begin{tikzpicture}[scale=.5]
\node at (0, 1.45) {};
\draw[line] (-.15,-.2)--(-.25, -.2) -- (-.25, 1.4)-- (-.15,1.4);
\draw[very thick] (0,0) rectangle (1.2,1.2);
\draw[line, densely dotted, thick] (.6,0)--(.6,1.2);
\draw[line] (1.35, -.2)--(1.45, -.2) -- (1.45, 1.4)--(1.35, 1.4);
\end{tikzpicture}
\end{gathered}\right).
    \end{equation}

\subsection{Lee-Yang Category}
Finally, we turn to the Lee-Yang model where the $S$ and $T$ matrices for Lee-Yang SymTFT are given by
    \begin{equation}
        S = \left(
\begin{array}{cccc}
 \frac{1}{10} \left(\sqrt{5}+5\right) & \frac{1}{10} \left(5-\sqrt{5}\right) & -\frac{1}{\sqrt{5}} & -\frac{1}{\sqrt{5}} \\
 \frac{1}{10} \left(5-\sqrt{5}\right) & \frac{1}{10} \left(\sqrt{5}+5\right) & \frac{1}{\sqrt{5}} & \frac{1}{\sqrt{5}} \\
 -\frac{1}{\sqrt{5}} & \frac{1}{\sqrt{5}} & \frac{1}{10} \left(-\sqrt{5}-5\right) & \frac{1}{10} \left(5-\sqrt{5}\right) \\
 -\frac{1}{\sqrt{5}} & \frac{1}{\sqrt{5}} & \frac{1}{10} \left(5-\sqrt{5}\right) & \frac{1}{10} \left(-\sqrt{5}-5\right) \\
\end{array}
\right)
    \end{equation}
and
\begin{equation}
    T = \textrm{Diag}\left(1,1,e^{\frac{2\pi i}{5}},e^{\frac{8 \pi i}{5}} \right)
\end{equation}
up to relabeling of line operators. The fusion table is presented in table \ref{Lee-Yang-fusion}
\begin{table}[!h]
    \footnotesize
    \centering
    \begin{tabular}{|c|c|c|c|c|}
    \hline
    $\otimes$& $W_0$ & $W_1$ &$W_2$&$W_3$\\
    \hline
    $W_0$&$W_0$&$W_1$&$W_2$&$W_3$\\
    \hline
    $W_1$&$W_1$&$W_0\oplus W_1 \oplus W_2 \oplus W_3$&$W_1 \oplus W_3$& $W_1 \oplus W_2$\\
    \hline
    $W_2$ &$W_2$ &$W_1 \oplus W_3$&$W_0 \oplus W_2$&$W_1$\\
    \hline
    $W_3$&$W_3$&$W_1 \oplus W_2$&$W_1$&$W_0 \oplus W_3$\\
    \hline
    \end{tabular}
    \caption{The fusion rule of Lee-Yang SymTFT}
    \label{Lee-Yang-fusion}
\end{table}


There is only one Lagrangian algebra given by 
    \begin{equation}
        \mathcal{L} = W_0 + W_1
    \end{equation}
with total quantum dimension $1$, and the corresponding vacuum is constructed as
    \begin{equation}
        |\Omega\rangle_{\mathcal{L}} = |0\rangle + |1\rangle = |1,1,0,0\rangle.
    \end{equation}
The action of line operators upon the topological boundary state $|\Omega\rangle_{\mathcal{L}}$ is shown in table \ref{Lee-Yang-L}.
        \begin{table}[!h]
        \centering
        \begin{tabular}{|c|c|c|}
        \hline
            Lines/cycles & $\Gamma_1$ & $\Gamma_2$ \\
            \hline
            $W_0$&$|1,1,0,0\rangle$ & $|1,1,0,0\rangle$\\
            \hline
            $W_1$ & $|\frac{1}{2} \left(3-\sqrt{5}\right),\frac{1}{2} \left(3+\sqrt{5}\right),0,0\rangle$ & $|1, 2, 1, 1\rangle$\\
            \hline
            $W_2$ & $|\frac{1}{2} \left(1-\sqrt{5}\right),\frac{1}{2} \left(1+\sqrt{5}\right),0,0\rangle$&$|0,1,1,1\rangle$\\
            \hline
            $W_3$ & $|\frac{1}{2} \left(1-\sqrt{5}\right),\frac{1}{2} \left(1+\sqrt{5}\right),0,0\rangle$&$|0,1,1,1\rangle$\\
            \hline
        \end{tabular}
        \caption{The action of line operators on the topological boundary state $|\Omega\rangle_{\mathcal{L}}$}
        \label{Lee-Yang-L}
        \end{table}
Therefore we have the identification 
    \begin{equation}
        W_2\sim W_3, \quad W_1 \sim W_0 + W_2 \sim W_0 + W_3
        \label{eq:LY boundary}
    \end{equation}
along the topological boundary. One may choose $W_0,W_2$ as two generators and verify the fusion rule of the Lee-Yang category
    \begin{equation}
        W_2 \times W_2 = W_0 + W_2.
    \end{equation}
We can identify
    \begin{equation}
        W_0 = I,\quad W_2 = \tau,
    \end{equation}
where $(I,\tau)$ are the identity and non-invertible TDLs in the Lee-Yang category.

\subsubsection{$F$-moves and self-duality in Lee-Yang Model}
\noindent
\textbf{$F$-moves}\\
We can obtain the $F$-coefficients and $R$-coefficients similar to the Ising case and here we will only illustrate the $F$-move. First, let's prepare a complete set of basis 
which consists of 4 linearly independent vectors,
\begin{align}
\begin{gathered}
\begin{tikzpicture}[scale=.5]
\node at (0, 1.2) {};
\draw[very thick] (0,0) rectangle (1.2,1.2);
\end{tikzpicture}
\end{gathered}
&:\, |0\rangle_{\mathcal{L}}\equiv|\Omega\rangle_{\mathcal L}=|1,1,0,0\rangle
\qquad\qquad\qquad\quad\quad\!\!\!
\begin{gathered}
\begin{tikzpicture}[scale=.5]
\node at (0, 1.2) {};
\draw[very thick] (0,0) rectangle (1.2,1.2);
\draw [line, thick] (0,.60) -- (1.2,.60); 
\end{tikzpicture}
\end{gathered}
:\, |1\rangle_{\mathcal{L}}\equiv W_{2}[\Gamma_1]|\Omega\rangle_{\mathcal L}=|-\xi^{-1},\xi,0,0\rangle\notag\\
\begin{gathered}
\begin{tikzpicture}[scale=.5]
\node at (0, 1.2) {};
\draw[very thick] (0,0) rectangle (1.2,1.2);
\draw [line, thick] (.60,0) -- (.60,1.20); 
\end{tikzpicture}
\end{gathered}
&:\, |2\rangle_{\mathcal{L}}\equiv W_{2}[\Gamma_2]|\Omega\rangle_{\mathcal L}=|0,1,1,1\rangle
\qquad\qquad
\begin{gathered}
\begin{tikzpicture}[scale=.5]
\node at (0, 1.2) {};
\draw[very thick] (0,0) rectangle (1.2,1.2);
\draw[thick]  (.6,0) arc (0:90:.6); 
\draw[thick] (.6,1.2) arc (-180:-90:.6);
\end{tikzpicture}
\end{gathered}
:\, |3\rangle_{\mathcal{L}}\equiv\hat{\mathbb T}\cdot W_{2}[\Gamma_2]|\Omega\rangle_{\mathcal L}=|0,1,\omega^2,\omega^8\rangle
\end{align}
where $\xi=\frac{\sqrt{5}+1}{2}$ and $\omega=e^{\frac{2\pi i}{10}}$. The solid line represents the non-invertible TDL $\tau$. Besides, we also need the state
\begin{align}
\begin{gathered}
\begin{tikzpicture}[scale=.5]
\node at (0, 1.2) {};
\draw[very thick] (0,0) rectangle (1.2,1.2);
\draw[thick]  (.6,0) arc (-180:-270:.6); 
\draw[thick] (.6,1.2) arc (0:-90:.6);
\end{tikzpicture}
\end{gathered}
:\, |4\rangle_{\mathcal{L}}\equiv\hat{\mathbb T}^{-1}\cdot W_{2}[\Gamma_2]|\Omega\rangle_{\mathcal L}=|0,1,\omega^8,\omega^2\rangle
\end{align}
Similar to the Ising case, we work out the $F$-move of the topological defect lines on the boundary. Notice that, on the boundary, we have
\begin{align}
\begin{gathered}
\begin{tikzpicture}[scale=1]
\node at (0, 1.2) {};
\draw[very thick] (0,0) rectangle (1.2,1.2);
\draw [line, thick] (.4,0) -- (.40,1.20); 
\draw [line, thick] (.8,0) -- (.80,1.20); 
\end{tikzpicture}
\end{gathered}
\quad=\quad
\begin{gathered}
\begin{tikzpicture}[scale=1]
\node at (0, 1.2) {};
\draw[very thick] (0,0) rectangle (1.2,1.2);
\end{tikzpicture}
\end{gathered}
\quad+\quad
\begin{gathered}
\begin{tikzpicture}[scale=1]
\node at (0, 1.2) {};
\draw[very thick] (0,0) rectangle (1.2,1.2);
\draw [line, thick] (.6,0) -- (.60,1.20); 
\end{tikzpicture}
\end{gathered}
\,,
\end{align}
by the fusion rule. On the other hand, F-move of the two $W_2$-lines implies
\begin{align}
\begin{gathered}
\begin{tikzpicture}[scale=1]
\node at (0, 1.2) {};
\draw[very thick] (0,0) rectangle (1.2,1.2);
\draw [line, thick] (.4,0) -- (.40,1.20); 
\draw [line, thick] (.8,0) -- (.80,1.20); 
\end{tikzpicture}
\end{gathered}
\quad=\quad \alpha\quad
\begin{gathered}
\begin{tikzpicture}[scale=1]
\node at (0, 1.2) {};
\draw[very thick] (0,0) rectangle (1.2,1.2);
\draw[thick] (.9,1.2) arc (0:-180:.3); 
\draw[thick] (.3,0) arc (180:0:.3); 
\end{tikzpicture}
\end{gathered}
\quad+\quad \beta\quad
\begin{gathered}
\begin{tikzpicture}[scale=1]
\node at (0, 1.2) {};
\draw[very thick] (0,0) rectangle (1.2,1.2);
\draw[thick] (.9,1.2) arc (0:-180:.3); 
\draw[thick] (.3,0) arc (180:0:.3); 
\draw[line, thick] (.6, .3) -- (.6, .9);
\end{tikzpicture}
\end{gathered}
\end{align}
We require the vevs of the $W_2$-bubbles to give the quantum dimension of the $W_2$-line on the boundary,
\begin{align}
\begin{gathered}
\begin{tikzpicture}[scale=1]
\node at (0, 1.2) {};
\draw[very thick] (0,0) rectangle (1.2,1.2);
\draw[thick] (.6,.6) circle (.25); 
\end{tikzpicture}
\end{gathered}&
\quad=\quad d_{\tau} \quad
\begin{gathered}
\begin{tikzpicture}[scale=1]
\node at (0, 1.2) {};
\draw[very thick] (0,0) rectangle (1.2,1.2);
\end{tikzpicture}
\end{gathered} \, .
\end{align}
Notice that for the junction vector space of three $W_2$-line, we can change the basis by multiplying any $\mathbb{C}$-number. We shall use that to fix the following diagram\footnote{In some literature it is convenient to choose a gauge such that
\begin{equation}
    \begin{gathered}
        \begin{tikzpicture}
    \draw[thick] (.6,.6) circle (.25); 
    \draw[line, thick] (.6,0) -- (.6, .35);
    \draw[line, thick] (.6,1.2) -- (.6, .85);
            \node at (0.6,-0.2) {$a$};
            \node at (0.6,1.4) {$b$};
            \node at (0.2,0.6) {$c$};
            \node at (1.0,0.6) {$d$};
            \node at (0.4,0.2) {$\mu$};
            \node at (0.4,0.9) {$\nu$};
        \end{tikzpicture}        
    \end{gathered}\quad = \quad \delta_{a,b} \delta_{\mu,\nu} \sqrt{\frac{d_c d_d}{d_a}} \quad \begin{gathered}
        \begin{tikzpicture}
                \draw[line, thick] (.6,0) -- (.6, 1.2);
                \node at (0.6,-0.2) {$a$};
        \end{tikzpicture}
    \end{gathered}
\end{equation}
where $\mu,\nu$ label the vector spaces of 3-junctions. Here we use a different gauge matching the conventions in \cite{Chang:2018iay}.}
\begin{align}
\begin{gathered}
\begin{tikzpicture}[scale=1]
\node at (0, 1.2) {};
\draw[very thick] (0,0) rectangle (1.2,1.2);
\draw[thick] (.6,.6) circle (.25); 
\draw[line, thick] (.6,0) -- (.6, .35);
\draw[line, thick] (.6,1.2) -- (.6, .85);
\end{tikzpicture}
\end{gathered}&
\quad=\quad d_{\tau} \quad
\begin{gathered}
\begin{tikzpicture}[scale=1]
\node at (0, 1.2) {};
\draw[very thick] (0,0) rectangle (1.2,1.2);
\draw [line,thick] (.60,0) -- (.60,1.20); 
\end{tikzpicture}
\end{gathered}\, .
\end{align}
It thus implies that
\begin{align}
\alpha=\beta=-\xi\,,
\end{align}
where we use the fact $d_{\tau} = - \xi^{-1}$. Using the $F$-move, we have
\begin{align}
\begin{gathered}
\begin{tikzpicture}[scale=1]
\node at (0, 1.2) {};
\draw[very thick] (0,0) rectangle (1.2,1.2);
\draw[thick]  (.6,0) arc (-180:-270:.6); 
\draw[thick] (.6,1.2) arc (0:-90:.6);
\end{tikzpicture}
\end{gathered}
\quad=\quad -\xi\quad
\begin{gathered}
\begin{tikzpicture}[scale=1]
\node at (0, 1.2) {};
\draw[very thick] (0,0) rectangle (1.2,1.2);
\draw[thick]  (.6,0) arc (0:90:.6); 
\draw[thick] (.6,1.2) arc (-180:-90:.6);
\end{tikzpicture}
\end{gathered}
\quad+\quad (-\xi)\quad
\begin{gathered}
\begin{tikzpicture}[scale=1]
\node at (0, 1.2) {};
\draw[very thick] (0,0) rectangle (1.2,1.2);
\draw[thick]  (.6,0) arc (0:90:.6); 
\draw[thick] (.6,1.2) arc (-180:-90:.6);
\draw[line, thick] (.424,.424)--(.776,.776);
\end{tikzpicture}
\end{gathered}
\,,
\end{align}
i.e.
\begin{align}
\label{eq:LY_fmove_1}
|4\rangle_{\mathcal{L}}=-\xi\,|3\rangle_{\mathcal{L}}-\xi\,|5\rangle_{\mathcal{L}}\,.
\end{align}
We thus solve
\begin{align}
\begin{gathered}
\begin{tikzpicture}[scale=.5]
\node at (0, 1.2) {};
\draw[very thick] (0,0) rectangle (1.2,1.2);
\draw[thick]  (.6,0) arc (0:90:.6); 
\draw[thick] (.6,1.2) arc (-180:-90:.6);
\draw[line, thick] (.424,.424)--(.776,.776);
\end{tikzpicture}
\end{gathered}
:\, |5\rangle_{\mathcal{L}}\equiv -|3\rangle_{\mathcal{L}}-\xi^{-1}|4\rangle_{\mathcal{L}}\,.
\end{align}
By an $S$-transformation, one further has
\begin{align}
\begin{gathered}
\begin{tikzpicture}[scale=.5]
\node at (0, 1.2) {};
\draw[very thick] (0,0) rectangle (1.2,1.2);
\draw[thick]  (.6,0) arc (180:90:.6); 
\draw[thick] (.6,1.2) arc (0:-90:.6);
\draw[line, thick] (.776,.424)--(.424,.776);
\end{tikzpicture}
\end{gathered}
:\, |6\rangle_{\mathcal{L}}\equiv \hat{\mathbb S}\cdot|5\rangle_{\mathcal{L}}=-|4\rangle_{\mathcal{L}}-\xi^{-1}|3\rangle_{\mathcal{L}}\,.
\label{eq:alpha_6}
\end{align}
Applying \eqref{eq:LY_fmove_1} to \eqref{eq:alpha_6}, we have
\begin{align}
|6\rangle_{\mathcal{L}}=-(-\xi\,|3\rangle_{\mathcal{L}}-\xi\,|5\rangle_{\mathcal{L}})-\xi^{-1}|3\rangle_{\mathcal{L}}=|3\rangle_{\mathcal{L}}+\xi\,|5\rangle_{\mathcal{L}}
\end{align}
Overall, we have computed the $F$-move
\begin{align}
\begin{pmatrix}
\begin{gathered}
\begin{tikzpicture}[scale=.5]
\node at (0, 1.2) {};
\draw[very thick] (0,0) rectangle (1.2,1.2);
\draw[thick]  (.6,0) arc (-180:-270:.6); 
\draw[thick] (.6,1.2) arc (0:-90:.6);
\end{tikzpicture}
\end{gathered}
\,\, {}\\
\begin{gathered}
\begin{tikzpicture}[scale=.5]
\node at (0, 1.2) {};
\draw[very thick] (0,0) rectangle (1.2,1.2);
\draw[thick]  (.6,0) arc (180:90:.6); 
\draw[thick] (.6,1.2) arc (0:-90:.6);
\draw[line, thick] (.776,.424)--(.424,.776);
\end{tikzpicture}
\end{gathered}
\,\,{}
\end{pmatrix}
\quad=\quad
\begin{pmatrix}
-\xi & -\xi
\,\,{}\\
1 & \xi\,\,{}
\end{pmatrix}
\cdot
\begin{pmatrix}
\begin{gathered}
\begin{tikzpicture}[scale=.5]
\node at (0, 1.2) {};
\draw[very thick] (0,0) rectangle (1.2,1.2);
\draw[thick]  (.6,0) arc (0:90:.6); 
\draw[thick] (.6,1.2) arc (-180:-90:.6);
\end{tikzpicture}
\end{gathered}
\,\,{}\\
\begin{gathered}
\begin{tikzpicture}[scale=.5]
\node at (0, 1.2) {};
\draw[very thick] (0,0) rectangle (1.2,1.2);
\draw[thick]  (.6,0) arc (0:90:.6); 
\draw[thick] (.6,1.2) arc (-180:-90:.6);
\draw[line, thick] (.424,.424)--(.776,.776);
\end{tikzpicture}
\end{gathered}
\,\,{}
\end{pmatrix}
\,,
\end{align}
or in terms of $F$-coefficients
\begin{equation}
    F^{WWW}_{W}(I,I) = F^{WWW}_W(I,W) = -\xi\, ,\quad F^{WWW}_W(W,I)=1\, ,\quad F^{WWW}_{W}(W,W) = \xi\, .
\end{equation}
\noindent
\textbf{Self-Duality}\\
The Lee-Yang model is actually self-dual with respect to gauging the whole fusion category $\mathcal C_{\rm LY}$. It can be observed by putting the whole Lagrangian algebra $\mathcal L=W_0+W_1$ on the boundary, which reduces to 
\begin{align}
\mathcal L\vert_{\rm bdy}=W_0+W_0+W_2\,,
\end{align}
from \eqref{eq:LY boundary}. According to \cite{Putrov:2024uor}, it implies that the gauging the identity line $W_0$ and the whole category $\mathcal C_{\rm LY}=\{W_0,\,W_2\}$ on the boundary are Morita equivalent. Therefore gauging the whole category gives back to the original Lee-Yang model.

Indeed, the duality in the Lee-Yang model can be verified on the level of the partition function in our setup. Gauging the $\mathcal C_{\rm LY}$, from the perspective of TDLs, is equivalent to paving the spacetime with a fine-enough meshed web of all lines in $\mathcal C_{\rm LY}$. In the case of a torus spacetime, after various $F$-moves, all non-trivial configurations of TDLs are given as follows:
\begin{align}
\begin{gathered}
\begin{tikzpicture}[scale=.75]
\node at (0, 1.2) {};
\draw[very thick] (0,0) rectangle (1.2,1.2);
\end{tikzpicture}
\end{gathered}
\,,\qquad
\begin{gathered}
\begin{tikzpicture}[scale=.75]
\node at (0, 1.2) {};
\draw[very thick] (0,0) rectangle (1.2,1.2);
\draw [line, thick] (0,.60) -- (1.2,.60); 
\end{tikzpicture}
\end{gathered}
\,,\qquad
\begin{gathered}
\begin{tikzpicture}[scale=.75]
\node at (0, 1.2) {};
\draw[very thick] (0,0) rectangle (1.2,1.2);
\draw [line, thick] (.60,0) -- (.60,1.20); 
\end{tikzpicture}
\end{gathered}
\,,\qquad
\begin{gathered}
\begin{tikzpicture}[scale=.75]
\node at (0, 1.2) {};
\draw[very thick] (0,0) rectangle (1.2,1.2);
\draw[thick]  (.6,0) arc (0:90:.6); 
\draw[thick] (.6,1.2) arc (-180:-90:.6);
\end{tikzpicture}
\end{gathered}
\,,\qquad
\begin{gathered}
\begin{tikzpicture}[scale=.75]
\node at (0, 1.2) {};
\draw[very thick] (0,0) rectangle (1.2,1.2);
\draw[thick]  (.6,0) arc (0:90:.6); 
\draw[thick] (.6,1.2) arc (-180:-90:.6);
\draw[line, thick] (.424,.424)--(.776,.776);
\end{tikzpicture}
\end{gathered}
\,.
\end{align} 
However, recall that the Drinfeld center $\mathcal Z$ is only four-dimensional. It implies that the five vectors associated with the above graphs must be \emph{linear dependent}. One can therefore have
\begin{align}
    |0\rangle_{\mathcal{L}}+\xi|1\rangle_{\mathcal{L}}+\xi|2\rangle_{\mathcal{L}}+\xi|3\rangle_{\mathcal{L}}+\xi^3|5\rangle_{\mathcal{L}}=0\,,
    \label{eq:linear_rel}
\end{align}
where we have normalized the coefficient of $\alpha_0$ to be unity. Further, notice that $\xi=\xi^2-1$. We multiply on both sides of \eqref{eq:linear_rel} by $\xi$, and recast it as
\begin{align}
|0\rangle_{\mathcal{L}}=\xi^2|0\rangle_{\mathcal{L}}+\xi^2|1\rangle_{\mathcal{L}}+\xi^2|2\rangle_{\mathcal{L}}+\xi^2|3\rangle_{\mathcal{L}}+\xi^4|5\rangle_{\mathcal{L}}\,,
\end{align}
or say diagrammatically,
\begin{align}
\begin{gathered}
\begin{tikzpicture}[scale=.75]
\node at (0, 1.2) {};
\draw[very thick] (0,0) rectangle (1.2,1.2);
\end{tikzpicture}
\end{gathered}
\quad=\quad \xi^2\,
\begin{gathered}
\begin{tikzpicture}[scale=.75]
\node at (0, 1.2) {};
\draw[very thick] (0,0) rectangle (1.2,1.2);
\end{tikzpicture}
\end{gathered}
\quad +\quad\xi^2\,
\begin{gathered}
\begin{tikzpicture}[scale=.75]
\node at (0, 1.2) {};
\draw[very thick] (0,0) rectangle (1.2,1.2);
\draw [line, thick] (0,.60) -- (1.2,.60); 
\end{tikzpicture}
\end{gathered}
\quad +\quad\xi^2\,
\begin{gathered}
\begin{tikzpicture}[scale=.75]
\node at (0, 1.2) {};
\draw[very thick] (0,0) rectangle (1.2,1.2);
\draw [line, thick] (.60,0) -- (.60,1.20); 
\end{tikzpicture}
\end{gathered}
\quad +\quad\xi^2\,
\begin{gathered}
\begin{tikzpicture}[scale=.75]
\node at (0, 1.2) {};
\draw[very thick] (0,0) rectangle (1.2,1.2);
\draw[thick]  (.6,0) arc (0:90:.6); 
\draw[thick] (.6,1.2) arc (-180:-90:.6);
\end{tikzpicture}
\end{gathered}
\quad +\quad\xi^4\,
\begin{gathered}
\begin{tikzpicture}[scale=.75]
\node at (0, 1.2) {};
\draw[very thick] (0,0) rectangle (1.2,1.2);
\draw[thick]  (.6,0) arc (0:90:.6); 
\draw[thick] (.6,1.2) arc (-180:-90:.6);
\draw[line, thick] (.424,.424)--(.776,.776);
\end{tikzpicture}
\end{gathered}
\,.
\label{eq:gauge_LY}
\end{align}
\eqref{eq:gauge_LY} is nothing but that gauging the full category $\mathcal C_{\rm LY}$ give back to the original Lee-Yang category. The coefficients in front of each graph on the RHS are known as the multiplication morphism of a trivalent junction $m\in{\rm Hom}_{\mathcal C_{\rm LY}}(A\otimes A,A)$, where $A$ is the to-be-gauged fusion category, here in our case $A=\mathcal C_{\rm LY}$ itself, see more details in \cite{Diatlyk:2023fwf}.



\subsection*{Acknowledgements}
 The work of QJ is supported by the National Research Foundation of Korea (NRF) Grant No. RS-2024-00405629 and Jang Young-Sil Fellow Program at the Korea Advanced Institute of Science and Technology. The work of JC is supported by the Fundamental Research Funds
for the Central Universities (No.20720230010) of China, and the National Natural Science
Foundation of China (Grants No. 12247103).
 
\appendix

\section{Modular data of 3d $\mathbb{Z}_N$ gauge theory}
In this appendix, we will calculate modular data of 3d $\mathbb{Z}_N$ gauge theory and construct the Hilbert space $\mathcal{H}(T^2)$ as an example. Let us consider the $2d$ $\mathbb{Z}_N$ symmetry with anomaly characterized by $H^3(\mathbb{Z}_N,U(1)) = \mathbb{Z}_N$, the SymTFT is the 3D $\mathbb{Z}_N$ gauge theory (Dijkgraaf-Witten theory)
    \begin{equation}
        S = \frac{N}{2\pi} \int \widetilde{A}\wedge d A  + \frac{2 k}{4\pi} \int A \wedge d A,
    \end{equation}
with $k=0,\cdots,N-1$. In particular, if $k=N$ we can eliminate the second term by shifting $\widetilde{A}\rightarrow \widetilde{A}-A$. The gauge transformations are
    \begin{equation}
        A \rightarrow A + d \Lambda,\quad \widetilde{A} \rightarrow \widetilde{A} + d \widetilde{\Lambda}
    \end{equation}
and we can consider the Wilson loop operators $W[\Gamma]$ and $\widetilde{W}[\Gamma]$ for $A$ and $\widetilde{A}$. The S-matrix elements between two kinds of line operators can be obtained by computing the correlation function
    \begin{equation}
        \langle W[\Gamma] \widetilde{W}[\Gamma'] \rangle = \int \mathcal{D}A \mathcal{D}\widetilde{A} e^{i \frac{N}{2\pi} \int \widetilde{A}\wedge d A + i \frac{2 k}{4\pi} \int A \wedge d A + i \int \eta(\Gamma) \wedge A + i \int \eta(\Gamma') \wedge \widetilde{A}},
    \end{equation}
where $\eta(\Gamma)$ is already defined in the last section. The phase on the exponent can be written as
    \begin{align}
         &\frac{N}{2\pi} \int \widetilde{A}\wedge d A +  \frac{k}{2\pi} \int A \wedge d A +\int \eta(\Gamma) \wedge A  + \int \eta(\Gamma') \wedge \widetilde{A} \nonumber\\
         =& \frac{N}{2\pi} \int \left(\widetilde{A} + \frac{2\pi}{N} \eta(D_{\Gamma}) - \frac{4\pi k}{N^2}\eta(D_{\Gamma'})\right)\wedge d \left(A + \frac{2\pi}{N} \eta(D_{\Gamma'}) \right) \nonumber\\
         &+  \frac{k}{2\pi} \int \left(A + \frac{2\pi}{N} \eta(D_{\Gamma'}) \right) \wedge d \left(A + \frac{2\pi}{N} \eta(D_{\Gamma'}) \right)  \nonumber \\
         &- \frac{2\pi}{N} \int \eta(D_{\Gamma}) \wedge d \eta(D_{\Gamma'})+\frac{2\pi k}{N^2} \int \eta(D_{\Gamma'}) \wedge d \eta(D_{\Gamma'})
    \end{align}
which indicates,
    \begin{equation}
        \langle W[\Gamma] \widetilde{W}[\Gamma'] \rangle = \langle W[\Gamma] \rangle \langle \widetilde{W}[\Gamma'] \rangle e^{-\frac{2\pi i}{N} \textrm{Link}(\Gamma,\Gamma')},
    \end{equation}
the second term gives the linking number between $\Gamma$ and $\Gamma'$. The spin of $\widetilde{W}[\Gamma]$ can be read from the last term as $e^{\frac{2\pi i k}{N^2}}$. Consider the insertion of two $\widetilde{W}$ Wilson loops as $\widetilde{W}[\Gamma]$ and $\widetilde{W}[\Gamma']$, one can similarly deduce that
    \begin{equation}
        \langle \widetilde{W}[\Gamma] \widetilde{W}[\Gamma'] \rangle = \langle \widetilde{W}[\Gamma] \rangle \langle \widetilde{W}[\Gamma'] \rangle e^{\frac{4\pi k i}{N^2} \textrm{Link}(\Gamma,\Gamma')}.      
    \end{equation}
Moreover, since $W^N$ link trivially with other line operators, one can identify
    \begin{equation}
    W^N[\Gamma] = 1.
    \end{equation}
On the other hand, $\widetilde{W}^N$ link trivially with $W$ operators but one has
    \begin{equation}
        \langle \widetilde{W}^N[\Gamma] \widetilde{W}[\Gamma'] \rangle = \langle \widetilde{W}^N[\Gamma] \rangle \langle \widetilde{W}[\Gamma'] \rangle e^{\frac{4\pi k i}{N} \textrm{Link}(\Gamma,\Gamma')},      
    \end{equation}
therefore one should identify
    \begin{equation}
        \widetilde{W}^N[\Gamma] = W^{-2k}[\Gamma].
    \end{equation}

Again, let us denote generic line operators using a pair of indices $\alpha,\tilde{\alpha} = 0,1,\cdots,N-1$ as
    \begin{equation}
        W_{(\alpha,\tilde{\alpha})} [\Gamma] = \exp \left(i \oint \alpha A + \tilde{\alpha} \widetilde{A} \right).
    \end{equation}
Each line operator has quantum dimension one so that the total quantum dimension is $N$. The elements of $S$ and $T$ matrices can be read from above as
    \begin{equation}
        S_{(\alpha,\tilde{\alpha}),(\beta,\tilde{\beta})} = \frac{1}{N}\omega^{-\alpha \tilde{\beta} - \tilde{\alpha} \beta + \frac{2k}{N} \tilde{\alpha} \tilde{\beta} },\quad T_{(\alpha,\tilde{\alpha}),(\beta,\tilde{\beta})} = \omega^{-\alpha \tilde{\alpha} + \frac{k}{N} \tilde{\alpha}^2} \delta_{\alpha,\beta} \delta_{\tilde{\alpha},\tilde{\beta}}
    \end{equation}
with $\omega = e^{\frac{2\pi i}{N}}$. The fusion coefficient is obtained via \eqref{Fusion-coefficient} as
    \begin{equation}
        \begin{split}
            N_{(\alpha,\tilde{\alpha}),(\beta,\tilde{\beta})}^{(\gamma,\tilde{\gamma})} =& \sum_{\sigma,\tilde{\sigma}} \frac{S_{(\sigma,\tilde{\sigma}),(\alpha,\tilde{\alpha})}^* S_{(\sigma,\tilde{\sigma}),(\beta,\tilde{\beta})}^* S_{(\sigma,\tilde{\sigma}),(\gamma,\tilde{\gamma})}}{S_{(0,0),(\sigma,\tilde{\sigma})}} \\
            =& \frac{1}{N^2} \sum_{\sigma,\tilde{\sigma}} \omega^{\sigma \tilde{\alpha} + \tilde{\sigma} \alpha - \frac{2k}{N}\tilde{\sigma}\tilde{\alpha}} \omega^{\sigma \tilde{\beta} + \tilde{\sigma} \beta - \frac{2k}{N}\tilde{\sigma}\tilde{\beta}} \omega^{-\sigma \tilde{\gamma} - \tilde{\sigma} \gamma + \frac{2k}{N}\tilde{\sigma}\tilde{\gamma}}\\
            =&\frac{1}{N} \sum_{\tilde{\sigma}} \delta_{\tilde{\alpha}+\tilde{\beta}-\tilde{\gamma},0} \omega^{ \tilde{\sigma} (\alpha+\beta-\gamma) - \frac{2k}{N}\tilde{\sigma}(\tilde{\alpha}+\tilde{\beta}-\tilde{\gamma})}
        \end{split}
    \end{equation}
where the delta function is defined mod $N$ such that
    \begin{equation}
        \delta_{k,0} = \left\{ \begin{array}{l}
            1 \quad k = 0\ \textrm{mod}\ N\\
            0 \quad \textrm{others}
        \end{array}\right. .
    \end{equation}
Notice that $-N<\tilde{\alpha}+\tilde{\beta}-\tilde{\gamma}<2N-1$, we can further sum over $\tilde{\sigma}$ and get
    \begin{equation}
        N_{(\alpha,\tilde{\alpha}),(\beta,\tilde{\beta})}^{(\gamma,\tilde{\gamma})} = \delta_{\gamma,\alpha+\beta - 2k \left[\frac{\tilde{\alpha} + \tilde{\beta}}{N}\right]} \delta_{\tilde{\gamma},\tilde{\alpha}+\tilde{\beta}},
    \end{equation}
where $[\cdots]$ is the floor function. The Hilbert space $\mathcal{H}(T^2)$ is spanned by $|(\alpha,\tilde{\alpha})\rangle$ and they satisfy
    \begin{equation}\label{ZN-states}
        \begin{split}
        W_{(\beta,\tilde{\beta})}[\Gamma_1] |(\alpha,\tilde{\alpha})\rangle =& \omega^{-\beta \tilde{\alpha} - \tilde{\beta} \alpha + \frac{2k}{N} \tilde{\alpha} \tilde{\beta}} |(\alpha,\tilde{\alpha})\rangle,\\
        W_{(\beta,\tilde{\beta})}[\Gamma_2] |(\alpha,\tilde{\alpha})\rangle =& |(\alpha+\beta-2k \left[\frac{\tilde{\alpha} + \tilde{\beta}}{N}\right],\tilde{\alpha}+\tilde{\beta})\rangle.
        \end{split}
    \end{equation}
Moreover, the state $|(\alpha,\tilde{\alpha})\rangle$ satisfies
    \begin{equation}
        |(\alpha+N,\tilde{\alpha})\rangle = |(\alpha,\tilde{\alpha})\rangle,\quad |(\alpha,\tilde{\alpha}+N)\rangle = |(\alpha-2k,\tilde{\alpha})\rangle
    \end{equation}
due to the fusion rule.

\bibliographystyle{JHEP}

\newpage
\bibliography{refs.bib}

\end{document}